\documentclass{jpp}
\usepackage{graphicx}
\usepackage{hyperref}
\usepackage{tikz}

\usepackage[utf8]{inputenc}
\usepackage[T1]{fontenc}
\usepackage{amsmath}
\usepackage{amssymb}
\usepackage{caption}
\usepackage{subcaption}
\usepackage{booktabs}
\usepackage{float}

\def\Cth{C_{\text{th}}}

\hypersetup{
    colorlinks=true,
    linkcolor=blue,
    urlcolor=blue,
    citecolor=blue}

    \shorttitle{Two-phase structure of cooling relativistic plasma}
\shortauthor{A. Wierzchucka et al.}

\title{Two-Phase Structure of Synchrotron-Cooling-Unstable Relativistic Plasma}

\author{A. Wierzchucka\aff{1,2}
  \corresp{\email{agnieszka.wierzchucka@merton.ox.ac.uk}},
  P. J. Bilbao\aff{1,3}, R. J. Ewart\aff{4}, D. A. Uzdensky\aff{1} \and A. A. Schekochihin\aff{1,2} }

\affiliation{\aff{1}Rudolf Peierls Centre for Theoretical Physics, University of Oxford, Oxford OX1 3PU, UK
\aff{2}Merton College, Oxford, OX1 4JD, UK
\aff{3}Lady Margaret Hall, Oxford, OX2 6QA, UK
\aff{4}Kavli Institute for Particle Astrophysics and Cosmology, Stanford University, Stanford CA, 94305, USA}

\newcommand{\B}[1]{{\mathbf{#1}}}

\begin{document}

\maketitle

\begin{abstract}
Using analytic theory, radiative particle-in-cell (PIC) simulations, and fluid simulations, we show that relativistic, synchrotron-cooling, collisionless, high-$\beta$ pair plasmas filament into a two-phase medium. This process occurs through the interplay of the synchrotron cooling instability (SCI) with the synchrotron firehose instability (SFHI). One phase has high plasma~$\beta$ and is infested with small-scale firehose fluctuations, which scatter particles and pin the pressure anisotropy to the firehose-marginal level. The other phase has much lower~$\beta$, causing the suppression of firehose modes and thus allowing large pressure anisotropies. We propose a fluid model for this two-phase plasma, which we use to study the linear and nonlinear evolution of the SCI and SFHI, and to predict the emergence time of the two-phase structure. 
\end{abstract}

\section{Introduction}
Plasmas in high-energy astrophysical environments are typically threaded by magnetic fields. Particles accelerated by strong electromagnetic fields emit synchrotron radiation, experiencing a radiation-reaction force and consequently losing energy. Synchrotron emission significantly affects plasmas in extreme relativistic astrophysical environments, such as pulsar magnetospheres \citep{Cerutti_Beloborodov-2017,Philippov_Kramer-2022}, pulsar-wind nebulae \citep{Rees_Gunn-1974, Kennel_Coroniti-1984b, Zanna_etal-2006, Kargaltsev_etal-2015}, magnetar magnetospheres \citep{Kaspi_Beloborodov-2017}, black-hole accretion flows \citep{Rees_etal-1982, Rees-1984, Narayan_Yi-1995, Quataert-2003, Yuan_Narayan-2014}, black-hole jets  \citep{Blandford_Konigl-1979, Begelman_etal-1984, Blandford_etal-2019}, and gamma-ray bursts \citep{Daigne_Mochkovitch-1998, Medvedev_Loeb-1999, Medvedev-2000, Mckinney_Uzdensky-2012}, as well as their afterglows \citep{Sari_etal-1998, Granot_Sari-2002, Piran-2005}. In particular, synchrotron losses in strongly magnetised relativistic plasmas can significantly modify the behaviour of key astrophysically important nonlinear collective plasma processes like magnetic reconnection  \citep{Lyubarskii-1996, Jaroschek_Hoshino-2009, Uzdensky_etal-2011, Cerutti_etal-2013, Uzdensky_Spitkovsky-2013, Uzdensky-2016, Cerutti_etal-2016, Hakobyan_etal-2019, Sironi_etal-2025}, turbulence \citep{Uzdensky-2018, Comisso_etal-2020, Comisso_Sironi-2021, Nattila_Beloborodov-2021, Lemoine_etal-2025}, and shocks \citep{Biermann_Strittmatter-1987, Hoshino_etal-1992, Sironi_Spitkovsky-2009, Kirk_Reville-2010, Sironi_etal-2013}. Laboratory plasma experiments are now capable of generating strong enough electromagnetic fields for the effects of radiation reaction to be directly observed \citep{Los_etal-2026}, motivating the study of basic physical processes in such regimes.

Many high-energy astrophysical plasma environments are rather dilute and therefore collisionless, as far as binary Coulomb collisions are concerned. Consequently, their distribution functions can be anisotropic, with different pressures parallel ($P_\parallel$) and perpendicular ($P_\perp$) to the magnetic field \citep{Kulsrud-1964}. Both negative and positive pressure anisotropies, $\Delta \equiv P_\perp / P_\parallel - 1$, occur naturally in such plasmas as a result of large-scale shearing or compressing motions, both in the relativistic and non-relativistic limits \citep{Chew_etal-1956, Wierzchucka_etal-2026, Ley_etal-2026}. However, in high-$\beta$ plasmas, where~$\beta$ is the ratio of the thermal to magnetic pressure, pressure anisotropies cannot grow without bound and are checked by kinetic processes, such as the firehose, ion-cyclotron, and mirror instabilities \citep{Chandrasekhar_etal-1958, Hasegawa-1969, Southwood_Kivelson-1993, Bott_etal-2024}. These generally arise when $|\Delta| \gtrsim 1/\beta$. Once small-scale magnetic-field fluctuations generated by anisotropy-driven instabilities reach sufficiently high levels, they begin to scatter particles, thereby pinning the pressure anisotropy to the respective instability's threshold, an effect observed \textit{in situ} by spacecraft in the solar wind \citep{Kasper_etal-2002, Bale_etal-2009, Chen_etal-2016} and easily reproduced in numerical simulations (e.g., \citealt{Kunz_etal-2014, Melville_etal-2016, Bott_etal-2025}). Thus, high-$\beta$ plasmas are often close to isotropy, while low-$\beta$ plasmas can sustain large pressure anisotropies.

In relativistic plasmas, anisotropic distributions can arise naturally as a direct result of synchrotron emission, which preferentially cools particles with large pitch angles (the angle between their velocity and the local magnetic-field direction), generating a negative pressure anisotropy ($\Delta < 0$). In a high-$\beta$ plasma, this anisotropy can become sufficiently strong to excite the firehose instability, in this context known as the synchrotron firehose instability (SFHI). The SFHI onset was studied by \cite{Zhdankin_etal-2023}, who also showed that, as expected, once the magnetic fluctuations generated by the SFHI become sufficiently strong, they scatter particles and drive the system back toward marginal firehose stability, $\Delta \sim -1/\beta$; continued cooling then re-establishes the anisotropy, rendering the plasma firehose-unstable again and repeating the cycle. The result is a slowly cooling plasma that remains close to pressure isotropy on timescales long compared to the firehose growth time. 

Relativistically hot plasmas undergoing synchrotron emission can also develop a population inversion with respect to perpendicular momentum, resulting in the formation of a ring structure in momentum space \citep{Bilbao_Silva-2023, Bilbao_etal-2024}. In low-$\beta$ environments, the ring momentum distribution can drive the electron-cyclotron maser instability (ECMI), which results in the generation of coherent radiation \citep{Bilbao_Silva-2023, Bilbao_etal-2025}. As the overall effect of synchrotron emission is to decrease the thermal pressure, plasmas with sufficiently high initial $\beta$ may experience the SFHI followed by the ECMI, as they transition from higher to lower $\beta$ \citep{Bilbao_etal-2024}. 

Synchrotron emission can also destabilise the plasma on larger, fluid scales. The synchrotron cooling instability (SCI) is excited in plasmas with $\beta$ larger than a certain critical value of order unity \citep{Simon_Axford-1967}. It corresponds to a radiation-cooling-destabilised entropy mode and thus represents the synchrotron-cooling analogue of the classic thermal instability \citep{Field-1965}. The SCI is triggered when compression generates a positive perturbation of the magnetic-field strength, creating a local region that cools more rapidly than its surroundings, driving a total pressure imbalance. In high-$\beta$ plasmas, the pressure-balanced nature of the instability means that the cooling-induced temperature decrease must then be compensated by an increase in density. The resulting compression leads to an increase in the magnetic-field strength via flux freezing, which in turn enhances synchrotron cooling, thereby establishing a positive feedback loop. On longer timescales, as the instability enters the nonlinear stage, $\beta$ decreases inside the localised perturbation. The total pressure then gains a significant contribution from the magnetic pressure, which stabilises the SCI at finite~$\beta$.  In this way, the SCI may lead to the formation of filamentary low-$\beta$ structures aligned with the magnetic field \citep{Eilek_Caroff_1979}. If there is no heating mechanism present, the perturbation eventually decays and the plasma returns to a homogeneous state \citep{Bodo_etal-1992}.

The original derivation of the (linear) SCI by \cite{Simon_Axford-1967} was limited to plasmas in which ultra-relativistic electrons provided the pressure, while much colder, heavier ions supplied the inertia. Since then, the SCI has been studied, primarily analytically, in a range of settings, including the effects of finite heat conduction and diffusion, which limit the smallest scale of the SCI \citep{Eilek_Caroff_1979, Rosso_Pelletier_1993}; power-law equilibrium distribution functions \citep{De_Gouveia_Dal_Pino_Reuven-1993}; and different plasma-heating mechanisms \citep{Bodo_etal-1990}. 
Like the original work, these studies adopted a fluid approach, focusing on plasmas composed of cold ions and hot electrons. 

In this paper, by building upon the foundational results of \cite{Simon_Axford-1967} and \cite{Zhdankin_etal-2023}, we present a comprehensive picture of a collisionless, relativistically hot, synchrotron-radiating plasma and show that such a plasma spontaneously filaments into a two-phase medium. We study both the linear development and the nonlinear interplay of the SCI  and the~SFHI, utilising analytic theory and numerical simulations. As a high-$\beta$ synchrotron-emitting plasma evolves, the SFHI is quickly excited, rendering the medium effectively collisional. The SCI grows on longer timescales  (comparable to the cooling time) in this firehose-infested environment. As the SCI develops, it causes the plasma to split into two phases. One phase has a high temperature and~$\beta$, while the other is cold and more strongly magnetised, with relatively low $\beta \sim 1$. The former is micro-turbulent and infested with firehose fluctuations, which pin the pressure anisotropy to the firehose threshold. The low $\beta$ of the latter results in the quenching of the SFHI, enabling a significant pressure anisotropy to build up. While these two phases have vastly different physical properties, they are approximately in pressure balance with each other. Once $\beta$ inside the laminar region becomes sufficiently small, the SCI is extinguished, and the plasma eventually returns to a homogeneous state, on a timescale much longer, by a factor of the initial~$\beta$, than the initial cooling time.

The structure of the paper is as follows. In Section~\ref{sec: Relativistic, Cooling Plasmas}, we derive a fluid description for collisionless, relativistically hot, pair plasmas undergoing optically thin synchrotron emission. In Section~\ref{sec: Synchrotron Firehose Instability}, we discuss how synchrotron emission can give rise to a negative pressure anisotropy, which excites the SFHI, and propose a simple model for a pair plasma infested with microscale firehose fluctuations. We use this model to predict the time and value of $\beta$ at which the SFHI will be shut off. In Section~\ref{sec: Synchrotron Cooling Instability}, we show that the SCI is excited in synchrotron-cooling plasmas and derive its linear evolution. In Section~\ref{sec: Particle-in-Cell Simulations}, we demonstrate the coexistence of the SCI and SFHI using PIC simulations. In Section~\ref{sec: Emergent Two-Phase Structure of Synchrotron Cooling Plasmas}, we discuss how the interplay between the two instabilities can lead to the development of a two-phase structure and propose an effective fluid model for this composite medium, in which the two phases are governed by different local equations of state. We then solve our two-phase fluid model numerically to study the full evolution of the SCI within this emergent structure. In Section \ref{sec: summary}, we summarise the emergent picture of synchrotron-cooling, collisionless plasmas. Finally, in Section \ref{sec: Discussion}, we place our results in the context of brightly glowing filamentary structures observed in the intracluster medium (ICM) \citep{Rudnick_etal-2022, Rajpurohit_etal-2022} and the Galactic Centre \citep{Yusef-Zadeh_etal-1984, Yusef-Zadeh_etal-2004, Yusef-Zadeh_etal-2022a} and briefly touch upon other astrophysical applications. We highlight the limitations of our theory and outline directions of future research in the same section. 

\section{Synchrotron-Cooling Collisionless Plasmas}
\label{sec: Relativistic, Cooling Plasmas}
We consider a species $s$ of collisionless particles with mass $m_s$ and charge $q_s$, undergoing classical synchrotron emission in an optically thin medium. The corresponding distribution function $f_s(t, \mathbf{x}, \mathbf{p})$, a Lorentz scalar, satisfies the Vlasov equation, which, written in a non-manifestly covariant form in the lab frame, is \citep{Hakim_Mangeney-1968, Hazeltine_Mahajan-2004}
\begin{equation}
    \label{eq: Vlasov + Rad}
    \frac{\partial f_s}{\partial t} + \mathbf{v} \cdot \nabla f_s + q_s \left(\mathbf{E} + \frac{\mathbf{v}}{c} \times \mathbf{B}\right)\cdot \frac{\partial f_s}{\partial \mathbf{p}}  + \frac{\partial}{\partial \mathbf{p}} \cdot (\mathbf{F}_{\text{rad}, s} f_s) = 0.
\end{equation}
The particle momentum $\mathbf{p}$ and velocity $\mathbf{v}$ are related by $\mathbf{p} = \gamma m_s \mathbf{v}$, where {${\gamma = \sqrt{1+(p/m_s c)^2}}$} is the Lorentz factor. The electric and magnetic fields are $\mathbf{E}$ and $\mathbf{B}$, and satisfy Maxwell's equations. Classical synchrotron emission is captured by the \cite{Landau_Lifshitz-1975} radiation-reaction force,
\begin{align}
\label{eq: rad react full}
&\qquad \qquad \qquad\mathbf{F}_{\text{rad}, s}
=
\frac{2q_s^3 \gamma}{3m_sc^3}\,
\left[
\left(\frac{\partial}{\partial t}+\mathbf v\cdot\nabla\right)\mathbf E
+\frac{\mathbf v}{c}\times
\left(\frac{\partial}{\partial t}+\mathbf v\cdot\nabla\right)\mathbf B
\right] \nonumber  + \\
&\frac{2q_s^4}{3m_s^2c^4}\!
\left \{
\mathbf E\times\mathbf B
+\frac{1}{c}\mathbf B\times(\mathbf B\times\mathbf v)
+\frac{1}{c}\mathbf E\,(\mathbf v\cdot\mathbf E) 
-
\gamma^2\left[
\left(\mathbf E+\frac{\mathbf v}{c}\times\mathbf B\right)^2
-\left(\frac{\mathbf v}{c}\cdot\mathbf E\right)^2
\right] \frac{\mathbf v }{c}
\right\}.
\end{align}

The full radiative kinetic system~\eqref{eq: Vlasov + Rad}-\eqref{eq: rad react full} is often difficult to study analytically and expensive to simulate numerically. In what follows, by taking moments of 
\eqref{eq: Vlasov + Rad} and making several simplifying assumptions, we derive a fluid model of an ultra-relativistically hot, collisionless, magnetised plasma, with a non-relativistic bulk flow, undergoing slow synchrotron emission. Despite our fluid model lacking kinetic-scale physics, in Section \ref{sec: Particle-in-Cell Simulations}, by means of kinetic simulations, we show that it captures the dynamics of synchrotron-cooling plasmas accurately. 

\subsection{Fluid Moments and Peculiar Variables}
The zeroth moment of~\eqref{eq: Vlasov + Rad} gives the continuity equation,
\begin{equation}
    \label{eq: Non-Covariant Continuty}
    \frac{\partial n_s}{\partial t} + \nabla \cdot (n_s \mathbf{u}_s) = 0,
\end{equation}
where
\begin{equation}
    n_s \equiv \int d \B{p} \  f_s , \quad \B{u}_s \equiv \frac{1}{n_s}\int d \B{p} \ \mathbf{v} f_s
\end{equation}
are the density and \cite{Eckart-1940}~bulk velocity, respectively, of the particles of species~$s$. Taking the first moment of~\eqref{eq: Vlasov + Rad}, one obtains the fluid momentum equation
\begin{equation}
    \label{eq: Momentum eq}
    \frac{\partial (n_s \B{p}_s )}{\partial t} + \nabla \cdot (n_s \mathbf{u}_s \B{p}_s ) = - \nabla \cdot \B{P}_s + q_s n_s \left( \B{E} + \frac{\B{u}_s\times \B{B}}{c} \right) + \bar{\B{F}}_{\text{rad}, s}.
\end{equation}
Here we have defined the average particle momentum, pressure tensor, and radiation-reaction force density (i.e., the radiation drag force per unit volume) as, respectively, 
\begin{align}
    \label{eq: moment definitions}
    \B{p}_s \equiv \frac{1}{n_s}\int d\B{p} \ \B{p} f_s, \quad \B{P}_s \equiv \int d\B{p} \ (\B{v} - \B{u}_s) (\B{p} - \B{p}_s) f_s, \quad \bar{\B{F}}_{\text{rad}, s} \equiv \int d\B{p} \  \B{F}_{\text{rad}, s} f_s.
\end{align}

The nonlinear relationship between the particle momentum and velocity, $\B{p} = \gamma m_s\B{v}$, implies that $\mathbf{p}_s$ is, in general, not parallel to~$\B{u}_s$. Therefore, in the frame moving with the bulk velocity~$\mathbf{u}_s$ the plasma does not necessarily have zero average momentum, reflecting the subtlety of defining the comoving frame in the relativistic case according to \cite{Landau_Lifshitz-1987} or \cite{Eckart-1940} recipes. We adopt the latter, i.e., we define our comoving frame to move with the bulk velocity~$\B{u}_s$. Here and in what follows, we label the quantities in this frame with an overtilde and refer to them as `peculiar'. 

To simplify the relationship between peculiar and laboratory-frame momentum significantly, we shall from now on assume that the bulk flow is non-relativistic, 
\begin{equation}
    \frac{u_s}{c} \ll 1.
\end{equation}
Then, to first order in $u_s/ c$, the Lorentz transformation between the laboratory-frame and peculiar momenta is
\begin{equation}
    \label{eq: Lorentz transform p}
    \B{p} = \tilde{\gamma}m_s\B{u}_s + \tilde{\B{p}},
\end{equation}
whereas for the velocities it is
\begin{equation}
    \label{eq: Lorentz transform v}
    \B{v} = \B{u}_s + \tilde{\B{v}}\left(1 - \frac{\mathbf{u}_s \cdot \mathbf{\tilde{v}}}{c^2} \right).
\end{equation}

\subsection{Gyrotropy and Parity}
\label{sec: Gyrotropy and Parity}
In highly magnetised plasmas, the particle gyration frequency $\Omega_s$ provides the shortest timescale, greatly exceeding the dynamical, cooling, and collision rates. Therefore, in analogy to non-relativistic kinetic MHD theory \citep{Kulsrud-1964}, we impose the following ordering\footnote{It can be shown that the condition $\tau_s \Omega_s \gg1 $ holds for relativistic particles in most astrophysical systems \citep{Uzdensky-2016}. In the case of ultra-relativistic electrons $\tau_e \Omega_e \sim \theta_e^{-2} {(B/B_\text{cl})}^{-1}$, where $B_\text{cl} \equiv e/r_e^2 \simeq 6 \times 10^{15} \, \rm{G}$ is the critical classical magnetic field, $\theta_e \equiv T_e/m_e c^2$, and $T_e$ is the electron temperature. As $B/B_\text{cl}$ is usually tiny (e.g., for a typical interstellar medium field of $10^{-6}\, \rm{G}$, $B/B_\text{cl} \sim 10^{-22}$), the condition $\tau_s \Omega_s \gg1 $ is easily satisfied even for highly relativistic particles.}
\begin{equation}
    \label{eq: ordering}
    \tau_s^{-1} \sim \omega \sim k u_s \ll \Omega_s,
\end{equation}
where $\omega$ and $\mathbf{k}$ are the characteristic frequency and wavevector of the plasma motions, and $\tau_s$ is the cooling timescale associated with synchrotron emission. We also assume 
\begin{equation}
    \label{eq: rho ordering}
    k \rho_s \ll 1,
\end{equation}
where $\rho_s \equiv v_\text{th,s}/\Omega_s$ is the Larmor radius, $v_\text{th,s}$ being the thermal speed defined as the width of the velocity distribution.

Under the orderings~\eqref{eq: ordering} and~\eqref{eq: rho ordering}, the momentum equation~\eqref{eq: Momentum eq} at zeroth order in~$\omega/\Omega_s$ and $k \rho_s$ is simply the ideal-MHD condition,
\begin{equation}
    \label{eq: ideal E}
    \B{E} = - \frac{\B{u}_s\times \B{B}}{c},
\end{equation} 
implying that the perpendicular bulk velocity is the same for all species:
\begin{equation}
    \label{eq: drift velocity}
    \mathbf{u}_{\perp s} = \mathbf{u}_\perp = c\frac{\mathbf{E} \times \mathbf{B}}{B^2}.
\end{equation}
The assumption of a non-relativistic bulk flow paired with~\eqref{eq: ideal E} implies that $|\B{E}| \ll |\B{B}|$. Thus, the magnetic field in the laboratory frame is, to lowest order in $u_s/c$, equal to that in the comoving frame, $\mathbf{B} = \tilde{\mathbf{B}}$. 

At the same order in $\omega/\Omega_s$ and $k \rho_s$, the kinetic equation~\eqref{eq: Vlasov + Rad} becomes
\begin{equation}
    \label{eq: lowest vlasov}
    (\tilde{\mathbf{p}} \times \mathbf{B})\cdot \frac{\partial f_s}{\partial \tilde{\mathbf{p}}} = 0.
\end{equation}
To interpret~\eqref{eq: lowest vlasov}, it is useful to adopt a cylindrical coordinate system for the peculiar momentum with the polar ($z$) axis parallel to $\mathbf{b} \equiv \mathbf{B}/B$:
\begin{equation}
    \tilde{\mathbf{p}} = \tilde{p}_\parallel \mathbf{b} + \tilde{p}_\perp( \hat{\mathbf{x}}\cos \phi + \hat{\mathbf{y}}\sin\phi).
\end{equation}
In these new variables,~\eqref{eq: lowest vlasov} reduces to
\begin{equation}
    \frac{\partial f_s}{\partial \phi} = 0,
\end{equation}
which implies gyrotropy of the distribution function in the comoving frame, viz.,
\begin{equation}
    \label{eq: gyrotropy}
     f_s = f_s(t, \mathbf{x}, \tilde{p}_\parallel, \tilde{p}_\perp).
\end{equation}

Next, to ignore all non-ideal effects apart from synchrotron cooling, e.g., any dynamics that arise as a result of parallel heat fluxes, we assume parity symmetry with respect to the parallel peculiar momentum \citep{Wierzchucka_etal-2026}, 
\begin{equation}
    \label{eq: parity}
    f_s(t, \mathbf{x}, \tilde{p}_\parallel, \tilde{p}_\perp) =  f_s(t, \mathbf{x}, -\tilde{p}_\parallel, \tilde{p}_\perp).
\end{equation}
Combined with~\eqref{eq: moment definitions},~\eqref{eq: Lorentz transform p}--\eqref{eq: Lorentz transform v}, and~\eqref{eq: gyrotropy}, this implies that, to lowest order in~$u_s/c$, the pressure tensor is diagonal:
\begin{equation}
    \label{eq: pressure gyro}
    \B{P}_s = P_{\perp s}( \mathbf{I} - \mathbf{b}\mathbf{b}) + P_{\parallel s}\mathbf{b}\mathbf{b},
\end{equation}
where the perpendicular and parallel pressures are respectively given by
\begin{equation}
    \label{eq: perp and parallel pressure def}
    P_{\perp s}= \frac{1}{2m_s}\int d\tilde{\mathbf{p}} \ \frac{\tilde{p}_\perp^2}{\tilde{\gamma}}f_s,  \quad P_{\parallel s} = \frac{1}{m_s}\int d\tilde{\mathbf{p}}  \ \frac{\tilde{p}_\parallel^2}{\tilde{\gamma}}f_s.
\end{equation}
It is also useful to define the average pressure, $P_s \equiv (2P_{\perp s} +  P_{\parallel s})/3 \equiv n_s T_s$, where $T_s \equiv \theta_s m_s c^2$ is the temperature and $\theta_s$ the dimensionless temperature. Note that, in contrast to the non-relativistic regime, where gyrotropy~\eqref{eq: gyrotropy} is sufficient for the pressure tensor to reduce to the form~\eqref{eq: pressure gyro}, in the relativistic limit $\mathbf{P}_s$ includes extra terms representing energy diffusion, which arise as a result of working in the Eckart instead of Landau frame, and disappear under the parity-symmetry assumption \citep{Gedalin_Oiberman-1995, Most_etal-2022, Wierzchucka_etal-2026b}. 

We will restrict our analysis to plasmas that are ultra-relativistically hot, i.e., $\tilde{p} \gg m_s c$, and so $T_s \gg m_s c^2$. In such plasmas, the characteristic gyration frequency is given by $\Omega_s = |q_s|B c/ 3 T_s$ and the thermal velocity becomes $v_\text{th,s} = c$. To ensure that finite-$u_s/c$ effects become important before first-order $k\rho_s$ corrections, we assume the ordering 
\begin{equation}
    \label{eq: small gyroradius}
    k \rho_s \ll \frac{u_s}{c} \ll 1,
\end{equation}
which, recalling~\eqref{eq: ordering}, is equivalent to
\begin{equation}
    \label{eq: small param}
    \frac{\omega}{\Omega_s} \ll \frac{u_s^2}{c^2}.
\end{equation}
The condition of ultra-relativistic temperatures results in several simplifications in our system. First, combining this assumption with the parity assumption~\eqref{eq: parity} allows the average momentum $\mathbf{p}_s$ to be evaluated from the distribution function in terms of $\mathbf{u}_s$, $n_s$, and the pressures~\eqref{eq: perp and parallel pressure def}, resulting in the following simple relationship:
\begin{equation}   
    \label{eq: momentum value}
    \mathbf{p}_s =  \frac{(3P_{\perp s} + P_{\parallel s})} {n_sc^2} {\mathbf{u}_s}
    + \frac{(P_{\parallel s} - P_{\perp s})} {n_sc^2} ({\mathbf{u}_s} \cdot \mathbf{b}) \mathbf{b},
\end{equation}
where we have kept terms to lowest order in $u_s/c$. Secondly, the assumption of ultra-relativistic temperatures greatly simplifies the form of the radiation-reaction force (Section \ref{sec: Radiation-Reaction Force in Relativistically Hot, Magnetised Plasmas}) and, thirdly, it allows us to express the pressure evolution in terms of generalised adiabatic indices (Section~\ref{sec: Pressure Evolution}).

\subsection{Radiation-Reaction Force}
\label{sec: Radiation-Reaction Force in Relativistically Hot, Magnetised Plasmas}
When the particles are ultra-relativistic and $\omega \ll \Omega_s$, the radiation-reaction force~\eqref{eq: rad react full} takes a much simpler form,
\begin{equation}
    \label{eq: F_rad intermediate}
    \mathbf{F}_\text{rad,s} = - \frac{2}{3} r_s ^2 \gamma^2\left [ \left(\mathbf{E} + \frac{\mathbf{v}}{c} \times \mathbf{B} \right )^2 - \left( \frac{\mathbf{v} \cdot \mathbf{E}}{c} \right)^2\right] \frac{\mathbf{v}}{c},
\end{equation}
where $r_s \equiv q_s^2/m_s c^2$ is the classical radius of the species $s$. This expression can be further simplified by recalling the ideal Ohm's law~\eqref{eq: ideal E}. 
Hence, to first order in $u_s/c$ and zeroth order in $\omega/\Omega_s$, one obtains the familiar formula for the classical synchrotron radiation reaction:
\begin{equation}
    \label{eq: Rad Force}
    \mathbf{F}_\text{rad,s} = - \frac{2}{3} {r_s^2 B^2 \tilde{\gamma}^2} \sin ^2\theta \ \frac{\mathbf{v}}{c} = 
    - \sigma_{T s } \, \frac{B^2}{4\pi c}\, \tilde{\gamma}^2 \sin ^2\theta \, {\mathbf{v}}, 
\end{equation}
where $\theta = \arcsin(\tilde{p}_\perp/ \tilde{p})$ is the particle's comoving pitch angle, $\sigma_{Ts} \equiv 8\pi r_s^2/3 $ is the Thomson scattering cross-section, and $\mathbf{v}$ is related to the peculiar velocity by~\eqref{eq: Lorentz transform v}. Alternatively,~\eqref{eq: Rad Force} can be expressed as 
\begin{equation}
    \mathbf{F}_\text{rad,s} = 
    -\frac{\tilde{p}^2_\perp}{3 T_s\tau_s } \, {\mathbf{v}} 
\end{equation}
in terms of the characteristic cooling timescale \citep{Zhdankin_etal-2023},
\begin{equation}
    \label{eq: cooling time}
    \tau_s \equiv \frac{m_s^2 c^3}{2 r_s^2 B^2 T_s },
\end{equation}
defined as the cooling time of a particle with $\theta = \pi/2$ and $\tilde{\gamma} = 3T_s/m_s c^2$\footnote{Note that the classical model of synchrotron emission employed here holds in the limit of~$\chi_\text{QED} \ll1$, where $\chi_\text{QED}$ is the invariant quantum parameter \citep{DiPiazza_etal-2012}. For an ultra-relativistic electron~$\chi_\text{QED} = p_\perp B/B_{\rm QED} m_e c\approx (\alpha_\text{fs}  \tau_e\Omega_e \theta_e)^{-1}$, where $B_{\rm QED} = m_e^2 c^3/e \hbar$ is the Schwinger critical field and $\alpha_\text{fs}$ is the fine-structure constant. Thus, for a slowly cooling ($\tau_e \Omega_e \gg 1$), ultra-relativistically hot plasma, use of the classical radiation force~\eqref{eq: rad react full} is justified.}.

The dependence of~\eqref{eq: Rad Force} on $\theta$ highlights a key difference between synchrotron emission and other radiation processes, like inverse Compton scattering: as synchrotron cooling reduces the energy of the system, it primarily cools particles with large $\theta$, generating negative pressure anisotropy \citep{Zhdankin_etal-2023, Bilbao_Silva-2023}. We will return to this property of synchrotron emission in Section~\ref{sec: Synchrotron Firehose Instability}.

Using~\eqref{eq: parity} and~\eqref{eq: Rad Force}, we can now easily calculate the radiation-reaction force density defined in~\eqref{eq: moment definitions}:
\begin{equation}
    \label{eq: rad moment}
    \bar{\mathbf{F}}_\text{rad,s} = - \frac{2 r_s^2 B^2 }{3 m_s^2 c^2}\frac{\mathbf{u}_s}{c}  \int d\tilde{\mathbf{p}}\  \tilde{p}_\perp^2 f_s.
\end{equation}
Strictly speaking, this quantity cannot be written rigorously in terms of any previously defined fluid moments. To do this approximately, we first consider how synchrotron cooling affects a plasma that has a drifting Maxwell--Jüttner distribution, which in the ultra-relativistic limit considered here takes the form
\begin{equation}
    \label{eq: MJ}
    f_s(\B{p}) = \frac{n_s}{8 \pi \theta^3_s m_s^3 c^3} e^{-\tilde{p}/\theta_s m_s c}.
\end{equation}
Evaluating~\eqref{eq: rad moment} then gives
\begin{equation}
    \bar{\mathbf{F}}_\text{rad,s} = - \frac{16 r_s^2}{3 m_s^2 c^4} \frac{\mathbf{u}_s}{c} \frac{B^2 P_s^2}{n_s}.
\end{equation}
For a general distribution function,~\eqref{eq: rad moment} can be written in the form
\begin{equation}
    \label{eq: F rad final}
    \bar{\mathbf{F}}_\text{rad,s} = - \alpha_m \frac{2 r_s^2}{ m_s^2 c^4} \frac{\mathbf{u}_s}{c} \frac{B^2 P_{\perp s}^2}{n_s},
\end{equation}
where the dimensionless parameter 
\begin{equation}
    \alpha_m \equiv \frac{n_sc^2}{3 P_{\perp s}^2} \int d\tilde{\mathbf{p}} \ \tilde{p}_\perp^2 f_s
\end{equation}
accounts for the fact that $f_s$ is not necessarily a thermal distribution. In what follows, we postulate that~$\alpha_m$ is a constant and, when considering specific examples, take its value to be that corresponding to the Maxwell--Jüttner distribution, $\alpha_m = \alpha_{m, \text{MJ}} =8/3$. In fact, in Section~\ref{sec: Synchrotron Cooling Instability}, we will find that the fastest-growing SCI mode is pressure-balanced, and so the exact form of $\bar{\mathbf{F}}_\text{rad,s}$ is not important in~\eqref{eq: Momentum eq} on the scales of interest. Nevertheless, the radiation force does play an important role in the pressure evolution, as we will explain in the following sections. 

\subsection{Pressure Evolution}
\label{sec: Pressure Evolution}
In an ultra-relativistically hot, collisionless, magnetised plasma without radiative cooling, the evolution of the perpendicular and parallel pressures is described by the double-adiabatic equations \citep{Wierzchucka_etal-2026}
\begin{align}
   \label{eq: pressure evolv}
   &P_{\perp s}\frac{d}{dt_s} \ln\left[  \frac{P_{\perp s}}{ B^{2} } \left( \frac{B^3}{n_s^2}\right)^{2 - \Gamma_\perp} \right] = 0, \quad 
    P_{\parallel s}\frac{d}{dt_s} \ln \left[  \frac{P_{\parallel s} B}{ n_s^{2} } \left( \frac{B^3}{n_s^2}\right)^{1 - \Gamma_\parallel} \right] = 0,
\end{align}
where $d/dt_s \equiv \partial/ \partial t + \mathbf{u}_s \cdot \nabla$. Here $\Gamma_\perp$ and $\Gamma_\parallel$ are the effective perpendicular and parallel relativistic adiabatic indices, which themselves depend on the pressure anisotropy: e.g., for an almost isotropic distribution  $(\Gamma_\perp, \Gamma_\parallel) = (8/5, 4/5)$, while when $P_\parallel \gg P_\perp$, $(\Gamma_\perp, \Gamma_\parallel) = (2, 1)$ \citep{Wierzchucka_etal-2026}. The two equations can be combined to give the (ultra-relativistic) equation for the conservation of energy:
\begin{equation}
    \label{eq: energy cons}
    \frac{dP_s}{dt_s} = 
    \frac{(3P_s+ P_{\parallel s})}{3n_s} \frac{d n_s}{d t_s} + \frac{(P_{\perp s} - P_{\parallel s})}{3 B}\frac{d B}{d t_s},
\end{equation}
which reduces to the familiar adiabatic law $P \propto n^{4/3}$ in the isotropic case, when $P_{\perp s} = P_{\parallel s}$.

To account for synchrotron emission,~\eqref{eq: pressure evolv} should include appropriate radiative-cooling terms on their right-hand sides. The reduction of pressure due to synchrotron cooling can be found by taking the $P_\perp$ and $P_\parallel$ moments of the radiation-reaction term in the kinetic equation~\eqref{eq: Vlasov + Rad} and recalling~\eqref{eq: Rad Force}:
\begin{align}
    \label{eq: rad pressure 1}
     \left. \frac{\partial P_{\perp s}}{\partial t} \right\vert_\text{SC} = -\frac{r_s^2 B^2}{3m_s^4 c^3 }\int d\tilde{\mathbf{p}} \frac{\tilde{p}_\perp^4}{\tilde{\gamma}^2}f_s, \quad
     \left. \frac{\partial P_{\parallel s}}{\partial t} \right\vert_\text{SC} = -\frac{2r_s^2 B^2}{3m_s^4 c^3 }\int d\tilde{\mathbf{p}}\frac{\tilde{p}_\parallel^2 \tilde{p}_\perp^2}{\tilde{\gamma}^2}f_s.
\end{align}
Similarly to~\eqref{eq: rad moment},~\eqref{eq: rad pressure 1} cannot, in general, be expressed in terms of the fluid moments that we have introduced so far. As in Section \ref{sec: Radiation-Reaction Force in Relativistically Hot, Magnetised Plasmas}, to approximate~\eqref{eq: rad pressure 1}, we first calculate the corresponding expressions for the case of a Maxwell--Jüttner distribution~\eqref{eq: MJ}:
\begin{align}
    \label{eq: rad pressure MJ}
     \left. \frac{\partial P_{\perp s}}{\partial t} \right\vert_\text{SC} = -\frac{32r_s^2}{15m_s^2 c^3 } \frac{B^2 P_s^2}{n_s}, \quad
     \left. \frac{\partial P_{\parallel s}}{\partial t} \right\vert_\text{SC} = -\frac{16r_s^2}{15m_s^2 c^3 } \frac{B^2 P_s^2}{n_s}.
\end{align}
Then, as the energy closure, we adopt the ultra-relativistic double-adiabatic equations~\eqref{eq: pressure evolv} with non-ideal radiative cooling terms whose form is motivated by~\eqref{eq: rad pressure MJ}:
\begin{align}
   &P_{\perp s} \frac{d}{dt_s} \ln \left[  \frac{P_{\perp s}}{ B^{2} } \left( \frac{B^3}{n_s^2}\right)^{2 - \Gamma_\perp} \right] =- \alpha_\perp\frac{2r_s^2}{m_s^2 c^3 } \frac{B^2 P_{\perp s}^2}{n_s},
   \label{eq: pressure evolv perp}
   \\ 
   &P_{\parallel s} \frac{d}{dt_s} \ln \left[  \frac{P_{\parallel s} B}{ n_s^{2} } \left( \frac{B^3}{n_s^2}\right)^{1 - \Gamma_\parallel} \right] = - \alpha_\parallel\frac{2r_s^2}{m_s^2 c^3 } \frac{B^2 P_{\perp s} P_{\parallel s}}{n_s},
    \label{eq: pressure evolv parallel}
\end{align}
where, to account for the fact that $f_s$ does not in general have to be a Maxwell--Jüttner distribution, we have introduced the dimensionless parameters
\begin{equation}
    \label{eq: alpha def}
    \alpha_\perp = \frac{n_s}{6 m_s^2P_{\perp s}^2} \int d\tilde{\mathbf{p}} \ \frac{\tilde{p}_\perp^4}{\tilde{\gamma}^2}f_s, \quad \alpha_\parallel = \frac{n_s}{3 m_s^2P_{\perp s} P_{\parallel s} } \int d\tilde{\mathbf{p}} \ \frac{\tilde{p}_\perp^2\tilde{p}_\parallel^2}{\tilde{\gamma}^2}f_s.
\end{equation}
In specific illustrative examples considered below, we will assume that $\alpha_\perp$ and $\alpha_\parallel$ can be approximated by their values for a Maxwell--Jüttner distribution,~$\alpha_\perp = \alpha_{\perp, \text{MJ}} =16/15$ and  $\alpha_\parallel = \alpha_{\parallel, \text{MJ}} =8/15$. The evolution equations~\eqref{eq: pressure evolv perp} and~\eqref{eq: pressure evolv parallel} can also be combined to find a version of~\eqref{eq: energy cons} with radiation:
\begin{equation} 
    \label{eq: energy}
    \frac{dP_s}{dt_s} = \frac{(3P_s+ P_{\parallel s})}{3n_s} \frac{d n_s}{d t_s} + \frac{(P_{\perp s} - P_{\parallel s})}{3 B}\frac{d B}{d t_s} - \frac{2 r_s^2}{3m_s^2 c^3 } \frac{B^2 P_{\perp s}(2 \alpha_\perp P_{\perp s} +  \alpha_\parallel P_{\parallel s})}{n_s}.
\end{equation}

\subsection{Single-Fluid Model}
\label{sec: Single-Fluid Model}
We now combine key results from the above discussion to formulate a single-fluid theory. Naturally, we define species-summed quantities~{$n~\equiv~N\sum_s n_s m_s / \sum_s m_s$}, $n\mathbf{u}~\equiv~N\sum_s m_s n_s\mathbf{u}_s / \sum_s m_s$, and~$\mathbf{P} \equiv \sum_s \mathbf{P}_s$, where $N$ is the total number of species. We then proceed as in the non-relativistic derivation of kinetic MHD \citep{Kulsrud-1964}. Summing the continuity~\eqref{eq: Non-Covariant Continuty} and the momentum~\eqref{eq: Momentum eq} equations over species gives
\begin{gather}
    \frac{\partial n}{\partial t} + \nabla \cdot(n \mathbf u) = 0,
    \label{eq: continuity} \\
    \sum_s \left\{ \frac{\partial (n_s \B{p}_s )}{\partial t} + \nabla \cdot (n_s \B{u}_s \B{p}_s ) \right\}= - \nabla \cdot \mathbf{P} + \rho \mathbf{E} + \frac{{\mathbf{j}} \times \mathbf{B}}{c}  -  \alpha_m\sum_s  \frac{2 r_s^2}{ m_s^2 c^4} \frac{\mathbf{u}_s}{c} \frac{B^2 P_{\perp s}^2}{n_s},
    \label{eq: momentum mid}
\end{gather}
where we have made use of~\eqref{eq: F rad final}. In what follows, we will expand \eqref{eq: momentum mid} to order $(\omega/\Omega_s)(u_s/c)^2$, which, as a result of the ordering~\eqref{eq: ordering}, requires retaining terms to order $u_s/c$ in time derivatives and to order $u_s^2/c^2$ in spatial derivatives (see Appendix~\ref{Appendix: Derivation of Lowest-Order Momentum Equation} for a detailed discussion).

To determine the charge density $\rho = \sum_s q_s n_s$ and current density $\mathbf{j} = \sum_s q_s n_s \mathbf{u}_s$, as well as the evolution of the magnetic field, we insert the ideal MHD condition~\eqref{eq: ideal E} into Maxwell's equations. Faraday's law then gives us the induction equation,
\begin{equation}
    \label{eq: induction}
    \frac{\partial \mathbf{B}}{\partial t} = \nabla \times (\mathbf{u} \times \mathbf{B}),
\end{equation}
while Gauss's, and Ampère's laws respectively result in expressions for $\rho$ and $\mathbf{j}$:
\begin{align}
    \label{eq: current and charge}
    \rho = -\frac{1}{4 \pi c}\nabla \cdot (\mathbf{u} \times \mathbf{B}), \quad \mathbf{j} = \frac{c}{4\pi} \nabla \times \mathbf{B} + \frac{1}{4\pi c} \frac{\partial (\mathbf{u} \times \mathbf{B})}{\partial t}.
\end{align}
Note that as a result of \eqref{eq: ideal E}, the electric-field term, $\rho \mathbf{E}$, in~\eqref{eq: momentum mid} is not negligible at order $u_s^2/c^2$ (see Appendix~\ref{Appendix: Derivation of Lowest-Order Momentum Equation} for details).

The species-summed momentum equation is then obtained by substituting~\eqref{eq: momentum value}, ~\eqref{eq: induction}, and~\eqref{eq: current and charge} into~\eqref{eq: momentum mid}. In addition, second-order corrections in $u_s/c$ must be retained in the pressure tensor~\eqref{eq: pressure gyro}. The resulting species-summed momentum equation ~\eqref{eq: A full momentum} at order $\omega/\Omega_s$ and $u_s^2/c^2$, together with its derivation, is presented in Appendix \ref{Appendix: Derivation of Lowest-Order Momentum Equation}. In its most general form, it is highly non-trivial. However, for a simple setup with no flows parallel to the magnetic field (i.e., $\mathbf{u}_s \cdot \mathbf{b} = 0$),~\eqref{eq: A full momentum} simplifies to
\begin{gather}
    \sum_s \left \{
    \frac{\partial}{\partial t} \left[
    \frac{1}{c^2}\left(
    3 P_{\perp s} + P_{\parallel s} 
    \right) \mathbf{u}_s
    \right]
    + 
    \nabla \cdot \left[
    \frac{1}{c^2}\left(
    3 P_{\perp s} + P_{\parallel s} 
    \right) \mathbf{u}_s\mathbf{u}_s 
    \right]
    \right \}
    \nonumber \\
    + \frac{\partial}{\partial t}  \frac{B^2}{4\pi c^2} \mathbf{u} +
     \nabla \cdot \left( \frac{B^2}{4\pi c^2} \mathbf{u} \mathbf{u}\right)
    =
    - \nabla \left(
    P_{\perp} + \frac{B^2}{8\pi \gamma_u^2}
    \right)
    +
    \nabla \cdot \left[
    \mathbf{b}\mathbf{b}
    \left(
    P_{\perp} - P_{\parallel}
    + \frac{B^2}{4\pi \gamma_u^2}
    \right)
    \right]
    \nonumber \\
    -\alpha_m\sum_s
    \frac{2 r_s^2}{m_s^2 c^4}
    \frac{\mathbf{u}_s}{c}
    \frac{B^2 P_{\perp s}^2}{n_s},
    \label{eq: full momentum no parallel}
\end{gather}
where $\gamma_u \equiv 1/\sqrt{1 - u^2/c^2}$. This is the only case that we will require in what follows.

We now restrict our analysis to pair plasmas, where the electrons and positrons have equal temperatures and densities. Inserting this assumption along with~\eqref{eq: drift velocity} into the momentum equation~\eqref{eq: full momentum no parallel}, we find a radiative single-fluid MHD equation of motion  in the field-perpendicular direction, with anisotropic pressure:
\begin{align}
    \label{eq: KMHD 1}
    \frac{\partial}{\partial t} \left[ \frac{1}{c^2}\left(3 P_{\perp } + P_{\parallel } + \frac{B^2}{4 \pi} \right) \mathbf{u} \right]
    + 
    \nabla \cdot \left[ \frac{1}{c^2}\left(3 P_{\perp } + P_{\parallel } + \frac{B^2}{4 \pi} \right) \mathbf{u}\mathbf{u}  
    \right]  = \nonumber \\
    - \nabla \left( P_\perp + \frac{B^2}{8\pi \gamma_u^2} \right)
+ \nabla \cdot \left[ \mathbf{b}\mathbf{b} \left( P_\perp - P_\parallel + \frac{B^2}{4\pi \gamma_u^2} \right) \right]  -  \alpha_m \frac{2 r_e^2}{ m_e^2 c^4} \frac{\mathbf{u}}{c} \frac{B^2 P_{\perp}^2}{n}.
\end{align}
Limiting ourselves to a plasma composed of electrons and positrons with equal densities and pressures also means that the species indices in~\eqref{eq: pressure evolv perp}, ~\eqref{eq: pressure evolv parallel} and \eqref{eq: energy} can now be dropped.

\subsection{Summary of Fluid Model}
The result of the above discussion is a closed set of fluid equations for relativistic, cooling, collisionless pair plasmas: continuity~\eqref{eq: continuity}, induction~\eqref{eq: induction}, momentum~\eqref{eq: KMHD 1}, and the two pressure equations~\eqref{eq: pressure evolv perp} and \eqref{eq: pressure evolv parallel}.

A series of assumptions underlie this model: a non-relativistic bulk velocity, ultra-relativistic temperatures, parallel parity~\eqref{eq: parity}, a plasma consisting of electrons and positrons with equal densities and pressures. The momentum equation~\eqref{eq: KMHD 1} is, in fact, a simplified version of~\eqref{eq: A full momentum} for flows strictly perpendicular to the magnetic field; when all gradients are only in the direction perpendicular to the magnetic field, which is the case that we consider in Section \ref{sec: Synchrotron Cooling Instability},~\eqref{eq: A full momentum} and~\eqref{eq: KMHD 1} are identical and the constraint $\mathbf{u}_s \cdot \mathbf{b} = 0$ has no effect. Relaxing these assumptions significantly increases the complexity of the system. A general discussion of relativistic, collisionless plasmas can be found in \cite{Wierzchucka_etal-2026b} and \cite{Abhishek_Stone-2026}, who present a covariant derivation of relativistic kinetic MHD.

There are five dimensionless parameters in our fluid model: the generalised adiabatic indices $\Gamma_\perp$ and $\Gamma_\parallel$, as well as the radiation coefficients $\alpha_m$, $\alpha_\perp$ and $\alpha_\parallel$, which capture the effects of non-thermal characteristics of the distribution function on synchrotron emission. In the case of a distribution function close to isotropy, $(\Gamma_\perp, \Gamma_\parallel) = (8/5, 4/5)$. If the distribution is Maxwell--Jüttner, $(\alpha_m, \alpha_\perp, \alpha_\parallel) = (8/3, 16/15, 8/15)$. Naturally, the radiation coefficients would have different values for highly non-thermal distributions. However, here we are dealing with a plasma subject to synchrotron radiation, which preferentially affects high-energy particles, opposing non-thermal particle acceleration and bringing the distribution close to thermal \citep{Zhdankin_etal-2020}. As we discuss in Section~\ref{sec: Synchrotron Firehose Instability}, this process is also made more efficient by the SFHI (cf. \citealt{Zhdankin_etal-2023}).

The fluid model~\eqref{eq: pressure evolv perp},~\eqref{eq: pressure evolv parallel}, ~\eqref{eq: continuity},~\eqref{eq: induction} and~\eqref{eq: KMHD 1} closely resembles non-relativistic double-adiabatic MHD equations \citep{Kulsrud-1964, Chew_etal-1956}. As expected, the momentum equation includes total pressure forces and forces due to magnetic-field tension. However, ultra-relativistic temperatures have modified the inertial terms in~\eqref{eq: KMHD 1}, which are now dominated by pressure rather than density contributions. Additionally, synchrotron emission has given rise to both radiation drag in the momentum equation~\eqref{eq: KMHD 1} and energy-loss terms in the pressure equations,~\eqref{eq: pressure evolv perp} and~\eqref{eq: pressure evolv parallel}. In the next section, we discuss how these radiative terms can give rise to pressure anisotropy and consequently excite instabilities. 

\section{Synchrotron Firehose Instability}
\label{sec: Synchrotron Firehose Instability}
\subsection{Pressure Anisotropy in a Synchrotron-Cooling Plasma}
Synchrotron emission destabilises relativistic plasmas on kinetic scales \citep{Zhdankin_etal-2023, Bilbao_etal-2025}. To illustrate this, we consider an isotropic, thermal initial state, with temperature $\theta_0 = T_0/m_ec^2 \gg 1$, total density $n_0$, and a uniform background magnetic field~$\B{B}_0 = B_0 \hat{\B{z}}$. We denote the corresponding pressure by $P_0 = n_0 T_0$. Then there exists a solution to~\eqref{eq: pressure evolv perp},~\eqref{eq: pressure evolv parallel},~\eqref{eq: continuity},~\eqref{eq: induction}, and~\eqref{eq: KMHD 1} whereby $n = n_0$, $\mathbf{B} = \mathbf{B}_0$, $\mathbf{u} = 0$, and the pressures evolve as
\begin{equation}
    \label{eq: homogeneous anisotropic pressure evolution}
    P_\perp(t) = \frac{P_0}{1 + \alpha_\perp t/\tau_0}, \quad P_\parallel(t) = \frac{P_0}{(1 +  \alpha_\perp t/\tau_0)^{\alpha_\parallel/\alpha_\perp}},
\end{equation}
where the synchrotron-cooling time associated with the initial state [cf.~\eqref{eq: cooling time}] is
\begin{equation}
    \label{eq: cooling time 0}
    \tau_0
    =
    \frac{m_e^2 c^3}{2 r_e^2 B_0^2 T_0}.
\end{equation}
As the plasma cools, a pressure anisotropy develops:
\begin{equation}
    \label{eq: anisotropy}
    \Delta
    \equiv
    \frac{P_\perp}{P_\parallel} - 1
    =
    \left(
        1 + \frac{\alpha_\perp t}{\tau_0}
    \right)^{\alpha_\parallel/\alpha_\perp - 1}
    - 1 .
\end{equation}
At early times, $t \ll \tau_0$, this expression reduces to
\begin{equation}
    \label{eq: anisotropy short}
    \Delta
    \approx
    -(\alpha_\perp - \alpha_\parallel)\frac{t}{\tau_0}
    =
    -\frac{8}{15}\frac{t}{\tau_0},
\end{equation}
where, in the final step, we used $\alpha_\perp = 16/15$ and $\alpha_\parallel = 8/15$ \citep{Zhdankin_etal-2023}.

\subsection{Synchrotron Firehose Instability}
Pressure anisotropy in collisionless plasmas can lead to the excitation of microscopic instabilities, such as the firehose instability, which arises when
\begin{equation}
    \label{eq: firehose threshold}
    \Delta < -\frac{C_\text{th}}{\beta_\parallel},
\end{equation}
where $\beta_\parallel \equiv 8\pi P_\parallel/B^2$ and $C_\text{th}$ is a constant of order unity. The value of $C_\text{th}$ depends on both the plasma composition and the specific type of firehose instability that is excited, i.e., kinetic or fluid \citep{Chandrasekhar_etal-1958, Yoon_etal-1993, Hellinger_Matsumoto_2000, Gary_Nishimura_2003, Bott_etal-2024, Bott_etal-2025}. In high-$\beta$ relativistic pair plasmas, the pressure anisotropy generated by synchrotron cooling~\eqref{eq: anisotropy} can grow until the firehose threshold~\eqref{eq: firehose threshold} with $C_\text{th}=1.4$ is exceeded, triggering the oblique kinetic firehose instability \citep{Hellinger_Matsumoto_2000, Hellinger_Matsumoto_2001, Zhdankin_etal-2023}. This instability, referred to in this context as the synchrotron firehose instability (SFHI), is triggered at time
\begin{equation}
    \label{eq: onset time}
    t_\text{onset} = \frac{{C_\text{th}}}{{(\alpha_\perp - \alpha_\parallel)\beta_{0}}}\tau_0=\frac{{15 C_\text{th}}}{{8\beta_{0}}}\tau_0,
\end{equation}
where $\beta_0 = 8\pi P_0/B_0^2$.

As the growth rate of the firehose instability is proportional to $|\Delta|^{1/2}$, which itself depends on time, the SFHI initially grows super-exponentially on a timescale $t_\text{gr}\sim\tau_0^{1/3} \Omega^{-2/3}_0$, where~{$\Omega_{0} = eB_0 c/ 3 T_{0}$} is the relativistic Larmor frequency of the initial state. Note that the relative size of this timescale and \eqref{eq: onset time} is controlled by the ratio $(\Omega_{0} \tau_0)^{2/3}/\beta_0$. In the physical setup discussed in Section \ref{sec: Particle-in-Cell Simulations}, $ (\Omega_{0} \tau_0)^{2/3} > \beta_0 \gg 1$, which is the case for many astrophysical environments. The firehose fluctuations quickly reach amplitudes large enough to scatter particles, pinning the anisotropy to the firehose threshold~\eqref{eq: firehose threshold}. The process then repeats, with negative anisotropy generated by the cooling rapidly regulated by the firehose fluctuations excited by it. Thus, after the initial excitation, on timescales longer than $t_\text{gr}$, the pressure anisotropy is pinned to the firehose threshold~\eqref{eq: firehose threshold}, making high-$\beta$, radiating plasma almost isotropic \citep{Zhdankin_etal-2023}. 

We propose that high-$\beta$ synchrotron-emitting plasmas where the SFHI has been excited are effectively described by an adapted version of the fluid model~\eqref{eq: continuity},~\eqref{eq: induction},~\eqref{eq: KMHD 1}, \eqref{eq: pressure evolv perp}, and~\eqref{eq: pressure evolv parallel}. Motivated by the adiabatic evolution of pressure in isotropic ultra-relativistic plasmas, $P \propto n^{4/3}$, we postulate that instead of~\eqref{eq: pressure evolv perp} and~\eqref{eq: pressure evolv parallel}, the pressure evolution in SFHI-dominated plasmas is described by the energy equation~\eqref{eq: energy}, with the pressure anisotropy pinned to the firehose threshold~\eqref{eq: firehose threshold}, equivalently written as
\begin{equation}
    \label{eq: threshold}
    P_\parallel - P_\perp = C_\text{th}\frac{B^2}{8 \pi}.
\end{equation}

\subsection{Shut-Off of the Synchrotron Firehose Instability}
\label{sec: Shut-Off of the Synchrotron Firehose Instability}
Synchrotron-emitting plasma is not always dominated by SFHI fluctuations. Consider the state of a firehose-infested plasma at some time~$t$. As the firehose instability primarily induces magnetic-field perturbations perpendicular to the guide field, $n$ and $B$ remain constant. When pitch-angle scattering reduces the anisotropy to the threshold~\eqref{eq: threshold}, firehose fluctuations quickly decay away \citep{Melville_etal-2016} and the pressure evolution is described by the radiative collisionless evolution equations~\eqref{eq: pressure evolv perp} and~\eqref{eq: pressure evolv parallel}. The SFHI will be re-excited if, starting at the marginal state~\eqref{eq: threshold}, synchrotron cooling can result in condition~\eqref{eq: firehose threshold} being satisfied again. This will occur if 
\begin{equation}
    \label{eq: FH shut off 1}
    \frac{d}{dt}\left(P_\parallel - P_\perp -C_\text{th} \frac{B^2}{8\pi}\right)>0.
\end{equation}
Inserting~\eqref{eq: pressure evolv perp} and~\eqref{eq: pressure evolv parallel} into~\eqref{eq: FH shut off 1}, one finds that, in a constant magnetic field, the SFHI is only re-excited if
\begin{equation}
    \label{eq: shut down}
    \frac{P_\perp}{P_\parallel} > \frac{\alpha_\parallel}{\alpha_\perp}.
\end{equation}
Thus, the subsequent re-excitation of the instability is suppressed once the pressure anisotropy reaches $\Delta=\alpha_\parallel/\alpha_\perp - 1$, or, using the threshold-pinning condition~\eqref{eq: threshold}, when $\beta_\parallel$ drops below
\begin{equation}
    \label{eq: SFHI shut off}
    \beta_\parallel = \frac{\alpha_\perp }{\alpha_\parallel} \beta_\perp = \frac{\alpha_\perp \Cth}{\alpha_\perp - \alpha_\parallel},
\end{equation}
which yields $\Delta = -1/2$ and $\beta_\parallel = 2.8$, for a relativistic pair plasma with $\alpha_\perp = 16/15$ and $\alpha_\parallel = 8/15$. In the fully kinetic regime, where $\alpha_\perp$ and $\alpha_\parallel$ are no longer constant but are instead given by~\eqref{eq: alpha def}, the SFHI shut-off criterion, the breaking of~\eqref{eq: shut down}, becomes
\begin{equation}
\label{eq: full SFHI shut off}
2\int d\tilde{\mathbf{p}} \,\frac{\tilde{p}_\perp^2\tilde{p}_\parallel^2}{\tilde{\gamma}^2}f
>
\int d\tilde{\mathbf{p}} \,\frac{\tilde{p}_\perp^4}{\tilde{\gamma}^2}f.
\end{equation}
Note that this occurs before the naive shut-off, which happens when $P_\perp = 0$ and $\beta_\parallel = C_\text{th}$, i.e., the point beyond which the instability condition \eqref{eq: firehose threshold} can no longer be satisfied since $\Delta = -1$ but $\beta_\parallel$ continues to decrease as the plasma cools. 

To summarise, synchrotron cooling generates a negative pressure anisotropy, exciting the SFHI at $t \sim \tau_0/\beta_0$. The resulting SFHI fluctuations grow rapidly, scattering particles in pitch angle and thereby pinning the pressure anisotropy to the firehose threshold~\eqref{eq: threshold}. Continued cooling repeatedly drives the plasma back into the unstable regime, maintaining the threshold-pinned state until~\eqref{eq: SFHI shut off}, or, more generally, \eqref{eq: full SFHI shut off}, is satisfied. At this point, radiation cooling cannot drive negative pressure anisotropy fast enough to catch up with the decreasing~$\beta_\parallel$, which increases the firehose threshold, and so the SFHI is shut off. 

\section{Synchrotron Cooling Instability}
\label{sec: Synchrotron Cooling Instability}
Apart from the SFHI, synchrotron cooling can also destabilise plasmas on fluid scales \citep{Simon_Axford-1967}. As in high-$\beta$ plasmas the SFHI is excited on the fast timescale $t_\text{onset} \sim \tau_0/\beta_0$, we will study the macroscopic stability of the plasma after $t_\text{onset}$, when it is already infested with microscopic SFHI fluctuations. In this case, the plasma behaviour is captured by the fluid system~\eqref{eq: energy},~\eqref{eq: continuity},~\eqref{eq: induction},~\eqref{eq: KMHD 1} and~\eqref{eq: threshold}. For simplicity and illustration of key ideas, in this section we will consider the high-$\beta$ limit in which the firehose pinning results in nearly isotropic pressure, $P_\perp \approx P_\parallel \approx P \gg B^2/8\pi$. This amounts to setting~$\Cth = 0$ in~\eqref{eq: threshold}. In Appendix~\ref{Appendix: Synchrotron Cooling Instability in Firehose-Infested Plasmas}, we provide the more general calculation keeping the~$C_\text{th}/\beta$ corrections, which turn out to have minimal effect on the key properties of the~SCI. 

\subsection{Linear Theory of SCI}
In a firehose-infested medium, the homogeneous but time-dependent cooling state, found by solving~\eqref{eq: energy} with the condition $P_\perp = P_\parallel$, is
\begin{equation}
    \label{eq: isotropic homogeneous}
    P(t) = P_\perp(t) = P_\parallel(t) = \frac{P_0}{1 + \alpha t/\tau_0},
\end{equation}
where $\alpha = (2\alpha_\perp + \alpha_\parallel)/3$. For the case of a thermal distribution, $\alpha = 8/9$. 

We now introduce an infinitesimal non-uniform perturbation, considering solutions to time-dependent radiative fluid equations~\eqref{eq: energy},~\eqref{eq: continuity},~\eqref{eq: induction}, and~\eqref{eq: KMHD 1} in the form
\begin{equation}
    \label{eq: pert}
    n = n_0 + \delta n, \ P = P_\perp = P_\parallel = P_0 + \delta P, \ \mathbf{u} = \delta\mathbf{u}, \ \mathbf{B} = B_0 \hat{\mathbf{z}} + \delta \mathbf{B}.
\end{equation}
Let us first study the short-time behaviour of the perturbations by assuming that they evolve on timescales $t \ll \tau_0$, which allows us to neglect the evolution of the background state~\eqref{eq: isotropic homogeneous}. We seek solutions in the standard plane-wave form
\begin{equation}
    \label{eq: delta}
    \delta n, \ \delta P, \ \delta \mathbf{u}, \ \delta \mathbf{B} \propto e^{\chi t/\tau_0 + i \mathbf{k} \cdot \mathbf{x} },
\end{equation}
with growth rate $\chi$ normalised to the initial cooling rate, $\tau_0^{-1}$. We take wavevector $\mathbf{k} = k \hat{\B{x}}$, i.e., assume propagation perpendicular to the background magnetic field.

Inserting~\eqref{eq: pert} and~\eqref{eq: delta} into~\eqref{eq: energy},~\eqref{eq: continuity},~\eqref{eq: induction}, and~\eqref{eq: KMHD 1} and linearising the equations yields the following general dispersion relation for a plasma with an adiabatic index~$\Gamma$:
\begin{equation}
    \label{eq: disp rel}
   \chi^4\left\{ [(4\beta_0 + 2)\chi + \alpha_m\beta_0](\chi + 2\alpha)\chi + \frac{k^2}{k_0^2} \left[ (\Gamma \beta_0 + 2) \chi - \alpha(\beta_0 - 4)\right] \right\} =0.
\end{equation}
Here $k_0 \equiv 1/\tau_0 c$ and $\Gamma = 4/3$. Unsurprisingly, the structure of~\eqref{eq: disp rel} closely resembles the general dispersion relation of \cite{Simon_Axford-1967}. Note also that in the absence of cooling ($\alpha = \alpha_m= 0$),~\eqref{eq: disp rel} becomes the dispersion relation for ultra-relativistic fast magnetosonic waves, $i\chi/\tau_0 = \pm kc\sqrt{(\Gamma\beta_0 + 2)/(4\beta_0 + 2)}$.

The dispersion relation~\eqref{eq: disp rel} has seven solutions. The four $\chi=0$ modes are slow magnetosonic and Alfvén waves, whose frequencies vanish in the case of perpendicular propagation. Then there are two fast magnetosonic modes, which are damped as a result of the cooling. Finally, we have the entropy mode with a purely real $\chi$, which is positive~if 
\begin{equation}
    \label{eq: critical beta}
    \beta_0 > \beta_c = 4.
\end{equation}
It is this mode that corresponds to the synchrotron cooling instability (SCI). In the case of finite $\Cth$, the above condition becomes \eqref{eq: beta crit c}, which, when $\Cth = 1.4$, $\alpha_\perp = 16/15$, and $\alpha_\parallel = 8/15$, evaluates to a somewhat lower value $\beta_c \approx 3.33$ (see Appendix \ref{Appendix: Synchrotron Cooling Instability in Firehose-Infested Plasmas} for details).

\begin{figure}
\centering
\includegraphics[width=0.9\linewidth]{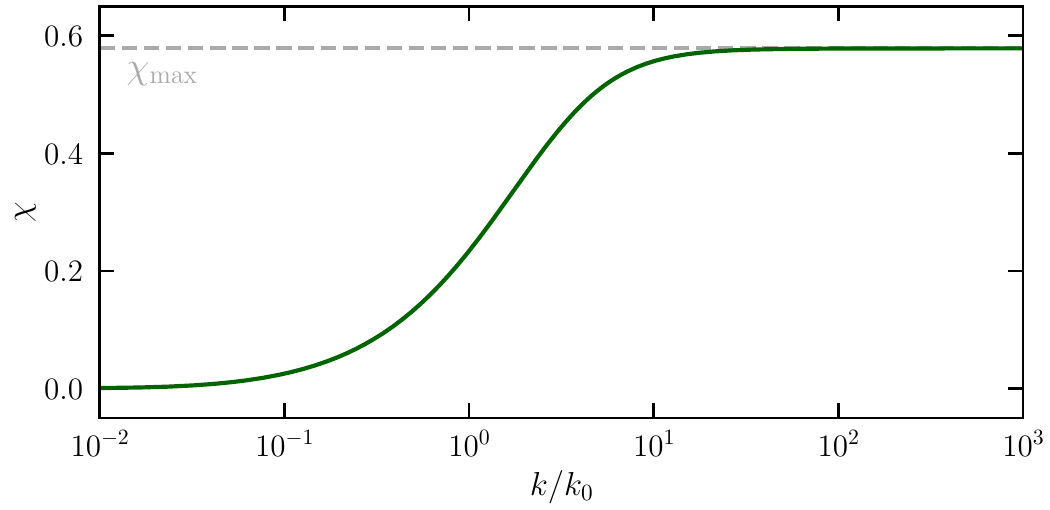}
\caption{The positive real root of~\eqref{eq: disp rel}, i.e., the growth rate of the SCI normalised to the cooling rate of the initial state $\tau_0^{-1}$, as a function of $k/k_0$, for $\Gamma = 4/3$, $\beta_0 = 40$, $\alpha = 8/9$, and $\alpha_m = 8/3$. It can be seen that at $k \gg k_0 = 1/\tau_0 c$, $\chi$ approaches the asymptotic value~\eqref{eq: SCI growth rate}, represented here by the grey dashed line.
 } 
\label{fig: chi vs k}
\end{figure}

The SCI results in a compressive flow perpendicular to the field, $\delta\mathbf{u} = \delta u \hat{\mathbf{x}}$, and does not change the direction of the magnetic field, only its magnitude. Consequently, the SCI perturbation is flux-frozen, satisfying $\delta n/n_0 = \delta B /B_0$, where $\delta B = \hat{\mathbf{z}}\cdot \delta \mathbf{B}$. The growth rate of the instability is shown in Figure \ref{fig: chi vs k} as a function of $k$. It is purely real and increases with $k$, asymptoting at $k \gg k_0$ to a plateau of
\begin{equation}
    \label{eq: SCI growth rate}
    \chi_\text{max}  = \alpha\frac{ \beta_0 -4} {\Gamma \beta_0 + 2}. 
\end{equation}
In this short-wavelength limit, the acoustic response is much faster than radiative cooling, and~\eqref{eq: KMHD 1} reduces to
\begin{equation}
\label{eq: Pressure Balance}
\nabla \left( P_\perp + \frac{B^2}{8 \pi} \right) = 0,
\end{equation}
implying that the fastest-growing modes of the SCI are pressure-balanced. When $\beta_0~\gg~1$,~\eqref{eq: SCI growth rate} becomes independent of $\beta_0$, and takes the value $\chi_\text{max} = \alpha/\Gamma = 2/3$ (using $\Gamma = 4/3$, $\alpha_\perp = 16/15$, and $\alpha_\parallel = 8/15$). When $\beta_0 <\beta_c$, the SCI is quenched and the perturbations decay. 

The physical mechanism behind the SCI is as follows. An initial compression produces a region of enhanced density and therefore, via flux freezing, enhanced magnetic field strength. The dependence of synchrotron emission on $B$ implies that this region cools faster than its surroundings. The plasma must, however, remain in total pressure balance. In the high-$\beta_0$ limit,~\eqref{eq: Pressure Balance} reduces to $\nabla P_\perp = 0$. The cooling-induced temperature decrease is then compensated by an increase in density, amplifying the initial perturbation and establishing a positive feedback loop.

When $\beta_0 \sim 1$, magnetic pressure contributes significantly to the total pressure. In this case, maintaining total pressure balance requires $P_\perp$ to decrease. The density response now depends on the competition between the temperature decrease due to synchrotron cooling and the increase in the magnetic pressure from the balance of total pressure. If $\beta_0 > \beta_c$, the temperature drop dominates, leading to a further increase in density and so continued growth of the perturbation. For $\beta_0 < \beta_c$, the opposite situation occurs and the density decreases, causing the perturbation to decay, so the~SCI is suppressed.

\subsection{Time-Evolving Linear Theory of SCI}
\label{sec: Time-Evolving Linear Theory of SCI}

Since $\chi \sim 1$, i.e., the growth rate is comparable to the rate at which the background state evolves, one cannot, in fact, ignore the effect of overall plasma cooling on the SCI. Thus, a time-dependent theory is required to characterise the evolution of the SCI in a quantitatively correct way, even in the linear regime.

We again introduce linear perturbations to the background homogeneous cooling state~\eqref{eq: isotropic homogeneous} with~\eqref{eq: pert}, but now relax the assumption $t/\tau_0 \ll 1$. Accordingly, we take 
\begin{equation}
    \delta n, \ \delta P, \ \delta \mathbf{u}, \ \delta \mathbf{B} \propto e^{i \mathbf{k} \cdot \mathbf{x} },
\end{equation}
now allowing for time-dependent amplitudes of the perturbations. As before, we assume that $\mathbf{k} = k \hat{\mathbf{x}}$. We focus on the fastest-growing SCI modes $k \gg k_0$, allowing the replacement of~\eqref{eq: KMHD 1} with the pressure-balance condition~\eqref{eq: Pressure Balance}. As the SCI results in a flow perpendicular to the magnetic field, we take $\delta \mathbf{u} = \delta u\hat{\mathbf{x}}$, considering a perturbation produced by an ideal-MHD compression and initially satisfying
\begin{equation}
    \frac{\delta n(t = 0)}{n_0} = \frac{\delta B(t =0)}{B_0}.
\end{equation}
Then integrating~\eqref{eq: continuity} and~\eqref{eq: induction} in time gives 
\begin{equation}
    \label{eq: flux freezing}
    \frac{\delta n(t)}{n_0} = \frac{\delta B(t)}{B_0}.
\end{equation}
Combining~\eqref{eq: flux freezing} with the linearised forms of~\eqref{eq: energy} and~\eqref{eq: Pressure Balance}, we obtain a linear evolution equation for $\delta B(t)$:
\begin{equation}
    \label{eq: delta B ode}
    \frac{\partial \delta B}{\partial t} = g(t) \delta B,
\end{equation}
where the time-evolving instantaneous linear growth rate is given by
\begin{equation}
    \label{eq: t B equation}
    g(t) = \frac{\alpha}{\tau_0}\frac{ (\beta_0 - 4) - 4\alpha t/\tau_0}{(2 + \Gamma\beta_0) + (4+ \Gamma \beta_0)\alpha t/\tau_0 + 2 (\alpha t/\tau_0)^2}.
\end{equation}
The first-order differential equation~\eqref{eq: delta B ode} has the solution
\begin{equation}
    \label{eq: B t evolution}
    \delta B(t) = \delta B(0) \exp \left[ \int_0^t dt' g(t')\right],
\end{equation}
whose exponent can be evaluated analytically:
\begin{equation}
    \label{eq: t exponent}
    \int_0^t dt' g(t') = \frac{1}{\Gamma } \left[\ln\left(1+\frac{\alpha t}{\tau_0} \right)-(2\Gamma + 1) \ln\left(1  + \frac{2}{2 + \Gamma \beta_0 }\frac{\alpha t}{\tau_0} \right) \right].
\end{equation}
It is easy to check that in the limit $t/\tau_0 \ll 1$,~\eqref{eq: t exponent} reduces to $\chi_\text{max} t/\tau_0$ with $\chi_\text{max}$ given by~\eqref{eq: SCI growth rate}. In general, substituting~\eqref{eq: t exponent} into~\eqref{eq: B t evolution} one finds
\begin{equation}
    \label{eq: full B evolution}
    \frac{\delta B(t)}{\delta B(0)} = \left( 1+ \frac{\alpha t}{\tau_0}\right)^{1/\Gamma} \left( 1 + \frac{2}{2 + \Gamma \beta_0} \frac{\alpha t}{\tau_0} \right)^{-(2\Gamma + 1)/\Gamma}, 
\end{equation}
so the linear SCI grows algebraically rather than exponentially. 

Naturally, \eqref{eq: full B evolution} applies only in the linear regime, $\delta B \ll B_0$; however, it still provides useful insight into various properties of the SCI. In an initially unstable plasma,~\eqref{eq: t B equation} changes sign at
\begin{equation}
    \label{eq: sat time}
    t_\text{sat} = \frac{(\beta_0 - 4)}{4\alpha}\tau_0,
\end{equation}
indicating that, after an initial growth period, the SCI saturates and the perturbation subsequently decays away. At times $t \gg t_{\text{sat}}$, this decay is $\delta B(t) \propto 1/t^2$, which is consistent with the preliminary discussion by \cite{Simon_Axford-1967} of the long-time asymptotic limit. 

The saturation time~\eqref{eq: sat time} coincides with the time at which the plasma $\beta$ of the homogeneous background state 
\eqref{eq: isotropic homogeneous},
\begin{equation}
    \label{eq: beta evolution}
    \beta(t) = \frac{\beta_0}{1+ \alpha t/\tau_0},
\end{equation}
reaches the marginal stability threshold $\beta(t) = \beta_c$ \eqref{eq: critical beta}. Evaluating~\eqref{eq: full B evolution} at~\eqref{eq: sat time} gives the maximum field amplification achievable by the linear SCI:
\begin{equation}
    \label{eq: sat scaling full}
    \frac{\delta B(t_\text{sat})}{\delta B(0)} = \left( \frac{\beta_0}{4}\right)^{1/\Gamma} \left[ \frac{( 2\Gamma + 1)\beta_0}{2(2+ \Gamma \beta_0)} \right]^{-(2\Gamma + 1)/\Gamma}.
\end{equation}
In the high-$\beta_0$ limit,~\eqref{eq: sat scaling full} becomes simply
\begin{equation}
    \label{eq: saturated scaling}
    \frac{\delta B(t_\text{sat})}{\delta B(0)} \sim \beta_0^{1/\Gamma}.
\end{equation}
Equating $\delta B(t_\text{sat}) \sim B_0$ gives the condition on $\beta_0$ for which a nonlinear theory is required to understand the full evolution of the SCI, $\beta_0 \gtrsim [B_0/\delta B(0)]^{\Gamma}$. We will return to this matter in Section~\ref{sec: Emergent Two-Phase Structure of Synchrotron Cooling Plasmas}, where we use fluid simulations to study the nonlinear evolution of the~SCI.

Let us summarise the evolution of the (linear) SCI, focusing on the case of high~$\beta_0$ and wavenumbers significantly exceeding $k_0$. Initially, for $t \ll \tau_0$, the SCI perturbation grows as $1+ \alpha t/\Gamma \tau_0$. When $ \tau_0 \ll t \ll \beta_0\tau_0$, the growth begins to slow, approaching~$\propto t^{1/{\Gamma}}$. The instability then saturates at time~$t_\text{sat} \approx \beta_0 \tau_0/4\alpha$, with maximum field contrast~\eqref{eq: saturated scaling}. Finally, at $t \gg \beta_0 \tau_0$ the SCI decays as $t^{-2}$, and the plasma eventually returns to a homogeneous state. In Appendix~\ref{Appendix: Synchrotron Cooling Instability in Firehose-Infested Plasmas}, we repeat the above calculation, relaxing the assumption $\Cth = 0$, and show that the same evolution occurs with order-unity corrections due to the finite value of $\Cth/\beta_0$. The equivalent of \eqref{eq: full B evolution} with finite $\Cth$ is \eqref{eq: cth delta B evolution}. Due to the assumption of ultra-relativistic plasma in the model developed in Section~\ref{sec: Relativistic, Cooling Plasmas} and used here to study the SCI, the above theory remains applicable only as long as the plasma temperature remains ultra-relativistic, i.e., $\theta = T/m_e c^2 \gg 1$. This condition can be recast~as
\begin{equation}
    \beta(t) \gg \frac{\beta_0}{\theta_0} = \frac{2}{\sigma_{c0}},
\end{equation}
where $\beta$ is given by~\eqref{eq: beta evolution} and $\sigma_{c0} \equiv {B_0^2}/{4 \pi n_0 m_e c^2}$ is the initial cold magnetisation parameter. 

We note that if for some reason the SFHI is suppressed and the plasma is truly collisionless, i.e., described by~\eqref{eq: pressure evolv perp},~\eqref{eq: pressure evolv parallel},~\eqref{eq: continuity},~\eqref{eq: induction}, and~\eqref{eq: KMHD 1}, all of the above calculations hold, with the replacement of $\Gamma$ with $\Gamma_\perp$ and $\alpha$ with~$\alpha_\perp$. 

\section{Particle-in-Cell Simulations}
\label{sec: Particle-in-Cell Simulations}
\begin{figure}
\center
\includegraphics[width=\linewidth]{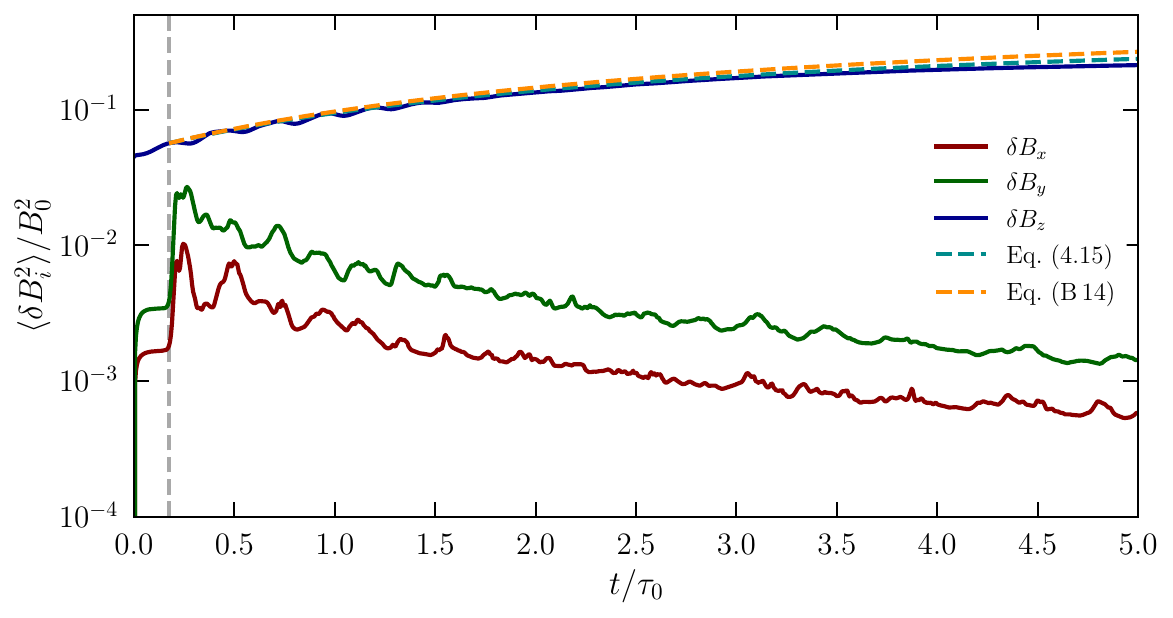}
\caption{Evolution of the box-averaged energy of the magnetic-field components, normalised to $B_0^2$, in the PIC simulation described in Section \ref{sec: Particle-in-Cell Simulations}. The SFHI is excited at $ t_\text{PIC}  = 0.17 \, \tau_0$ (grey dashed line), which is somewhat later than the predicted value~\eqref{eq: onset time},~$t_{\text{onset}} = 0.066 \, \tau_0$, manifesting as an increase in $\delta B_x$ (red line) and $\delta B_y$ (green line). Growth in $\delta B_z$ (blue line) is a result of the SCI, and agrees well with the time-dependent linear prediction~\eqref{eq: cth delta B evolution} with $C_\text{th} = 1.4$, $\alpha_\perp = 0.9$, $\alpha_\parallel = 0.48$ and $\Gamma = 4/3$ (orange dashed line) and with the simplified linear evolution \eqref{eq: full B evolution} with the same parameters but $C_\text{th} = 0$ (cyan dashed line). Note that we begin the orange and cyan lines at the time when the SFHI is excited, using the value of $\beta_\perp$ at that time.} 
\label{fig: B/B0}
\end{figure}
\subsection{Numerical Setup}
\label{sec: Numerical Setup}
We test our theoretical model of synchrotron-cooling collisionless plasmas, and specifically the co-existence of the SCI and SFHI by performing fully kinetic numerical simulations using the relativistic radiative particle-in-cell (PIC) code \texttt{OSIRIS} \citep{Fonseca_etal-2002}. Synchrotron emission is included through an implementation of the \cite{Landau_Lifshitz-1975} radiation-reaction force in the standard particle pusher \citep{Vranic_etal-2016}. 

Since the SFHI generates fluctuations with wavevectors both parallel and perpendicular to the magnetic field, we adopt a two-dimensional simulation setup in the $x$--$z$ plane with periodic boundary conditions in both directions. However, the particle velocities are fully three-dimensional. We initialise an electron-positron plasma with a uniform magnetic field $\mathbf{B}_0 = B_0 \hat{\mathbf{z}}$. Both species are initially described by uniform Maxwell--Jüttner distributions with density $n_0/2$ and temperature $T_0 = 100\,m_e c^2$. We select $n_0$ and $B_0$ so that $\beta_{e0} = \beta_{p0} = \beta_0/2 = 20$ and $\tau_0 \Omega_0 = 2 \times 10^3$. This set of parameters gives the invariant quantum parameter $\chi_\text{QED} \approx (\alpha_\text{fs} \tau_e\Omega_e \theta_e)^{-1}\approx 7 \times 10^{-4} \ll 1 $, justifying the use of the classical radiation-reaction force. All times and distances are normalised to the fiducial electron gyrofrequency, $\Omega_0 = e B_0 c / 3 T_0$, and the corresponding Larmor radius, $\rho_{e0} = c/\Omega_0$, respectively.

The simulation domain has size $L_x \times L_z = 300\,\rho_{e0} \times 37.5\,\rho_{e0} = 0.15\,c\tau_0 \times 0.019\,c\tau_0$ and consists of $N_x \times N_z = 13328 \times 1673$ cells, corresponding to a spatial resolution $\Delta x \approx \Delta z = 0.0225\,\rho_{e0}$. Our choice of $N_x$ and $N_z$ ensures that both the Larmor radius and the Debye length, 
\begin{equation}
    \label{eq: lambda de}
    \lambda_{\mathrm{D}s} = \sqrt{\frac{T_s}{4 \pi n_s e^2}} = {\rho_s} \sqrt{\frac{2} {9\beta_s}},
\end{equation}
remain resolved in both the high- and low-$\beta$ regions throughout the entire simulation. The large domain size and value of $\tau_0 \Omega_0$ ensure clear scale separation between the microscale SFHI and the macroscale SCI. We use a timestep $\Delta t = 0.01\,\Omega_0^{-1}$, which satisfies the Courant--Friedrichs--Lewy (CFL) condition throughout the simulation, and evolve the simulations until $5 \, \tau_0$. We use cubic interpolation for the particle shape and $400$ particles per cell per species. 

In addition to the uniform background magnetic field and density, we impose an initial seed perturbation
\begin{equation}
    \frac{\delta B_z}{B_0} = \frac{\delta n_s}{n_{0s}} = -a \cos(kx),
\end{equation}
where $k = 2\pi/L_x$. The chosen values of $L_x$ and $\tau_0 \Omega_0$ ensure that the condition for the fastest-growing SCI mode, $k/k_0 \approx 42 \gg 1$, is satisfied throughout the whole simulation. Our chosen box size also gives $k \rho_{e0} = 0.02$, which guarantees that the assumption of small gyroradius~\eqref{eq: rho ordering} holds for the perturbation. To capture the two-phase dynamics while maintaining a relatively small initial perturbation amplitude, we select $a = 0.3$. We initialise electron and positron (non-relativistic) counter-streaming bulk flows $\mathbf{u}_e = -\mathbf{u}_p \propto \hat{\mathbf{y}}$ to produce the current in~\eqref{eq: current and charge} corresponding to the initially imposed magnetic-field perturbation. Finally, to satisfy the pressure-balance condition~\eqref{eq: Pressure Balance} at $t=0$, we impose a corresponding temperature perturbation of the form
\begin{equation}
    \frac{\delta \theta_s}{\theta_{s0}}
    =
    \frac{B_0}{B_z}\left(1 + \frac{1}{2\beta_{s0}} \right)
    -
    \frac{1}{2\beta_{s0}}\frac{B_z}{B_0}
    - 1 \approx a\left(1 + \frac{1}{\beta_{s0}} \right)\cos(kx),
\end{equation}
where $\theta_{s0} \equiv T_0/m_e c^2$, and the last expression is obtained in the linear limit $a \ll1 $.

\subsection{SFHI Excitation}
In Figure~\ref{fig: B/B0}, we plot the evolution of the integrated energy of each of the magnetic-field components as a function of time. At $t  = 0.17 \, \tau_0 $, when the pressure anisotropy has reached $\Delta \approx -0.075 \approx -2.8/\beta_\parallel$ and $\beta_\perp \approx 34$, we observe a rapid increase in $\delta B_x$ and $\delta B_y$, characteristic of the SFHI. The measured excitation time is somewhat later than our theoretical prediction $t_\text{onset} = 0.066 \, \tau_0$ [see \eqref{eq: onset time}]. This discrepancy is likely the result of a finite value of $\tau_0 \Omega_0$, as scans with larger values of this parameter, and hence better scale separation between the slow cooling and fast SFHI, show the latter being excited closer to the theoretical prediction (see Appendix~\ref{Appendix: Effect of Particle Noise on PIC Simulations}). 

\begin{figure}
\centering
\includegraphics[width=\linewidth]{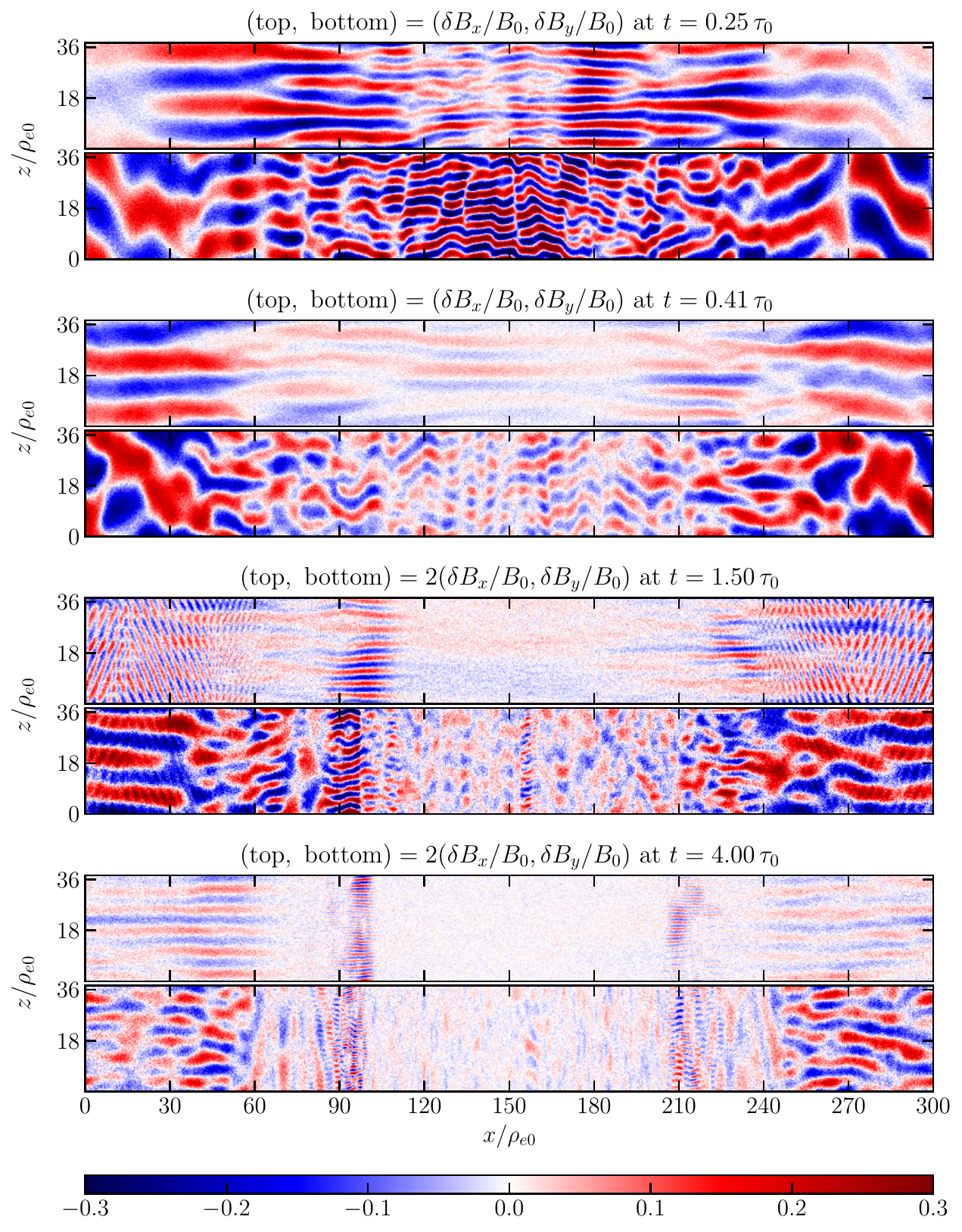}
\caption{Snapshots of $\delta B_x/B_0$ and $\delta B_y/B_0$ obtained from our PIC simulations at times $t = 0.25 \, \tau_0$, $t = 0.41 \, \tau_0$, $t = 1.5 \, \tau_0$, and $t = 4 \, \tau_0$. The panels demonstrate the evolution of the SFHI and its interplay with the SCI discussed in Section~\ref{sec: Particle-in-Cell Simulations}. As a result of the SCI, the SFHI fluctuations initially have a smaller scale in the middle of the box. At $t = 1.5 \, \tau_0$, a sub-dominant, oblique structure emerges on the sides, at smaller scales than the initial SFHI excitations. Eventually, the SFHI is shut off in the middle of the box due to the PIC noise (see Appendix~\ref{Appendix: Effect of Particle Noise on PIC Simulations}).} 
\label{fig: Delta B PIC}
\end{figure}

The SFHI predominantly generates magnetic fluctuations perpendicular to the mean magnetic field, as shown in Figure~\ref{fig: Delta B PIC}, where we give snapshots of $\delta B_x$ and $\delta B_y$ at four different points in time, illustrating the main stages of the system's evolution. As can be seen in the $t = 0.25 \, \tau_0$ snapshots, the unstable $\delta B_x$ perturbations have wavevectors parallel to the background magnetic field, while the $\delta B_y$ perturbations are oblique, which agrees with the observations of \cite{Zhdankin_etal-2023}. The growth of the $\delta B_y$ perturbations is characteristic of the oblique, kinetic firehose instability; the parallel modes are likely manifestations of the parallel firehose instability. 

\begin{figure}
\centering
    \includegraphics[width=\linewidth]{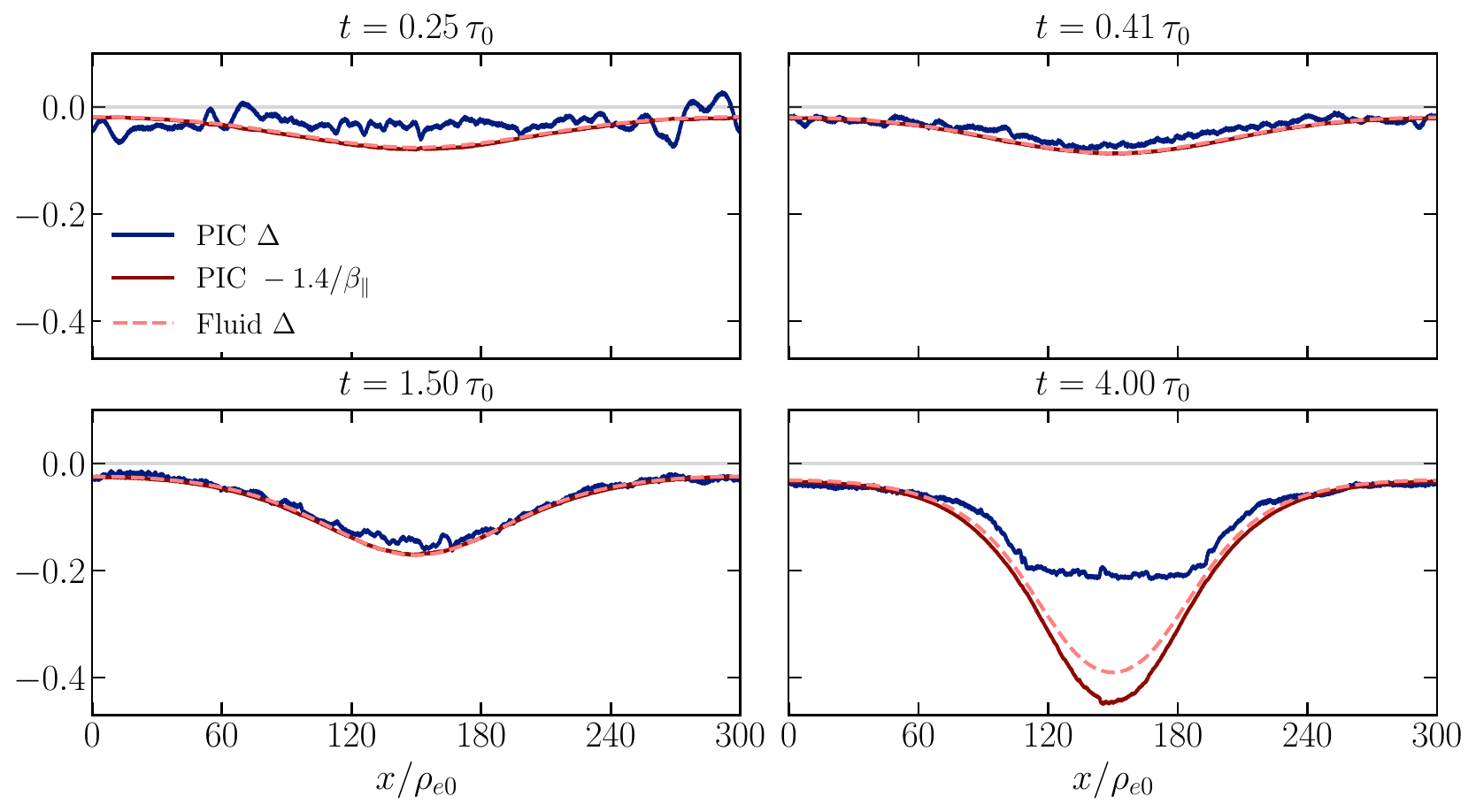}
\caption{
Variation of the $z$-averaged total pressure anisotropy with $x$ (blue line), along with the threshold \eqref{eq: threshold} for the firehose instability with $\Cth = 1.4$ (solid dark-red line), at the same times and in the same PIC simulations as Figure~\ref{fig: Delta B PIC}. Initially the anisotropy grows, but, once the SFHI is excited, it is pinned to the threshold~\eqref{eq: threshold} with $\Cth = 1.4$. As discussed in Section~\ref{sec: Fluid Simulations} and Appendix~\ref{Appendix: Effect of Particle Noise on PIC Simulations}, at later times, numerical particle collisions result in pressure isotropisation and shut-off of the SFHI in the middle of the box. The pressure anisotropy at the same times obtained in a simulation of the fluid model~\eqref{eq: energy},~\eqref{eq: continuity},~\eqref{eq: induction},~\eqref{eq: KMHD 1}, and~\eqref{eq: threshold} is shown by the dashed light-red lines and discussed in Section~\ref{sec: Fluid Simulations}. }
\label{fig: pressure anisotropy PIC}
\end{figure}

When the firehose perturbations grow to appreciable amplitude, they begin to scatter particles, which results in isotropisation and thus a decrease of $|\Delta|$ below the SFHI threshold~\eqref{eq: threshold}, thereby suppressing the instability. Continued synchrotron emission subsequently drives the pressure anisotropy beyond the threshold once again, re-exciting the SFHI. This repeated cycle of anisotropy build-up, instability growth, and nonlinear saturation produces the oscillatory behaviour of $\delta B_x$ and $\delta B_y$ shown in Figure~\ref{fig: B/B0}. Thus, the main consequence of the SFHI is the regulation of pressure anisotropy to the threshold~\eqref{eq: threshold}. In Figure~\ref{fig: pressure anisotropy PIC}, where we plot the measured pressure anisotropy and the threshold~\eqref{eq: threshold} as a function of $x$ at different times, we do indeed observe this pinning, with the expected value of $\Cth = 1.4$ in \eqref{eq: threshold}. Overall, the initial evolution of the SFHI agrees with \cite{Zhdankin_etal-2023}. 

\begin{figure}
\center
\includegraphics[width=\linewidth]{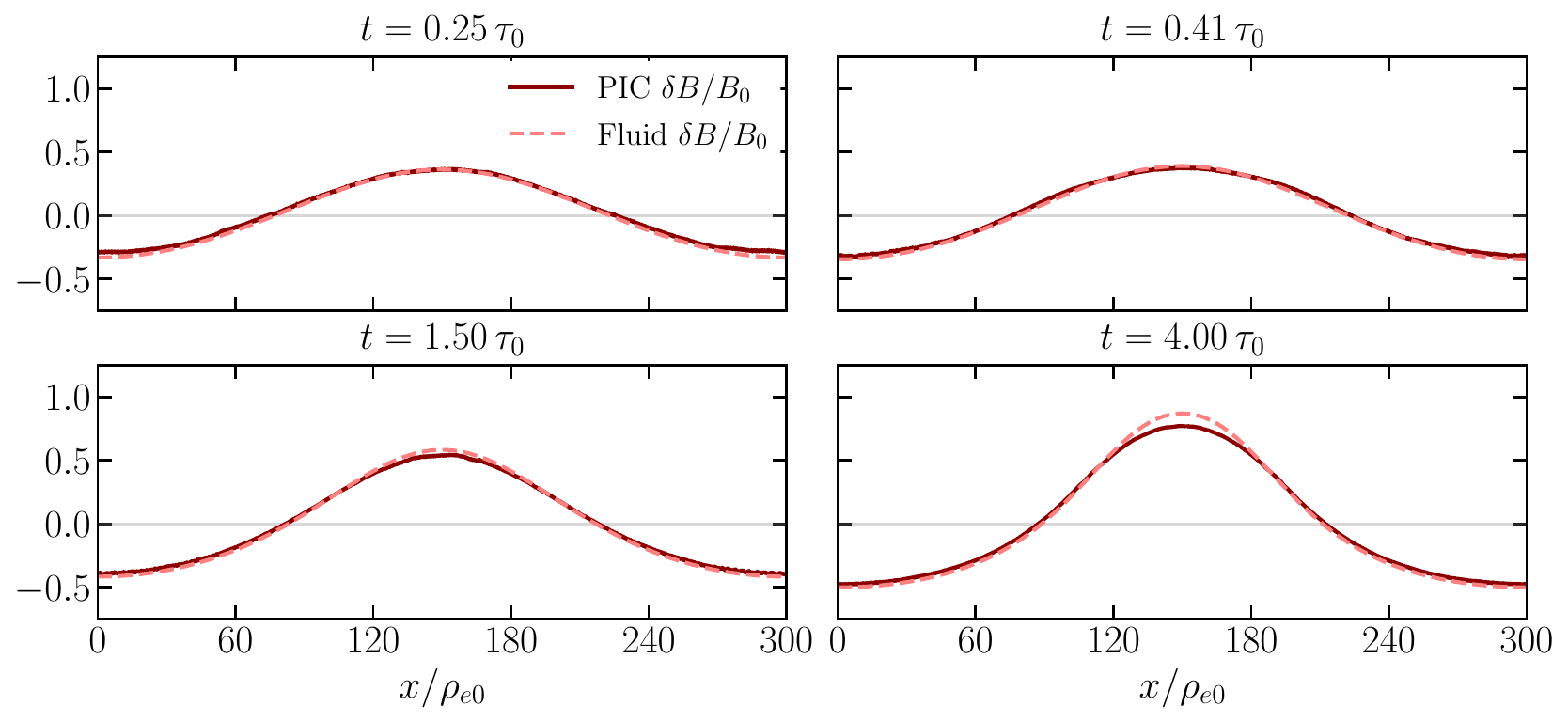}
\caption{The perturbed field strength $\delta B/B_0$ for the same PIC (solid dark-red lines) and fluid (dashed light-red lines) simulations, and at the same times as the plots of Figure~\ref{fig: pressure anisotropy PIC}. The density profile evolves in the analogous way, due to the condition of flux freezing. There is excellent agreement between the PIC (Section~\ref{sec: Particle-in-Cell Simulations}) and the fluid data (Section \ref{sec: Fluid Simulations}) until the SFHI is shut off in the middle of the box. Even after then, there remains good agreement outside the middle of the box.} 
\label{fig: delta B, delta n compare}
\end{figure}

\subsection{Linear SCI Growth in SFHI-Infested Medium}

\begin{figure}
\center
\includegraphics[width=\linewidth]{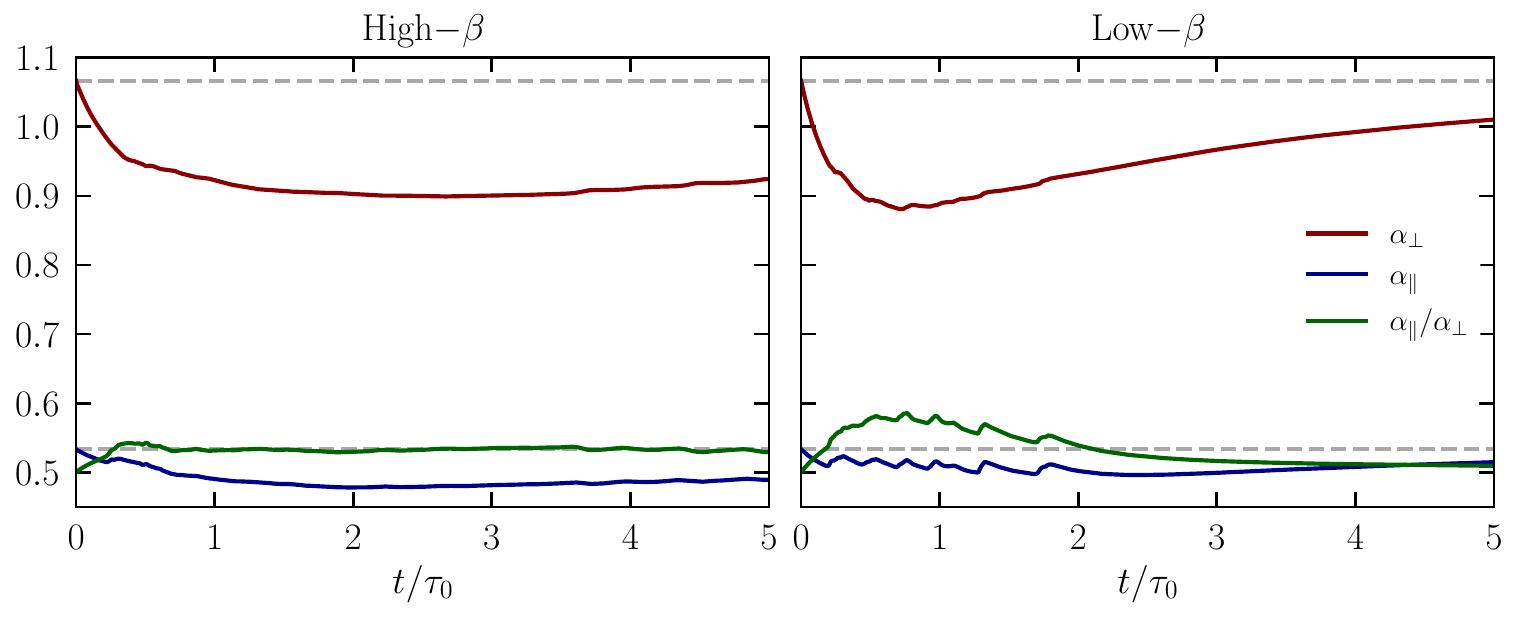}
\caption{The coefficients $\alpha_\perp$ (red line) and $\alpha_\parallel$ (blue line), defined in~\eqref{eq: alpha def}, at $x = 0$ (high-$\beta$ region) and $x = 150 \, \rho_{e0}$ (low-$\beta$ region), as functions of time, computed in the same PIC simulation as shown in Figures~\ref{fig: Delta B PIC}--\ref{fig: delta B, delta n compare}. The top and bottom grey dashed lines correspond respectively to the Maxwell--Jüttner values $\alpha_\perp = 16/15$ and $\alpha_\parallel = 8/15$. The ratio $\alpha_\parallel/\alpha_\perp$, which determines the pressure anisotropy \eqref{eq: shut down} when the SFHI is shut off, is shown by the green line and remains around $0.5$ in both regions. In the high-$\beta$ regions, the coefficients tend to the constant values $\alpha_\perp = 0.9$ and $\alpha_\parallel=0.48$, but eventually start to increase near the end of the simulation as a result of the finiteness of the number of computational particles per Debye area, as discussed in Appendix \ref{Appendix: Effect of Particle Noise on PIC Simulations}. }
\label{fig: alpha}
\end{figure}

On timescales of order~$\tau_0$, we start to observe a growing perturbation of the magnetic field's strength, i.e.,~$\delta B \approx \delta B_z$, in the centre of the domain. This occurs due to the SCI. In Figure~\ref{fig: delta B, delta n compare}, we plot z-averaged $\delta B$ as a function of $x$ at the same moments in time as the snapshots in Figure~\ref{fig: Delta B PIC}. The density changes simultaneously in proportion to $B$, maintaining the flux-freezing relation \eqref{eq: flux freezing}. The enhanced magnetic field in the centre of the perturbation causes this region to cool faster than its surroundings, thereby creating a temperature decrease. Consequently, the SCI results in the middle of the box becoming denser, cooler, and more strongly magnetised than the sides. Most interestingly, as illustrated in Figure~\ref{fig: pressure anisotropy PIC}, this means that the middle region has a substantially lower $\beta$ than the edges. Since the SCI develops within an SFHI background in which pressure anisotropy is pinned to the threshold \eqref{eq: threshold}, we observe that the low-$\beta$ region has a larger (negative) $\Delta$. Despite the developing strong inhomogeneities in the aforementioned quantities, we find the whole domain to be in approximate pressure balance. 

As the linear development of the SCI described in~Section~\ref{sec: Synchrotron Cooling Instability} depends on the parameters $\alpha_\perp$ and $\alpha_\parallel$ defined in \eqref{eq: alpha def} and assumed constant in our fluid theory, we measure their values in both the high- and low-$\beta$ regions, and show them as functions of time in Figure~\ref{fig: alpha}. We find that, in the high-$\beta$ regions (left panel of Figure~\ref{fig: alpha}), the two parameters do indeed tend to constant values, $\alpha_\perp = 0.9$ and $\alpha_\parallel = 0.48$, whence $\alpha = 0.76$. These differ slightly from the Maxwell--Jüttner values $\alpha_\perp = 16/15$ and $\alpha_\parallel = 8/15$, which is unsurprising as the underlying distribution function is not exactly thermal. Surprisingly though, both parameters eventually begin to increase in our simulations. In the low-$\beta$ region (right panel of Figure~\ref{fig: alpha}), this increase starts at around $t \approx  \tau_0$, while in the high-$\beta$ region, it begins much later, at $t \approx 3.5 \, \tau_0$. As we discuss in Appendix~\ref{Appendix: Effect of Particle Noise on PIC Simulations}, this behaviour is a numerical artifact associated with the finite number of computational particles per Debye area. The higher-resolution simulations presented there indicate that $\alpha_\perp$ and $\alpha_\parallel$ saturate at approximately constant values in SFHI-dominated plasmas, but further work is required to determine the exact values of these constants. Nonetheless, Figure~\ref{fig: alpha} shows that their ratio, which determines $\Delta$ \eqref{eq: shut down} when the SFHI is shut off, remains close to $0.5$ in both the high- and low-$\beta$ regions, similar to the case for a thermal distribution. 

In Figure~\ref{fig: B/B0}, we compare the linear evolution of the SCI in an SFHI-infested medium discussed in Section~\ref{sec: Synchrotron Cooling Instability} to the PIC simulation results. We let $\alpha_\perp = 0.9$, $\alpha_\parallel = 0.48$, and $\Gamma = 4/3$ in the analytic fluid model of the linear evolution, and find that both \eqref{eq: cth delta B evolution} (with $\Cth = 1.4$) and the simplified relation \eqref{eq: full B evolution} (with $\Cth = 0$) describe the growth of $\delta B_z$ well until $t \approx 2.2 \, \tau_0$. We will discuss the late-time departure from the linear theory and nonlinear growth of the SCI in Section~\ref{sec: Emergent Two-Phase Structure of Synchrotron Cooling Plasmas}. 

\subsection{Modification of SFHI Spatial Structure by SCI}
The growth of the SCI also modifies the spatial structure of the SFHI. At early times, the characteristic length scale associated with the SFHI, the Larmor radius $\rho_e$, is smaller in the middle of the box (where the magnetic field is stronger). This follows from the form of the ultra-relativistic Larmor radius, $\rho_e \propto T/B \propto P/nB$. Initially, $\beta$ is high, so pressure balance everywhere is dominated by the thermal pressure $P$, which remains approximately uniform; combined with flux freezing, which gives $n \propto B$, this yields $\rho_e \propto 1/B^2$. The SFHI structures therefore have a smaller scale in the middle of the box, as evident in the first panel of Figure~\ref{fig: Delta B PIC}.

Later, at $t \approx 1.5 \, \tau_0$, Figure~\ref{fig: Delta B PIC} shows the emergence of oblique modes in both $\delta B_x$ and $\delta B_y$ superimposed on the SFHI, at the sides of the box. They also have a $\delta B_z$ component (not shown), grow on scales smaller than those of the SFHI and fade away by $t \approx 3.7 \, \tau_0$. The appearance of these modes does not affect our overall picture. They appear to be the result of a secondary, Larmor-scale instability (cf. \citealt{Bott_etal-2024, Bott_etal-2025}). In preliminary homogeneous simulations of the SFHI, we have found that this secondary instability occurs only if the parameter $\tau_e \Omega_e / \beta_e$, where $\tau_e$ is defined in~\eqref{eq: cooling time}, is sufficiently small. Pressure balance and flux freezing give $\tau_e \Omega_e / \beta_e \propto 1/T^{3}$, so this parameter is smallest in the hotter, high-$\beta$ regions at the sides of the box, making the secondary instability more likely to be excited there. As the plasma cools, $\tau_e \Omega_e/\beta_e$ increases, quenching the secondary instability, as observed. 

\medskip
To summarise, by means of PIC simulations, we have demonstrated that in a collisionless, synchrotron-cooling plasma, the SFHI is quickly excited and provides an effectively collisional background, in which the SCI develops. The SCI creates regions of high and low $\beta$, which in turn modify somewhat the spatial structure of the SFHI. The initial growth of the SCI is well described by the linear theory developed in Section~\ref{sec: Synchrotron Cooling Instability}, up to $t \approx 2.2\, \tau_0$. In the next section, we consider the evolution of both instabilities on longer timescales, on which the SCI reaches nonlinear amplitudes and saturates.

\section{Emergent Two-Phase Structure of Synchrotron-Cooling Plasmas}
\label{sec: Emergent Two-Phase Structure of Synchrotron Cooling Plasmas}

Figure~\ref{fig: pressure anisotropy PIC} shows that, as the SCI evolves, the contrast in $\beta$ becomes larger. Eventually, the lowest $\beta$ will drop down to the value \eqref{eq: SFHI shut off} at which the SFHI is shut off. From that moment on, the plasma becomes effectively two-phase. The first phase has a high $\beta$ and is filled with the small-scale SFHI fluctuations. As a result, the pressure anisotropy there remains pinned to the instability threshold~\eqref{eq: threshold} throughout the macroscopic dynamics. The second phase has $\beta = \mathcal{O}(1)$, resulting in the suppression of the SFHI as described in Section \ref{sec: Shut-Off of the Synchrotron Firehose Instability}. This low-$\beta$ phase is effectively laminar, with its pressure anisotropy growing freely, no longer regulated by SFHI-induced particle scattering. 

\subsection{Two-Phase Model of Synchrotron-Cooling Plasmas}
\label{sec: Two-Phase Model of Synchrotron Cooling Plasmas}
We propose the following effective fluid description of the nonlinear evolution of the SCI, explicitly incorporating  the two-phase nature of the medium into our model. The plasma as a whole satisfies the continuity~\eqref{eq: continuity}, momentum~\eqref{eq: KMHD 1}, and magnetic induction~\eqref{eq: induction} equations. For the fastest-growing SCI modes ($k\gg k_0$) the momentum equation reduces to the perpendicular pressure balance~\eqref{eq: Pressure Balance}. The two-phase nature of our model enters through the different particular choices of the equation of state for each phase. 

Prior to the onset of the SFHI, the plasma evolves according to the double-adiabatic equations~\eqref{eq: pressure evolv perp} and~\eqref{eq: pressure evolv parallel}, with $\alpha_\perp = 16/15$ and $\alpha_\parallel = 8/15$. Once the SFHI is triggered and causes efficient particle scattering, these are replaced with the energy equation~\eqref{eq: energy} and the firehose marginality condition~\eqref{eq: threshold}. Note that our PIC simulations in Section~\ref{sec: Particle-in-Cell Simulations} show that once the SFHI is excited, $\alpha_\perp = 0.9$ and $\alpha_\parallel = 0.48$. After the emergence of the two-phase state, the low-$\beta$ regions satisfying the SFHI shut-off condition~\eqref{eq: SFHI shut off} revert back to the double-adiabatic description, while the high-$\beta$ regions continue to evolve under the SFHI-regulated closure. 

As long as the high-$\beta$ regions have $\beta \gg 1$, scattering of particles by firehose fluctuations results in nearly isotropic (firehose-marginal) pressure. In this limit, the energy equation~\eqref{eq: energy} reduces to
\begin{equation}
    P\frac{d}{dt} \ln\left( \frac{P}{n^{\Gamma}}\right) = - {\alpha}\frac{2 r_e^2}{m_e^2 c^3 } \frac{B^2 P^2}{n},
\end{equation}
where $\Gamma = 4/3$, and we have used the fact that $B/B_0 = n/n_0$ as the SCI evolves. By contrast, in the low-$\beta$ regions, the perpendicular pressure, which is the only relevant pressure in the pressure-balanced limit of perpendicular compression, evolves according to~\eqref{eq: pressure evolv perp}, which we rewrite as follows:
\begin{equation}
    P_\perp\frac{d}{dt} \ln\left( \frac{P_\perp}{n^{\Gamma_\perp}}\right) = - \alpha_\perp\frac{2 r_e^2}{m_e^2 c^3 } \frac{B^2 P_\perp^2}{n},
\end{equation}
with $\Gamma_\perp$ anisotropy-dependent and given by \cite{Wierzchucka_etal-2026}. For example, as discussed in Section~\ref{sec: Pressure Evolution}, $\Gamma_\perp = 8/5$ when $P_\perp \approx P_\parallel$. Thus, the high- and low-$\beta$ phases are characterised by different (perpendicular) adiabatic indices and values of~$\alpha_\perp$.

\subsection{Fluid Simulations}
\label{sec: Fluid Simulations}
To illustrate this two-phase description and demonstrate how it emerges as a result of the SCI in the nonlinear regime, we solve the fluid model introduced in Section \ref{sec: Two-Phase Model of Synchrotron Cooling Plasmas} numerically. We use the same parameter set as that of the PIC-simulation setup discussed in Section~\ref{sec: Particle-in-Cell Simulations}, except now we allow all quantities to vary only in the $x$-direction. Thus, our fluid simulations do not directly include the SFHI, but instead model its effect by automatically enforcing the threshold condition \eqref{eq: threshold} in the regions where the dynamics attempt to cross into the SFHI-unstable regime. Full details of these two-phase-fluid simulations are provided in Appendix~\ref{Appendix: Details of Fluid Simulations}. Note that, in our fluid simulations, we artificially delay the onset of the SFHI, i.e., its effect of pinning pressure anisotropy to the threshold, to the time at which the SFHI is excited in the PIC simulations, $t_\text{PIC}  = 0.17 \, \tau_0$. Additionally, after this time, we change the values of $(\alpha_\perp, \alpha_\parallel)$ from the Maxwell--Jüttner values $(16/15, 8/15)$ to the values $(0.9, 0.48)$ measured in the PIC simulations. These two adaptations ensure that we can directly compare our fluid picture to the PIC simulations despite the latter not having perfect scale separation.

\subsubsection{Comparison Between PIC and Fluid Simulations}
\label{sec: Comparison Between PIC and Fluid Simulations}
\begin{figure}
\centering
\includegraphics[width=0.9\linewidth]{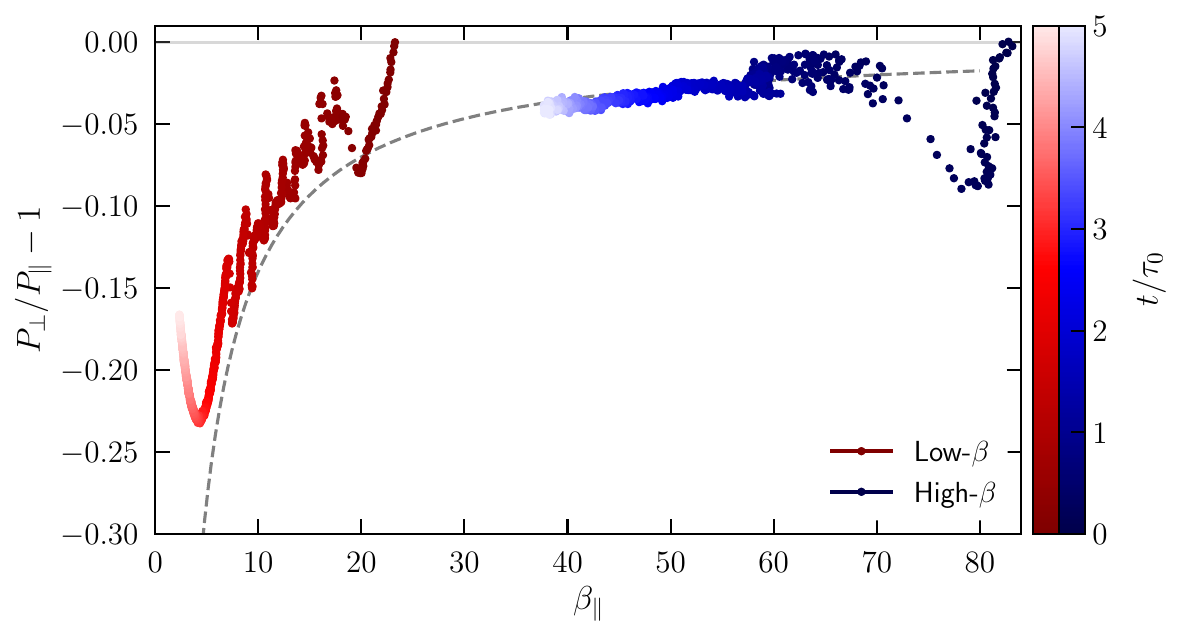}
\caption{
Evolution of the pressure anisotropy as a function of (total) $\beta_{\parallel}$ at $x = 150 \, \rho_{e0}$ (red points) and at $x = 0 \, \rho_{e0}$ (blue points). The firehose threshold, \eqref{eq: threshold} with $C_\text{th} = 1.4$, is shown by the grey dashed line. The shade of the colour of the points indicates the time~$t/\tau_0$. At $t \approx 2.2 \, \tau_0$, numerical isotropisation effects discussed in Appendix~\ref{Appendix: Effect of Particle Noise on PIC Simulations} start to influence the dynamics, eventually dominating over synchrotron cooling at $t \approx 3.3 \, \tau_0$, and leading to $\Delta$ increasing in the low-$\beta$ region.}
\label{fig: marginality}
\end{figure}

We compare our fluid model at early and intermediate timescales to the PIC simulation in Figures~\ref{fig: pressure anisotropy PIC} and~\ref{fig: delta B, delta n compare}, where we plot the pressure anisotropy and magnetic-field profiles, respectively, at four moments in time. The pressure anisotropy in the fluid model quickly gets pinned to the threshold value, while the initial magnetic-field perturbation grows on longer timescales due to the SCI, alongside a corresponding perturbation in plasma's density. The two approaches agree within about $7\%$ for $t \lesssim 2.2\, \tau_0$ (the linear regime), after which they start to diverge slightly in the middle of the box. For example, in Figure~\ref{fig: delta B, delta n compare} at $t = 4 \, \tau_0$ and $x \approx 150 \, \rho_{e0}$, the magnetic field obtained from the fluid simulations exceeds the PIC-simulation value by about $14\%$, while at the same location and time, Figure~\ref{fig: pressure anisotropy PIC} shows the low-$\beta$-region $|\Delta|$ decreasing in the PIC simulations. This pressure isotropisation can be seen more clearly in Figure~\ref{fig: marginality}, where we show $\Delta$ in both the high- and low-$\beta$ regions as functions of $\beta$ throughout the entire PIC simulation. As a result of this isotropisation, the PIC simulations show that the SFHI shuts off in the low-$\beta$ region at $t \approx 2.2 \, \tau_0$, when $\beta_\perp = 4.5$,  $\beta_\parallel = 5.9$, and $\Delta = -0.24$. This can be seen most clearly in the $t = 4\, \tau_0$ snapshot in Figure~\ref{fig: Delta B PIC}, where SFHI fluctuations have completely vanished from the middle of the domain. However, this extinguishing process is not the one described in Section \ref{sec: Shut-Off of the Synchrotron Firehose Instability}, but a numerical artifact that occurs due to the finiteness of the number of computational particles per Debye area, resulting in artificial isotropisation of the particle distribution at low temperatures. We discuss this matter further in Appendix~\ref{Appendix: Effect of Particle Noise on PIC Simulations}. For these reasons, from $t = 2.2 \, \tau_0$ onwards, we use only the two-phase-fluid simulations to study the subsequent evolution of the SCI on much longer timescales. 

\subsubsection{Long-Time SCI Evolution in Fluid Simulations}
\begin{figure}
\centering
\includegraphics[width=\linewidth]{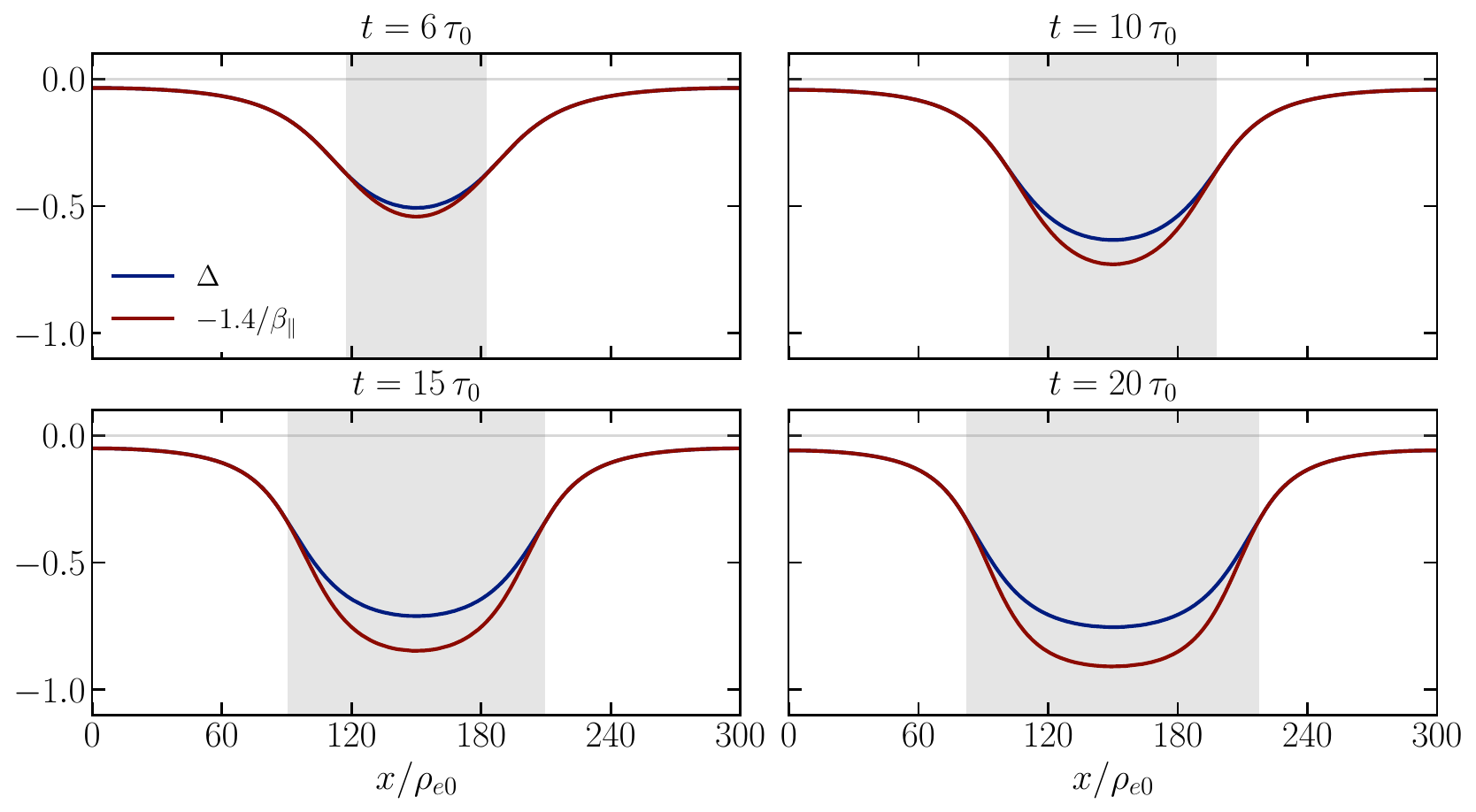}
\caption{The profiles of the total pressure anisotropy with $x$, and the firehose threshold~\eqref{eq: threshold} at times $t = 6 \, \tau_0$, $t = 10 \, \tau_0$, $t = 15 \, \tau_0$, and $t = 20 \, \tau_0$. The data is obtained from simulations of the fluid model~\eqref{eq: energy},~\eqref{eq: continuity},~\eqref{eq: induction},~\eqref{eq: KMHD 1}, and~\eqref{eq: threshold}. The regions where the anisotropy stops being pinned to the threshold~\eqref{eq: threshold}, i.e., where the SFHI is no longer active, are shaded in grey.}
\label{fig: pressure anisotropy fluid}
\end{figure}

\begin{figure}
\center
\includegraphics[width=\linewidth]{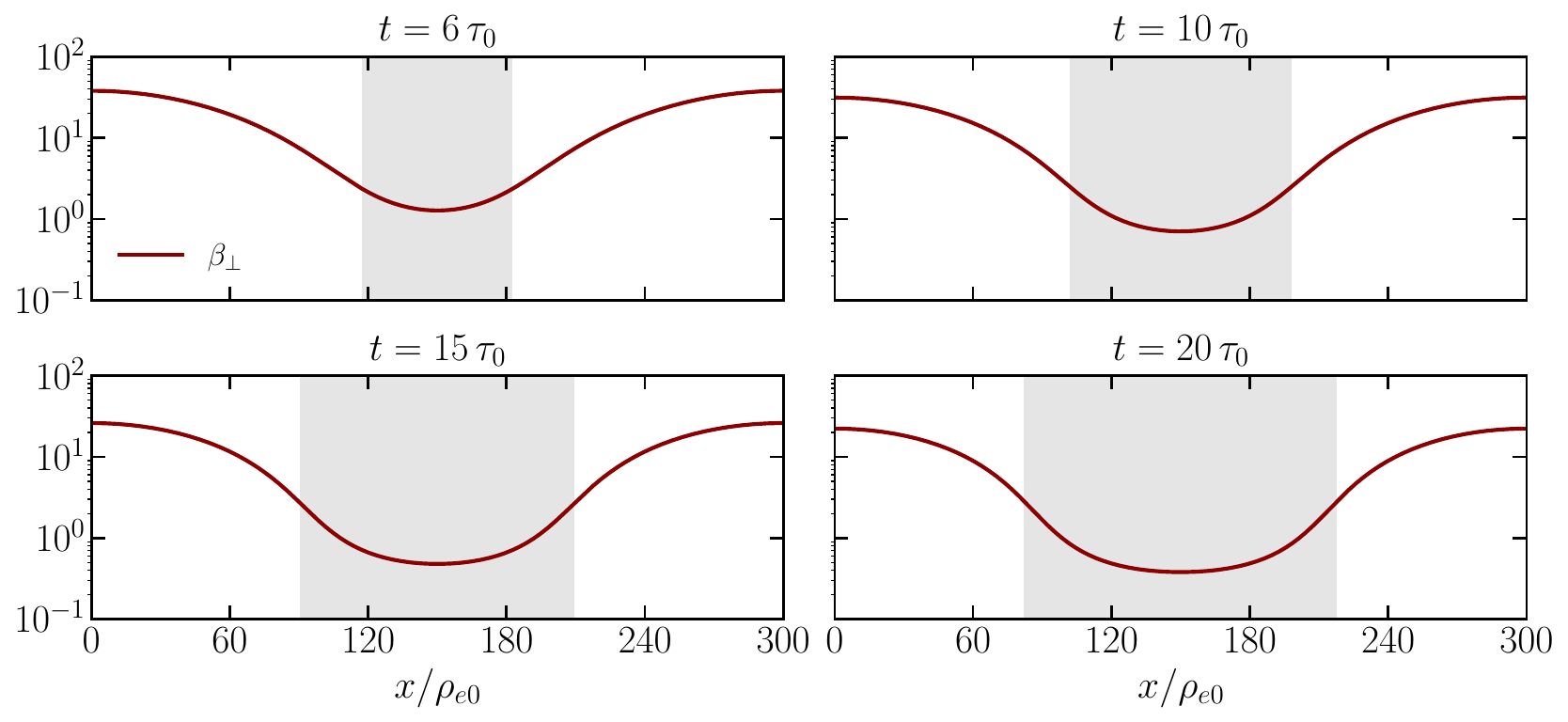}
\caption{The profiles of $\beta_{\perp}(x)$ from the same simulations and at the same times as in Figure~\ref{fig: pressure anisotropy fluid}. At saturation of the SCI, i.e., at $ t \approx 10 \, \tau_0$, $\beta_\perp = 0.7$ in the middle and $\beta_\perp = 31.3$ on the sides of the box.}
\label{fig: fluid beta}
\end{figure}

In Figure~\ref{fig: pressure anisotropy fluid}, we show the profiles of the pressure anisotropy and the firehose threshold~\eqref{eq: threshold}, obtained in our fluid simulations, at times beyond $t = 2.2 \, \tau_0$. Figure~\ref{fig: fluid long delta B, delta n compare} and Figure~\ref{fig: fluid beta} show the corresponding profiles of the magnetic-field strength and $\beta_\perp$, respectively.

We find that $\Delta$ stops being pinned to the threshold \eqref{eq: threshold} in the middle of the low-$\beta$ region at $t \approx 3.5 \, \tau_0$, when the pressure anisotropy is $\Delta = -0.35$, $\beta_\perp = 2.6$, and $\beta_\parallel = 4.0 $, which signifies the quenching of the SFHI there. This pressure anisotropy is slightly smaller in magnitude than the predicted value for SFHI shut-off \eqref{eq: shut down}, which, for $\alpha_\perp = 0.9$ and $\alpha_\parallel = 0.48$, is $\Delta = -0.47$. This modest discrepancy is likely due to the compressing flow generated by the SCI perturbation, which is still growing at this time, driving positive pressure anisotropy in the low-$\beta$ region according to the non-radiative terms of~\eqref{eq: pressure evolv perp} and \eqref{eq: pressure evolv parallel}. The emergence of the two-phase structure can be seen in Figure~\ref{fig: pressure anisotropy fluid}, where $\Delta$ is pinned to the threshold \eqref{eq: threshold} in the high-$\beta$ regions but is no longer constrained in the low-$\beta$ regions (light-grey shaded areas). 

Sometime after the SFHI begins to shut off, at $t \approx 10 \, \tau_0$, we observe the saturation of the SCI, with a maximum amplitude of $\delta B(x)/B_0 = 0.9$ at $x = L_x/2 = 150 \, \rho_{e0}$ (see Figure~\ref{fig: fluid long delta B, delta n compare}), and corresponding values of $\beta_\perp = 0.7$ and $\beta_\parallel = 1.9$ (see Figure~\ref{fig: fluid beta}). The saturation time is slightly earlier than the prediction \eqref{eq: saturation time}, here equal to $t_\text{sat} = 13.7 \, \tau_0$. The value of $\beta_\perp$ at saturation is also smaller than the linear-theory estimate \eqref{eq: beta crit c}, which for our setup is $\beta_{\perp c} = 3.29$. This discrepancy is likely due to nonlinear effects associated with the finite size of the SCI perturbations. Note that we evaluated \eqref{eq: beta crit c} and \eqref{eq: saturation time} using $\alpha_\perp = 0.9$, $\alpha_\parallel = 0.48$, and $\beta_{\perp 0} = 34$, i.e., the value of $\beta_\perp$ when the SFHI is first activated. Determining precisely whether SFHI shut-off or SCI saturation occurs first requires further studies of the pressure evolution in synchrotron-cooling plasmas, in particular, of the exact values of $\alpha_\perp$ and~$\alpha_\parallel$.

Post-saturation, the SCI begins to decay and the region free of SFHI fluctuations begins to expand outwards from the middle of the box, as can be seen in the plots of pressure anisotropy and the firehose threshold \eqref{eq: threshold} in Figure~\ref{fig: pressure anisotropy fluid}. The magnetic-field perturbation flattens in the middle and the plasma eventually returns to a homogeneous state (see Figure~\ref{fig: fluid long delta B, delta n compare}).  Due to the assumption of ultra-relativistic temperature, the fluid model solved here becomes no longer valid after about $t \approx 20 \, \tau_0$, at which point $T_\perp/m_e c^2 \approx 2$. To study the final evolution of the SCI, one would need to include the finite-$\theta$ corrections to the equations developed in Section \ref{sec: Relativistic, Cooling Plasmas}.

\begin{figure}
\center
\includegraphics[width=\linewidth]{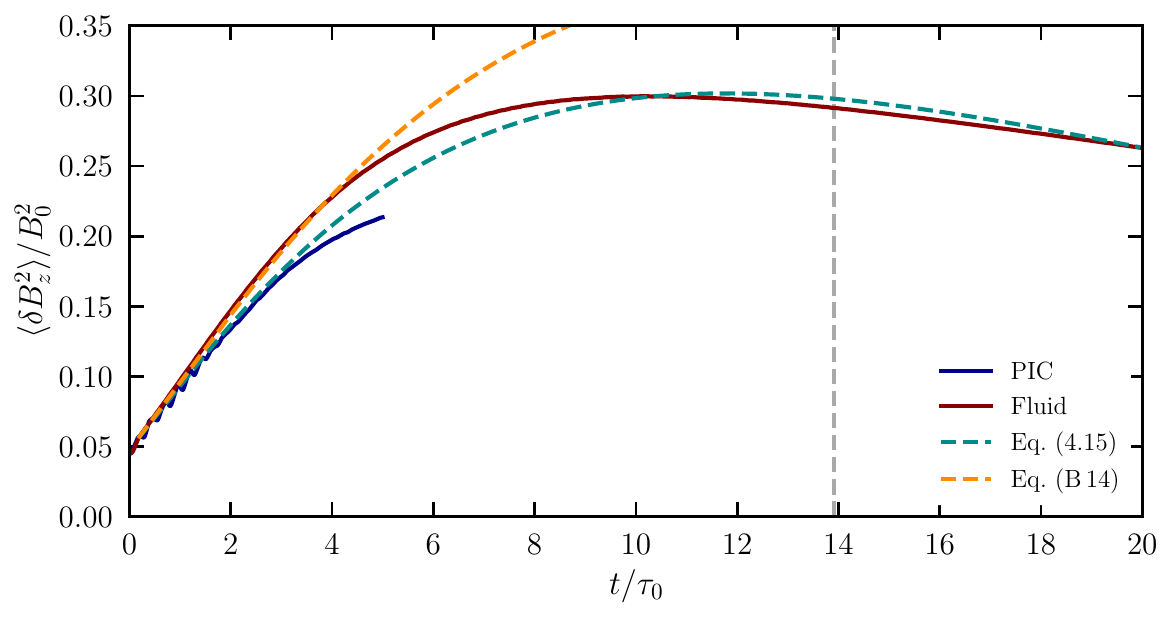}
\caption{Evolution of the mean (box-averaged) square of the magnetic-field component $\delta B_z/B_0$, whose growth is due purely to the SCI, obtained from simulations of the two-phase-fluid model described in Section \ref{sec: Two-Phase Model of Synchrotron Cooling Plasmas} (red line) and from the PIC simulation in Section~\ref{sec: Particle-in-Cell Simulations} (blue line). The linear prediction~\eqref{eq: cth delta B evolution} with $\alpha_\perp = 0.9$, $\alpha_\parallel = 0.48$, and $\Gamma = 4/3$ is shown by the orange dashed line, the simplified linear prediction~\eqref{eq: full B evolution} with analogous parameters but $\Cth = 0$ by the cyan line. Note that we begin the orange and cyan lines at the time when the SFHI is excited, using the value of $\beta_\perp$ at that time. The SCI saturates at $t \approx 10 \, \tau_0$, which is slightly earlier than the sum of the predicted saturation time \eqref{eq: saturation time}, $t_\text{sat}= 13.7 \, \tau_0$, and the PIC SFHI-onset time $t_\text{PIC} = 0.17 \, \tau_0$, here shown by the grey dashed line.} 
\label{fig: fluid B-B0}
\end{figure}

In Figure~\ref{fig: fluid B-B0}, the full evolution of the magnetic field obtained from our 1D fluid and 2D PIC simulations is summarised and compared to the linear theory from Section~\ref{sec: Synchrotron Cooling Instability} and Appendix~\ref{Appendix: Synchrotron Cooling Instability in Firehose-Infested Plasmas}. The figure shows that both sets of numerical results agree well with each other and with the linear prediction \eqref{eq: cth delta B evolution} until $t \approx 2.2\, \tau_0$, when the effects of numerical particle isotropisation become significant in the PIC simulations (see Section \ref{sec: Comparison Between PIC and Fluid Simulations} and Appendix~\ref{Appendix: Effect of Particle Noise on PIC Simulations}). After this time and until $t \approx 4 \, \tau_0$, while the fluid-simulation result is still well captured by~\eqref{eq: cth delta B evolution}, the PIC-simulation result, now no longer physical, is better described by the simplified evolution equation \eqref{eq: full B evolution}. This is likely due to the fact that the numerical isotropisation does not pin the anisotropy to the firehose threshold but rather to perfect isotropy, which corresponds to $\Cth = 0$, as assumed in the derivation of \eqref{eq: full B evolution}. The departure of the fluid simulation from the linear evolution~\eqref{eq: cth delta B evolution} at late times is expected, because nonlinear effects become important when the SCI reaches large amplitudes.

\begin{figure}
\center
\includegraphics[width=\linewidth]{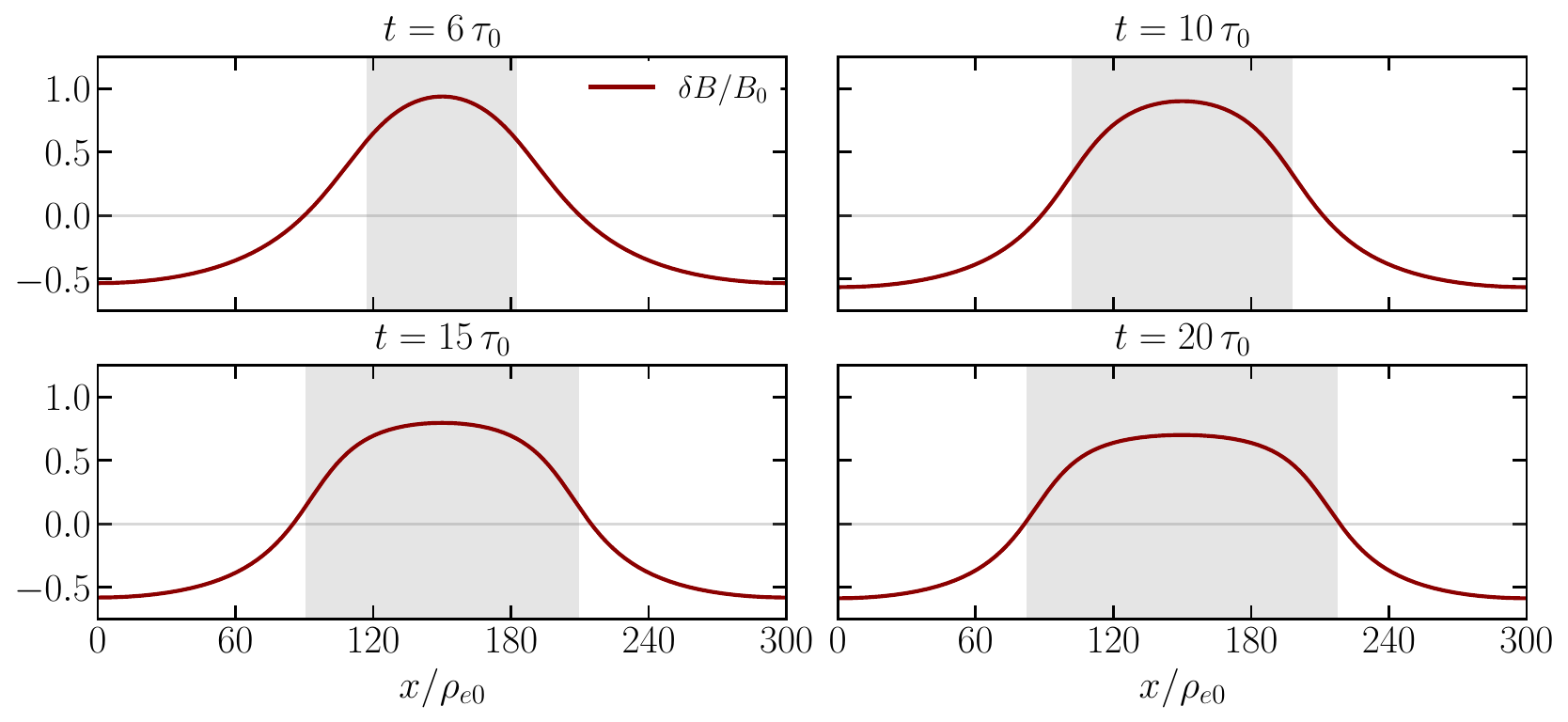}
\caption{The profiles of $\delta B(x)/B_0$ from the same simulation and at the same times as shown in Figure~\ref{fig: pressure anisotropy fluid}. The saturation occurs at $t \approx 10 \, \tau_0$, with $\delta B/B_0 = 0.9$ in the middle and $\delta B/B_0 = -0.57$ on the sides of the box. The density profile evolves in the analogous way, due to the condition of flux freezing, $\delta B/B_0 $ = $\delta n /n_0$.} 
\label{fig: fluid long delta B, delta n compare}
\end{figure}

\section{Summary}
\label{sec: summary}
By means of analytical calculations and numerical simulations, we have found that, as a high-$\beta$, magnetised, relativistic, collisionless, synchrotron-cooling plasma evolves, it filaments into a two-phase medium. This occurs due to the interaction of two instabilities, the SCI and the SFHI. We summarise the evolution of the plasma below.

As a high-$\beta$ plasma with an initial cooling time $\tau_0$ cools via synchrotron emission, negative pressure anisotropy is generated. This leads to the excitation of the microscopic firehose instability at time $t \sim \tau_0/\beta_0$. On the timescale $\sim\tau_0^{1/3} \Omega_0^{-2/3}$ after the excitation, when the firehose fluctuations become large enough, they start to scatter particles, and keep the pressure anisotropy pinned to the firehose threshold~\eqref{eq: threshold} \citep{Zhdankin_etal-2023}. 

A synchrotron-cooled plasma is also macroscopically unstable to the SCI, which develops on timescales $\tau_0$, creating regions of higher and lower~$\beta$. At some point, the SFHI shuts off in the low-$\beta$ regions but continues to exist in the high-$\beta$ regions. The plasma thus becomes a two-phase medium, with the two phases in total-pressure balance, but with different densities, temperatures, and microscopic dynamics, the latter manifesting macroscopically as different equations of state. One phase has high $\beta$ and is populated by firehose fluctuations, which scatter particles and so keep the pressure anisotropy pinned to the firehose threshold. The second phase has low $\beta$, so the firehose instability is quenched and pressure anisotropy can grow unchecked. The low-$\beta$ phase is more strongly magnetised, colder, and denser than its high-$\beta$ surroundings. 

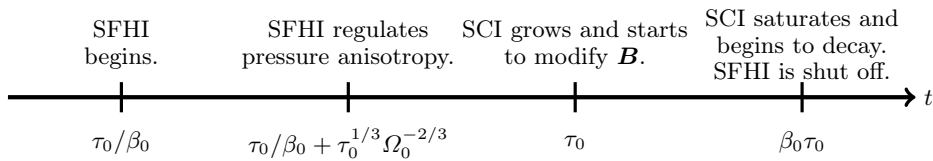
\begin{figure}
\centering
\begin{tikzpicture}
    \draw[ultra thick, ->] (0,0) -- (12,0);
    \node[anchor=west] at (12,0) {$t$};

    \foreach \x in {1.5, 4.5, 7.5, 10.5} {
        \draw[very thick] (\x, -0.2) -- (\x, 0.2);
    }

    \node at (1.5, -0.6) {$\tau_0 / \beta_0$};
    \node at (4.5, -0.6) {$\tau_0/\beta_0 + \tau_0^{1/3} \Omega_{0}^{-2/3}$};
    \node at (7.5, -0.6) {$\tau_0$};
    \node at (10.5, -0.6) {$\beta_0\tau_0$};

    \node[font=\small, align=center] at (1.5, 0.7) {SFHI \\begins.};
    \node[font=\small, align=center] at (4.5, 0.7) {SFHI regulates\\pressure anisotropy.};
    \node[font=\small, align=center] at (7.5, 0.7) {SCI grows and starts\\to modify $\boldsymbol{B}$.};
    \node[font=\small, align=center] at (10.5, 0.7) {SCI saturates and\\ begins to decay. \\ SFHI is shut off.};

\end{tikzpicture}
\caption{Chronological ordering of key processes in relativistic, synchrotron-cooling plasma along with their timescales. Note that determining if the SCI saturation or SFHI shut-off occurs first requires further investigation of $\alpha_\perp$ and $\alpha_\parallel$ in the nonlinear regime. }
\label{fig: timescales summary}
\end{figure}

The evolution of both phases is captured by the fluid equations~\eqref{eq: continuity},~\eqref{eq: induction}, and~\eqref{eq: KMHD 1}, but the evolution of $P_\perp$ and $P_\parallel$ in the two regions differs. The low-$\beta$ phase is laminar and described well by double-adiabatic collisionless equations~\eqref{eq: pressure evolv perp} and~\eqref{eq: pressure evolv parallel}. In contrast, the high-$\beta$ phase is firehose-turbulent and so modelled by~\eqref{eq: energy} and~\eqref{eq: threshold}, which capture the pinning of the anisotropy to the firehose threshold. On the timescale $\beta_0 \tau_0$, the SCI saturates and the inhomogeneity that has built up due to it begins to decay, eventually bringing the plasma into a homogeneous, low-$\beta$ state. We summarise the key timescales of this evolution in Figure~\ref{fig: timescales summary}.

\section{Discussion}
\label{sec: Discussion}
The filamentation of synchrotron-cooling plasmas could provide a tentative general paradigm for the formation of coherent structures in astrophysical environments. The most noteworthy examples are the recently observed, high-energy, long, thin non-thermal filamentary structures in the ICM \citep{Rudnick_etal-2022, Rajpurohit_etal-2022} and the Galactic Centre \citep{Yusef-Zadeh_etal-1984, Yusef-Zadeh_etal-2004, Yusef-Zadeh_etal-2022a}. Similar structures are also found in the cores of galaxy groups \citep{Owen_etal-2000, Rudnick_etal-2022} and in radio relics \citep{Bridle_Fomalont-1976}. Several models have been proposed to explain the formation of these structures; however, no consensus has yet emerged.

Non-thermal filaments commonly emit synchrotron radiation at radio wavelengths; in cases where the emitting particles attain sufficiently high Lorentz factors, e.g., in Galactic-Centre filaments powered by pulsar-wind nebulae (PWN), the synchrotron emission can extend into the X-ray band \citep{Wang_etal-2002, Bykov_etal-2017, Churazov_etal-2024, Yusef-Zadeh_etal-2024}. The ultra-relativistic electrons and positrons injected by PWN into the ambient magnetised plasma could provide the correct composition for the SCI to develop in such environments and lead to subsequent filament formation. In the picture of a synchrotron-cooling plasma presented above, the low-$\beta$ region generated by the SCI forms the filament core, while the turbulent, high-$\beta$ plasma is the surrounding medium. The laminar region at the centre of the proposed filament, with suppressed kinetic instabilities, provides an environment through which energetic particles can stream along the magnetic field \citep{Churazov_etal-2026}. Such a scenario provides an efficient channel for the rapid transport of high-energy particles and may offer insight into the propagation of cosmic rays.

The SCI has previously been invoked as a possible origin of filaments embedded in the nuclear wind, particularly those not associated with interactions involving obstacles. Examples include the Harp and the Radio Arc groupings of multiple filaments \citep{Yusef-Zadeh_etal-2022a}. However, in this case, as in the original derivation of the SCI by \cite{Simon_Axford-1967}, the plasma is not thought to be a pair plasma, but instead one in which an ultra-relativistic electron population supplies the pressure while a cold ion species supplies the inertia. The $\beta$ contrast attainable by the SCI is then limited by the pressure of the cold, non-cooling particles. We still expect such a plasma to filament into a two-phase medium, with the dynamics of the SFHI adjusted by the electron-ion composition: e.g., in such a plasma the SFHI is triggered later than in pair plasmas \citep{Zhdankin_etal-2023, Ghorbanalilu-2026}, so the medium will likely remain effectively collisionless, and the pressure anisotropy unregulated, for longer than in our model.

In the ICM, pressure is supplied mostly by non-relativistic thermal ions and electrons, which do not cool via synchrotron emission. The contribution of the relativistic lepton sub-population is too small for the SCI to be able to create a significant $\beta$ contrast. However, the SCI could be important in radio lobes and AGN-inflated bubbles, where the thermal plasma has been depleted and so the pressure is supplied predominantly by relativistic particles \citep{Owen_etal-2000, Birzan_etal-2004, Rudnick_etal-2022, Churazov_etal-2026}.

An important natural question is how the evolution of the SCI in three spatial dimensions differs from the results presented here. A valuable insight into this can be obtained from radiative MHD studies of thermal instabilities driven by cooling processes other than synchrotron emission. Current fluid simulations suggest that, in three dimensions, the filamentary structures produced by thermal instabilities remain aligned with the magnetic field \citep{Sharma_etal-2010, Kunz_etal-2012}. These studies also indicate that the parallel and perpendicular dimensions of a filament are limited by thermal conductivity to the corresponding Field length \citep{Sharma_etal-2010}, similarly to the SCI \citep{Eilek_Caroff_1979}. Since thermal conductivity perpendicular to the magnetic field is much smaller than in the parallel direction, these structures tend to be elongated along the field lines. In existing fluid simulations of filament formation driven by cooling, thermal conductivity is introduced as an additional term in the fluid equations \citep{Sharma_etal-2010, Kunz_etal-2012, Choi_Stone-2012, Jennings_Li-2021}. However, in the kinetic picture of the SCI presented here, the SFHI fluctuations may self-consistently regulate thermal conductivity through particle scattering \citep{Bott_etal-2021, Bott_etal-2025}. As a result, the SFHI may also self-consistently limit the perpendicular and parallel extent of SCI-generated filaments.

It is also interesting to consider how heating processes affect the SCI. The finite lifetime of the instability predicted by our linear theory and observed in our simulations is a consequence of the absence of an energy source that would balance the cooling losses. Fluid studies of both the SCI and other thermal instabilities have shown that, when heating is present, the plasma can evolve into a state in which cooling-driven filamentary structures become long-lived stable features \citep{Bodo_etal-1992, Sharma_etal-2010, Kunz_etal-2012}.

Overall, our work motivates several avenues for future research. Two- and three-dimensional studies of the SCI are needed to investigate the realistic morphology and statistical properties of the filaments that it produces, as well as their stability to processes such as the Kelvin--Helmholtz instability. It would also be valuable to examine how non-ideal effects, such as thermal conduction or collisions, which act to isotropise the particle distribution, modify the picture presented here. How (and whether) our two-phase picture applies in systems with external forcing and how radiative cooling shapes the turbulence that develops remain open questions. Recent studies of synchrotron cooling in ideal MHD are suggestive in this regard: driven turbulence there does give rise to filaments, which glow brightly in synchrotron emission \citep{Sun_etal-2026}. What those studies do not address is the character of large-scale turbulence driven by the SCI itself. Finally, the saturated values of the radiation coefficients $\alpha_\perp$ and $\alpha_\parallel$ in SFHI-dominated plasmas require further investigation, especially their dependence on key physical parameters such as~$\beta$.

In the wider context of extreme plasma physics, the fluid model developed here represents a step towards a deeper analytical understanding of collisionless relativistic plasmas. The evolution equations that we have found could be applied to a broad range of high-energy astrophysical environments in which relativistic electron-positron pair plasmas provide an important, or even dominant, component. Examples include pulsar winds, PWN, relativistic jets, and accreting black-hole coronae.

Naturally, on account of the assumptions made in this paper, namely that the bulk flows are non-relativistic and the heat fluxes are negligible, we have omitted physical processes that may be important in more complex scenarios. Relaxing these assumptions requires a full theory of relativistic kinetic MHD \citep{Abhishek_Stone-2026, Wierzchucka_etal-2026b}. However, the simple set-up considered here already suggests that collisionless magnetised plasmas behave differently in the relativistic, radiative regime than in the non-relativistic limit, e.g., through modified double-adiabatic equations \citep{Wierzchucka_etal-2026, Ley_etal-2026}, the generation of negative pressure anisotropy by synchrotron emission and spontaneous filamentation of an initially homogeneous medium. As increasingly sophisticated kinetic and fluid descriptions become available, the physics of relativistic collisionless plasmas is beginning to emerge as a coherent theoretical framework rather than a collection of isolated phenomena.

\section*{Acknowledgements}
We are grateful to F. Bacchini, A. Boone, A. F. A. Bott, B. Cerutti, E. Churazov, J. Edmiston, I. Khabibullin, J. M. Mehlhaff, M. L. Nastac, L. Sironi and A. G. R. Thomas for fruitful discussions. Simulations were performed at MareNostrum 5 (Spain), funded by EuroHPC-JU extreme scale access project KEEP: Kinetic Electrodynamics in Extreme Plasmas (EHPC-EXT-2025E02-066). Claude AI (Opus 5) assisted the authors in proof-reading the manuscript. A. W. was supported by a Clarendon Scholarship and the Merton College Tira Wannamethee Scholarship. P. J. B. was supported by a Leverhulme–Peierls Fellowship. R.~J.~E. was supported by the Simons Foundation grant MP-SCMPS-00001470. The work of A. A. S. and P. J. B. was supported in part by the UK STFC grant ST/W000903/1. A.~A.~S. and A. W. were also supported in part by the Simons Foundation via the Simons Investigator Award to A. A. S. This research was supported in part by the grant NSF PHY-2309135 to the Kavli Institute for Theoretical Physics (KITP).

\section*{Declaration of Interests}
The authors report no conflict of interest.

\appendix

\section{Derivation of Lowest-Order Momentum Equation}
\label{Appendix: Derivation of Lowest-Order Momentum Equation}
In this Appendix, we present a derivation of the species-summed momentum equation~\eqref{eq: KMHD 1} starting from~\eqref{eq: momentum mid}. One of the assumptions underlying the fluid model of Section \ref{sec: Relativistic, Cooling Plasmas} is that the bulk flow is non-relativistic, $u_s/c \ll 1$. Accordingly, in the derivation below, we expand the momentum equation to second order in~$u_s/c$. As a result of the ordering~\eqref{eq: ordering}, this requires retaining terms up to first order in~$u_s/c$ in the time derivatives and to second order in the terms involving spatial derivatives. 

Expanding the Lorentz transformation to second order in $u_s/c$, we find the following relations between the laboratory and peculiar variables:
\begin{equation}
    \label{eq: momentum 2nd}
    \mathbf{p} = \tilde{\mathbf{p}} + \left( \tilde{\gamma}m_s + \frac{\tilde{\mathbf{p}} \cdot \mathbf{u}_s}{2 c^2}\right)\mathbf{u}_s, \quad 
    \gamma = \left( 1 + \frac{u_s^2}{2 c^2}\right) \tilde{\gamma} + \frac{\mathbf{u}_s \cdot \tilde{\mathbf{p}}}{m_s c^2}.
\end{equation}
It is also useful to note that the Jacobian for the change of variables is 
\begin{equation}
    \label{eq: Jacobian}
    \left| \frac{\partial \mathbf{p}}{\partial \tilde{\mathbf{p}}}\right| = \frac{\gamma}{\tilde{\gamma}}
\end{equation}
to all orders of $u_s/c$. 

The distribution function is gyrotropic in the frame moving with the bulk velocity $\mathbf{u}_s$. Therefore, in general, the peculiar momenta should be split into parallel and perpendicular components:
\begin{equation}
    \tilde{\mathbf{p}} = \tilde{p}_\parallel \tilde{\mathbf{b}} + \tilde{p}_\perp( \hat{\mathbf{x}}\cos \phi + \hat{\mathbf{y}}\sin\phi),
\end{equation}
where $\tilde{\mathbf{b}} = \tilde{\mathbf{B}}/\tilde{B}$ and $\tilde{\mathbf{B}}$ is the magnetic field in the co-moving frame. Recalling the ideal-MHD condition~\eqref{eq: ideal E} and the Lorentz transformation for electromagnetic fields, one can show that
\begin{equation}
    \label{eq: B tilde}
    \tilde{\mathbf{B}} = \left( 1 - \frac{u_s^2}{2 c^2} \right) \mathbf{B} +  \frac{\mathbf{u}_s \cdot \mathbf{B}}{c} \frac{\mathbf{u}_s}{2 c}.
\end{equation}
Consequently, 
\begin{equation}
    \label{eq: gyro int}
    \int d \tilde{\mathbf{p}} \ \tilde{\mathbf{p}} \tilde{\mathbf{v}} f_s = P_{\perp s}( \mathbf{I} - \tilde{\mathbf{b}}\tilde{\mathbf{b}}) + P_{\parallel s}\tilde{\mathbf{b}}\tilde{\mathbf{b}},
\end{equation}
with $P_{\perp s}$ and $P_{\parallel s}$ defined in~\eqref{eq: perp and parallel pressure def}.

We now turn to the fluid equations. As expected, the continuity equation~\eqref{eq: Non-Covariant Continuty} is unchanged. Using~\eqref{eq: Jacobian}, one can also show that, to first order in $u_s/c$,
\begin{equation}
    n_s = \int d \tilde{\B{p}} f_s.
\end{equation}
Expanding the momentum equation~\eqref{eq: momentum mid} is more involved. We first focus on the pressure tensor~$\mathbf{P}_s$, defined in~\eqref{eq: moment definitions}. Inserting~\eqref{eq: momentum 2nd} and~\eqref{eq: Jacobian} into this definition, we find 
\begin{equation}
    \label{eq: pressure tensor mid}
    \mathbf{P}_s = \int d \tilde{\mathbf{p}} \ \tilde{\mathbf{p}} \tilde{\mathbf{v}} f_s + \frac{P_{\parallel s} - P_{\perp s}}{2 c^2} (\mathbf{u}_s \cdot \mathbf{b} ) \left( \mathbf{b}\mathbf{u}_s - \mathbf{u}_s \mathbf{b}\right).
\end{equation}
The first term of~\eqref{eq: pressure tensor mid} is given by~\eqref{eq: gyro int}, which, when written to second order in $u_s/c$, turns~\eqref{eq: pressure tensor mid} into
\begin{align}
    \label{eq: pressure 2nd final}
    \mathbf{P}_s = P_{\perp s}\mathbf{I} + {(P_{\parallel s} - P_{\perp s})} \left[ 1 - \frac{(\mathbf{u}_s \cdot \mathbf{b})^2}{c^2} \right] \mathbf{b} \mathbf{b} 
    + \frac{P_{\parallel s} - P_{\perp s}}{c^2} (\mathbf{u}_s \cdot \mathbf{b} ) \mathbf{b} \mathbf{u}_s,
\end{align}
where we have made use of~\eqref{eq: B tilde}.

Next, we focus on calculating the Lorentz force,
\begin{equation}
    \mathbf{F}_{\rm{L}} = \rho \mathbf{E} + \frac{{\mathbf{j}} \times \mathbf{B}}{c}.
\end{equation}
Using the induction equation \eqref{eq: induction}, Gauss's law, and Ampère's law~\eqref{eq: current and charge}, we obtain
\begin{align}
    \label{eq: Lorentz Force}
     \mathbf{F}_{\rm{L}} &=  - \frac{\partial}{\partial t} \left[ \frac{B^2}{4 \pi c^2} \left( \mathbf{u}  - \mathbf{u} \cdot \mathbf{b} \mathbf{b}\right) \right] - \nabla \left[ \frac{B^2}{8 \pi\gamma_u^2} + \frac{(\mathbf{u}\cdot \mathbf{B})^2}{8 \pi c^2} \right]
     \nonumber
     \\
     &- \nabla \cdot \left \{  \frac{B^2}{4\pi c^2} \left[\mathbf{u} \mathbf{u} -\mathbf{u} \cdot \mathbf{b} (\mathbf{b} \mathbf{u} + \mathbf{u} \mathbf{b})\right] - \frac{B^2}{4\pi \gamma_u^2 } \mathbf{b} \mathbf{b}  \right \},
\end{align}
where $\mathbf{u}$ is the species-averaged flow defined in Section \ref{sec: Single-Fluid Model} and $\gamma_u = 1/\sqrt{1-u^2/c^2}$ is the associated Lorentz factor. The expression~\eqref{eq: Lorentz Force} is simply the divergence of the Maxwell stress tensor minus the time derivative of the electromagnetic momentum density, with the electric field given by the ideal-MHD condition~\eqref{eq: ideal E}. 

Inserting~\eqref{eq: momentum value},~\eqref{eq: pressure 2nd final}, and~\eqref{eq: Lorentz Force} into the momentum equation~\eqref{eq: momentum mid}, and completing the species summation, we find
\begin{gather}
    \frac{\partial}{\partial t} \left[ \frac{1}{c^2}\left(3 P_{\perp} + P_{\parallel} + \frac{B^2}{4 \pi} \right) \mathbf{u} + \frac{1}{c^2}\left(P_{\parallel} - P_{\perp} - \frac{B^2}{4\pi}\right) {\mathbf{u}} \cdot \mathbf{b} \mathbf{b}  \right]
    + 
    \nonumber \\
    \nabla \cdot \left[ \frac{1}{c^2}\left(3 P_{\perp} + P_{\parallel} + \frac{B^2}{4 \pi} \right) \mathbf{u}\mathbf{u}  
    +
    \frac{1}{c^2}\left(P_{\parallel} - P_{\perp} - \frac{B^2}{4\pi}\right) {\mathbf{u}} \cdot \mathbf{b} \left( \mathbf{b}\mathbf{u} + \mathbf{u} \mathbf{b}\right)
    \right] = 
    \nonumber \\
     - \nabla \left[ P_\perp + \frac{B^2}{8\pi \gamma_u^2} + \frac{(\mathbf{u}\cdot \mathbf{B})^2}{8 \pi c^2} \right]
     +
     \nabla \cdot \left[\left(P_{\perp} - P_{\parallel}' +  \frac{B^2}{4 \pi \gamma_u^2}\right)\mathbf{b} \mathbf{b} - \frac{P_{\perp} - P_{\parallel}}{c^2} (\mathbf{u} \cdot \mathbf{b})^2\mathbf{b} \mathbf{b} \right ] \nonumber
     \\
     - \alpha_m \frac{2 r_e^2}{ m_e^2 c^4} \frac{\mathbf{u}}{c} \frac{B^2 P_{\perp}^2}{n},
     \label{eq: A full momentum}
\end{gather}
where we have employed~\eqref{eq: drift velocity} and defined
\begin{equation}
    \label{eq: P parallel prime}
    P_\parallel' \equiv P_\parallel + \frac{P_\perp + P_\parallel}{c^2}\sum_s [(\mathbf{u} - \mathbf{u}_s) \cdot \mathbf{b}]^2.
\end{equation}
Note that as in the main text, we assume a pair plasma with equal densities and pressures, with the species-summed pressures $P_\perp = 2P_{\perp s}$ and $P_{\parallel} = 2P_{\parallel s}$. In the case of $\mathbf{u}_s \cdot \mathbf{b} = 0$,~\eqref{eq: A full momentum} reduces to~\eqref{eq: KMHD 1}. In the general case, however, the kinetic correction in~\eqref{eq: P parallel prime} means that the fluid system is not closed and the kinetic equation is required.

The calculation presented here can also be carried out by starting from the conservation of the energy-momentum tensor for relativistic, magnetised, collisionless plasmas and expanding in $u_s/c$ \citep{Gedalin_Oiberman-1995, Wierzchucka_etal-2026b}.

\section{Synchrotron Cooling Instability in Firehose-Infested Plasma}
\label{Appendix: Synchrotron Cooling Instability in Firehose-Infested Plasmas}

The simple derivation of the SCI in Section \ref{sec: Synchrotron Cooling Instability} assumed that the SFHI pinned the pressure anisotropy at the threshold~\eqref{eq: threshold} with $\Cth = 0$. Below, we generalise that derivation by allowing for a non-zero $\Cth$ and by investigating the stability of a synchrotron-cooling plasma in a firehose-infested medium described by equations~\eqref{eq: energy},~\eqref{eq: continuity},~\eqref{eq: induction},~\eqref{eq: KMHD 1}, and~\eqref{eq: threshold}.

We consider an anisotropic state with temperature $T_0$, perpendicular and parallel pressures $P_{\perp 0}$ and $P_{\parallel 0}$, total density $n = n_0$, and a uniform background magnetic field $\mathbf{B}_0 = B_0 \hat{\mathbf{z}}$. The threshold condition~\eqref{eq: threshold} also requires 
\begin{equation}
    P_0 = P_{\perp0} + \frac{C_\text{th}}{3} \frac{B_0^2}{8\pi} = P_{\parallel0} - \frac{C_\text{th}}{3} \frac{B_0^2}{4\pi}
\end{equation}
to hold. Then there exists a solution to~\eqref{eq: energy},~\eqref{eq: continuity},~\eqref{eq: induction},~\eqref{eq: KMHD 1}, and~\eqref{eq: threshold} whereby $n = n_0$, $\mathbf{B} = \mathbf{B}_0 $, $\mathbf{u} = 0$, and the perpendicular pressure satisfies 
\begin{equation}
    \label{eq: cooling equation C}
    \frac{\partial P_\perp}{\partial t} = - \frac{\alpha}{\tau_{\perp 0}} \frac{P_\perp}{P_{\perp 0}} \left( P_\perp + C_\text{th}\frac{\alpha_\parallel }{3\alpha} \frac{B_0^2}{8\pi} \right),
\end{equation}
where $\alpha = (2\alpha_\perp + \alpha_\parallel)/3$. Here we have redefined the initial synchrotron cooling time~\eqref{eq: cooling time 0} in terms of the perpendicular temperature, $T_{\perp} \equiv P_{\perp}/n$, as follows:
\begin{equation}
    \tau_{\perp 0}
    =
    \frac{m_e^2 c^3}{2 r_e^2 B_0^2 T_{\perp 0}} = \frac{\beta_0}{\beta_{\perp0}} \tau_0.
\end{equation}
Solving~\eqref{eq: cooling equation C} yields the following evolution of $P_\perp$:
\begin{equation}
    \label{eq: evolution P C}
    \frac{P_\perp(t)}{P_{\perp 0}} = \frac{1}{(1 + {3\alpha \beta_{\perp 0}}/{\alpha_\parallel C_\text{th}})e^{tC_\text{th}\alpha_\parallel/3\beta_{\perp 0} \tau_{\perp0}} - {3\alpha\beta_{\perp 0}}/{\alpha_\parallel C_\text{th}}} \equiv P'_\perp(t).
\end{equation}
Note that, in the case of vanishing $\Cth$, and so $P_{\perp 0} = P_{\parallel 0} = P_0$,~\eqref{eq: evolution P C} reduces to~\eqref{eq: isotropic homogeneous}.

\subsection{Linear Theory}
\label{sec: Linear Theory}
We now introduce infinitesimal non-uniform perturbations, considering solutions to the fluid system~\eqref{eq: energy},~\eqref{eq: continuity},~\eqref{eq: induction},~\eqref{eq: KMHD 1}, and~\eqref{eq: threshold} in the form 
\begin{equation}
    \label{eq: pert C}
    n = n_0 + \delta n, \ P_\perp = P_{\perp 0} + \delta P_\perp, \ P_\parallel = P_{\parallel 0} + \delta P_\parallel, \ \mathbf{u} = \delta\mathbf{u}, \ \mathbf{B} = B_0 \hat{\mathbf{z}} + \delta \mathbf{B}.
\end{equation}
Again, as in the main text, we first study the short-time behaviour of the perturbations by assuming that they evolve on timescales $t \ll \tau_{\perp 0}$, allowing us to neglect the evolution of the background state. We seek solutions of the form
\begin{equation}
    \label{eq: delta C}
    \delta n, \ \delta P_\perp, \ \delta P_\parallel, \ \delta \mathbf{u}, \ \delta \mathbf{B} \propto e^{\chi t/\tau_{\perp 0}+ i \mathbf{k} \cdot \mathbf{x}},
\end{equation}
where the growth rate $\chi$ is now normalised to $\tau_{\perp 0}^{-1}$. We limit ourselves to propagation perpendicular to the magnetic field, viz., $\mathbf{k} = k \hat{\mathbf{x}}$.

Combining~\eqref{eq: pert C} and~\eqref{eq: delta C} with~\eqref{eq: energy},~\eqref{eq: continuity},~\eqref{eq: induction},~\eqref{eq: KMHD 1}, and~\eqref{eq: threshold}, and linearising the resulting system, we find the dispersion relation 
\begin{gather}
    \chi^4\left\{\left[(4 \beta_{\perp 0} + 2 + \Cth)\chi + \alpha_m \beta_{\perp 0}
   \vphantom{\frac{\alpha_\parallel \Cth}{3 \beta_{\perp 0}}}\right]
   \left( \chi + 2 \alpha + \frac{ \alpha_\parallel \Cth}{3 \beta_{\perp 0}}\right)\chi
   +  \right.
   \nonumber
   \\
     \frac{k^2}{k_{\perp 0
   }^2}\left.\left[  \left(\Gamma\beta_{\perp 0} + 2 - \frac{\Cth}{3}\right) \chi - \alpha\left(\beta_{\perp 0} -4 + \frac{\alpha_\parallel C_\text{th}}{\alpha} - \frac{2\alpha_\parallel C_\text{th}}{3\alpha\beta_{\perp 0}}\right) \right] \right\} =0,
   \label{eq: disp rel C correction }
\end{gather}
where $k_{\perp 0} = 1/\tau_{\perp0}c$, $\beta_{\perp 0} = 8\pi P_{\perp 0}/B_0^2$, and, as in \eqref{eq: disp rel}, $\Gamma = 4/3$ and $\alpha = (2\alpha_\perp + \alpha_\parallel)/3$. Like in Section \ref{sec: Synchrotron Cooling Instability}, the dispersion relation has seven modes one of which is the cooling-modified entropy mode. When 
\begin{equation}
\label{eq: beta crit c}
\beta_{\perp 0} > \beta_{\perp c}= 2 - \frac{\alpha_\parallel \Cth}{2\alpha} + \sqrt{\left(\frac{\alpha_\parallel\Cth}{2\alpha}\right)^2 - \frac{4\alpha_\parallel\Cth}{3\alpha} + 4},
\end{equation}
this mode is destabilised, giving rise to the SCI. For $\Cth = 1.4$, i.e., for the value corresponding to the kinetic firehose instability  \citep{Chandrasekhar_etal-1958, Yoon_etal-1993, Hellinger_Matsumoto_2000, Gary_Nishimura_2003, Bott_etal-2024, Bott_etal-2025}, $\beta_{\perp c} \approx 3.33$, close to $\beta_{\perp c} = 4$ obtained for $\Cth = 0$ in~\eqref{eq: critical beta}. 

\begin{figure}
\centering
\includegraphics[width=0.9\linewidth]{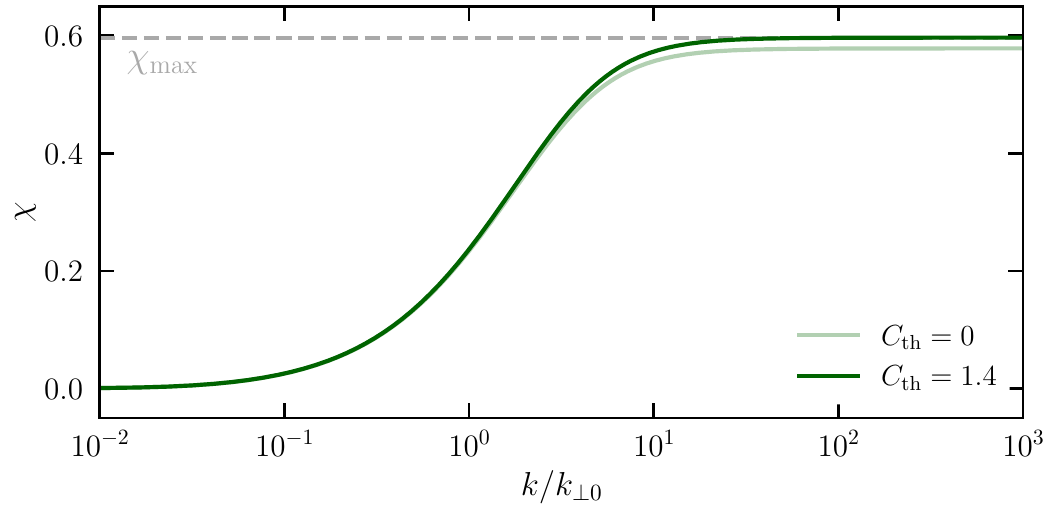}
\caption{The positive real root of~\eqref{eq: disp rel C correction }, i.e., the growth rate of the SCI, normalised to the cooling rate of the initial state $\tau_{\perp 0}^{-1}$, as a function of $k/k_{\perp 0}$, for $\Gamma = 4/3$, $\beta_{\perp 0} = 40$, $\alpha_\perp = 16/15$, $\alpha_\parallel = 8/15$ and $\alpha_m = 8/3$. The dark and light green curves correspond to $\Cth = 1.4$ and $\Cth = 0$, respectively. The two solutions are in close agreement over the entire range of $k/k_{\perp 0}$, differing only slightly at large values. For $k\gg k_{\perp 0} = 1/\tau_{\perp 0}c $, the $\Cth = 1.4$ case approaches the asymptotic value~\eqref{eq: SCI growth rate with C}, represented here by the grey dashed line.}
\label{fig: chi vs k C}
\end{figure}

The growth rate of the SCI is purely real, as shown in Figure \ref{fig: chi vs k C} where we evaluate it for both $\Cth = 0$ and $\Cth = 1.4$, increases with $k$, and at $k \gg k_{\perp 0}$ asymptotes to the maximum value 
\begin{equation}
    \label{eq: SCI growth rate with C}
    \chi_\text{max}  = {\alpha}\, \frac{ \beta_{\perp 0} -4 + \alpha_\parallel C_\text{th}/\alpha - 2\alpha_\parallel C_\text{th}/3 \alpha \beta_{\perp 0}} {2 -\Cth/3 + \Gamma \beta_{\perp 0}}.
\end{equation}
In this limit, the momentum equation~\eqref{eq: KMHD 1} again reduces to pressure balance~\eqref{eq: Pressure Balance}. When $\beta_{\perp0} \gg 1$, we find that~\eqref{eq: SCI growth rate with C} becomes $\chi_\text{max} =\alpha/\Gamma$, the same value as in the case $\Cth = 0$ [see~\eqref{eq: SCI growth rate}].

Note that in the derivation and discussion above, we assumed $\Cth < 6$, so the term $\Gamma\beta_{\perp 0} + 2 - \Cth/3$ was positive for all $\beta_{\perp 0 }$. Were this not satisfied, the criteria for the SCI would change because, when $\beta_{\perp 0}<(\Cth/3 - 2)/\Gamma$, the denominator of \eqref{eq: SCI growth rate with C} would be negative. Thus, not only would there be an instability when $\beta_{\perp 0} \gg 1$, but also at $\beta_{\perp 0} \ll 1$. We will not consider the case $\Cth>6$, as current literature suggests that $\Cth > 2$ is not physically relevant \citep{Chandrasekhar_etal-1958, Yoon_etal-1993, Hellinger_Matsumoto_2000, Gary_Nishimura_2003, Bott_etal-2024, Bott_etal-2025}.

\begin{figure}
\centering
\begin{subfigure}{\linewidth}
    \centering
\includegraphics[width=0.9\linewidth]{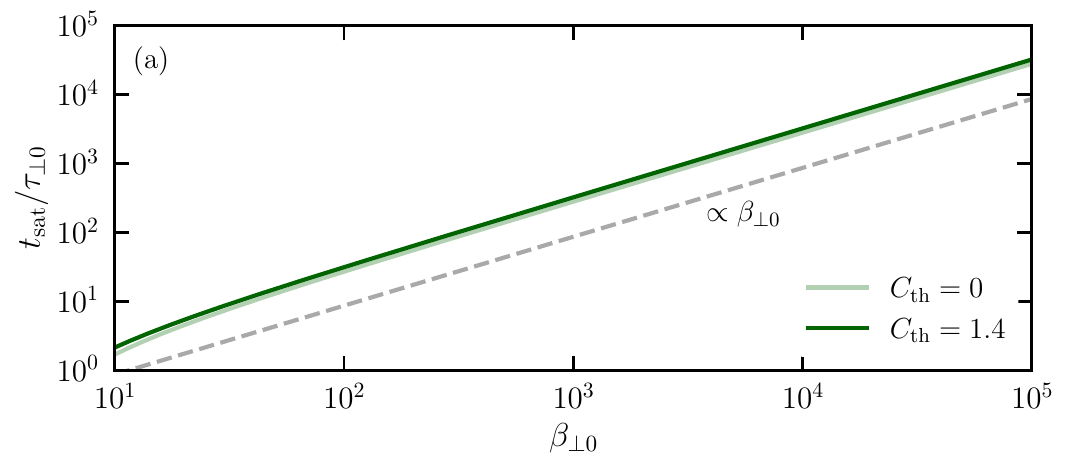}
\end{subfigure}
\\[1ex]
\begin{subfigure}{\linewidth}
    \centering
\includegraphics[width=0.9\linewidth]{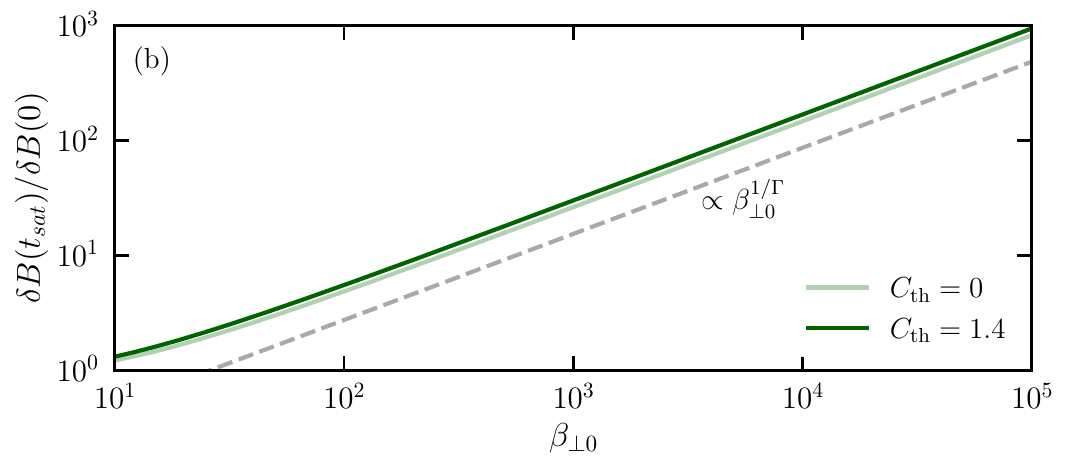}
\end{subfigure}
\caption{(a) The saturation time~\eqref{eq: saturation time} normalised by $\tau_{\perp 0}$, as a function of $\beta_{\perp 0}$. For large values of $\beta_{\perp0}$, $t_\text{sat} \propto \beta_{\perp0}$ (grey dashed line). (b) The maximum magnetic-field perturbation $\delta B(t_\text{sat})$ of the SCI, normalised to the initial value $\delta B(0)$. For $\beta_{\perp 0} \gg 1$, the saturation amplitude satisfies the simple scaling \eqref{eq: sat scaling full} with $\beta_0$ replaced by $\beta_{\perp 0}$, indicated by the grey dashed line. In both panels, we use $\alpha_\perp = 16/15$, $\alpha_\parallel = 8/15$, and $\Gamma = 4/3$. We show both quantities for $\Cth = 0$ (light green line) and $\Cth = 1.4$ (dark green line), finding that the two cases differ only slightly.}
\label{fig: saturation Cs}
\end{figure}

\subsection{Time-Evolving Linear Theory}
As $\chi \sim 1$, we must take into account the cooling of the background state, relaxing the assumption $t \ll \tau_{\perp 0}$. We again introduce linear perturbations
\begin{equation}
\delta n, \ \delta P_\perp, \ \delta P_\parallel, \ \delta \mathbf{u}, \ \delta \mathbf{B} \propto e^{i \mathbf{k} \cdot \mathbf{x} }
\end{equation}
to the homogeneous cooling state~\eqref{eq: evolution P C}, allow their amplitudes to be time-dependent, and assume that $\mathbf{k} = k \hat{\mathbf{x}}$ and $k \gg k_{\perp 0}$, thereby focusing on the fastest-growing SCI mode. This again allows us to replace~\eqref{eq: KMHD 1} with pressure balance~\eqref{eq: Pressure Balance} and find, in the same way as in Section~\ref{sec: Time-Evolving Linear Theory of SCI}, that the magnetic-field perturbation evolves according to \eqref{eq: delta B ode}, where now the instantaneous growth rate is 
\begin{equation}
    \label{eq: t B equation with C}
    g(t) = \frac{\alpha} {\tau_{\perp 0}}\frac{\beta_{\perp 0}P_\perp'^2 - (4-\alpha_\parallel C_\text{th}/\alpha)P_\perp' - 2\alpha_\parallel C_\text{th}/3\alpha\beta_{\perp 0}}{2 - \Cth/3+ \Gamma \beta_{\perp 0} P_\perp'},
\end{equation}
with $P_\perp'(t) = P_\perp/P_{\perp 0}$ given by~\eqref{eq: evolution P C}. As in Section \ref{sec: Time-Evolving Linear Theory of SCI},~\eqref{eq: delta B ode} is solved by \eqref{eq: B t evolution}, where the exponent is now
\begin{equation}
    \label{eq: t exponent C}
    \int_0^t dt' g(t') = a\ln P'_\perp + b\ln \left(\frac{\beta_{\perp 0}P_\perp' + \alpha_\parallel \Cth/3\alpha}{\beta_{\perp0}+ \alpha_\parallel \Cth/3\alpha} \right) -d\ln \left(\frac{\Gamma\beta_{\perp 0}P_\perp' + 2- \Cth/3}{\Gamma\beta_{\perp0}+ 2-\Cth/3} \right)
\end{equation}
with the coefficients
\begin{equation}
    \label{eq: a b d}
    a = \frac{2}{2-\Cth/3}, \quad b = \frac{1-\alpha_\parallel \Cth/3\alpha}{1-\Cth/6 - \Gamma \alpha_\parallel \Cth/6\alpha}, \quad d = \frac{1+ (a+b)\Gamma}{\Gamma}.
\end{equation}
Substituting~\eqref{eq: t exponent C} into~\eqref{eq: B t evolution}, we find
\begin{equation}
    \label{eq: cth delta B evolution}
    \frac{\delta B(t)}{\delta B(0)} = P^{\prime a}_\perp \left(\frac{\beta_{\perp 0}P_\perp' + \alpha_\parallel \Cth/3\alpha}{\beta_{\perp0}+ \alpha_\parallel \Cth/3\alpha} \right)^b \left(\frac{\Gamma\beta_{\perp 0}P_\perp' + 2- \Cth/3}{\Gamma\beta_{\perp0}+ 2-\Cth/3} \right)^{-d}.
\end{equation}
When $\Cth = 0$, \eqref{eq: a b d} yields $a=1$, $b=1$, $d=2+1/\Gamma$, and this expression reduces to~\eqref{eq: full B evolution}, as it should.

The sign of the instantaneous growth rate~\eqref{eq: t B equation with C} changes when $\beta_\perp = 8\pi P_\perp/B_0^2$ reaches the critical value~\eqref{eq: beta crit c}. An explicit expression for the time at which this occurs is obtained by substituting $\beta_\perp = \beta_{\perp c}$ into~\eqref{eq: evolution P C}:
\begin{equation}
\label{eq: saturation time}
t_{\rm sat}
= \frac{3 \beta_{\perp 0}\tau_{\perp 0}}{\alpha_\parallel \Cth}
\ln \left[
\frac{\beta_{\perp 0}(\alpha_\parallel\Cth + 3\alpha\beta_{\perp c})}
{\beta_{\perp c}(\alpha_\parallel\Cth + 3\alpha\beta_{\perp 0})}
\right].
\end{equation}
After this time, the perturbation decays. Similarly to the case of $\Cth = 0$, when $\beta_{\perp 0} \gg 1$, we find that $t_{\rm sat} \propto \beta_{\perp 0}$. Figure~\ref{fig: saturation Cs}a shows the saturation time~\eqref{eq: saturation time} as a function of $\beta_{\perp 0}$ for $\Cth = 0$ and $\Cth = 1.4$, demonstrating that finite $\Cth$ has only a modest effect. In Figure~\ref{fig: saturation Cs}b, we also show the saturation amplitude of the magnetic-field perturbation $\delta B(t_\text{sat})$ as a function of $\beta_{\perp 0}$, found by substituting \eqref{eq: saturation time} into~\eqref{eq: cth delta B evolution}. When $\beta_{\perp 0} \gg 1$, the saturation amplitude satisfies~\eqref{eq: saturated scaling} with $\beta_0$ replaced by~$\beta_{\perp 0}$, independently of the value of $\Cth$, as expected.

Thus, finite-$\Cth$ corrections do not significantly affect the linear SCI evolution. However, one important difference between the finite-$\Cth$ and $\Cth = 0$ cases is the evolution of $\delta B$ at late times, when $t \gg t_\text{sat}$. With $\Cth \neq 0$, the perturbation decays as
\begin{equation}
    \label{eq: long_time decay}
    \delta B(t) \propto e^{-2t\Cth\alpha_\parallel/\beta_{\perp 0}\tau_{\perp 0}(6 - \Cth)},
\end{equation}
rather than following the power-law behaviour $\delta B \propto t^{-2}$ found when $\Cth = 0$\footnote{This result holds when $t \gg 3\beta_{\perp 0}\tau_{\perp 0}/\alpha_\parallel\Cth$. In contrast, in the intermediate limit $3\beta_{\perp 0} \tau_{\perp 0}/\alpha_\parallel \Cth \gg t \gg t_\text{sat}$, which exists for sufficiently small $\Cth$, $\delta B \propto t^{-2}$ like in the case of $\Cth = 0$. Recall also that, as discussed at the end of Appendix~\ref{sec: Linear Theory}, we do not consider the case with $\Cth >6$, so \eqref{eq: long_time decay} always decays.}. However, as discussed in Section~\ref{sec: Emergent Two-Phase Structure of Synchrotron Cooling Plasmas}, the SFHI is suppressed around $t=t_{\rm sat}$ in the low-$\beta$ regions generated by the~SCI. Consequently, the single-phase theory developed here is unlikely to remain exactly applicable at such late times. 

\section{Effect of Particle Noise on PIC Simulations}
\label{Appendix: Effect of Particle Noise on PIC Simulations}
\subsection{Finite-Particle-Number Effects in PIC Simulations}
\label{Appendix: Finite-Particle-Number Effects in PIC Simulations}

In Sections~\ref{sec: Particle-in-Cell Simulations} and~\ref{sec: Comparison Between PIC and Fluid Simulations}, we noted that in the colder, low-$\beta$ regions, the increase in $|\Delta|$ due to synchrotron emission started to slow down at around $t \approx 2.2 \, \tau_0$, and $|\Delta|$ eventually started to decrease. This behaviour is a numerical artifact associated with the finite number of computational particles per Debye area, 
\begin{equation}
    \label{eq: N_D definition}
    N_\mathrm{D} = N_\text{ppc} \frac{\lambda_{\mathrm{D}e}^2}{\Delta x\Delta z},
\end{equation}
in our PIC simulations, where $\Delta x$ and $\Delta z$ are respectively the resolution in the $x$ and $z$ directions, $N_\text{ppc}$ is the number of computational particles of each species per cell, and $\lambda_{\mathrm{D}e}$ is the Debye length~\eqref{eq: lambda de}. We explain this behaviour below. 

Finite-particle effects (often characterised as particle noise) are an intrinsic feature of PIC codes and, similarly to collisions, lead to isotropisation of the distribution function even when key spatial scales like $\lambda_{\mathrm{D}e}$ are spatially resolved \citep{Langdon_Birdsall-1970, Touati_etal-2022}. In the non-relativistic limit, for electrostatic simulations, the effective frequency of numerical collisions is given by
\begin{equation}
    \label{eq: non-rel freq}
    \nu_e = \frac{\omega_{\text{pe}}\ln \Lambda}{2 \pi N_\mathrm{D}}
\end{equation}
where $\ln \Lambda$ is the PIC Coulomb logarithm and $\omega_{\text{pe}} = \sqrt{4 \pi n_e e^2/m_e}$ the non-relativistic electron plasma frequency  \citep{Langdon_Birdsall-1970, Touati_etal-2022}. Unlike in physical collisions, $\ln \Lambda$ depends on the particle-shape function and resolution of the code. To get rough estimates of how $\nu_e$ scales, we (very crudely) let $\ln \Lambda \sim 1$ and assume that \eqref{eq: non-rel freq} holds in the relativistic electromagnetic regime with $\omega_{\text{pe}}$ replaced by its relativistic counterpart, $\omega_{\text{pe,r}} = \sqrt{4 \pi n_e e^2/3\theta_e m_e}$. This gives
\begin{equation}
    \label{eq: PIC collisons}
    \nu_e \tau_e \sim \frac{\omega_\text{pe,r}}{\Omega_e}\frac{\tau_e \Omega_e}{2\pi N_\mathrm{D}} = \sqrt\frac{3\beta_e}{2} \frac{\tau_e \Omega_e}{2\pi N_\mathrm{D}}.
\end{equation}
Thus, dominance of numerical isotropisation over synchrotron cooling can be mitigated in two main ways: decreasing $\tau_e \Omega_e$ (i.e., increasing the cooling rate) or increasing $N_\mathrm{D}$. From~\eqref{eq: PIC collisons}, it is clear that at a fixed resolution, $\nu_e \tau_e \propto n^{3/2}/B^2 T^{5/2} N_\text{ppc} $. Consequently, the PIC-simulation dynamics of a species that emits synchrotron radiation will eventually be dominated by numerical effects as the plasma cools.

In the macroscale simulations of Section~\ref{sec: Particle-in-Cell Simulations}, recalling the pressure-balance and flux-freezing conditions, we find that, in the high-$\beta$ regions, $\nu_e \tau_e \propto 1/T$. Note that in the derivation of this scaling, we used the fact that, as a result of the compressing flow generated by the SCI, $N_\text{ppc} \propto n$, i.e., there are more particles per cell in regions of the plasma that have been compressed. Therefore, numerical effects will become important in the colder, low-$\beta$ regions first. In fact, while our fiducial macroscale simulations (with $N_\text{ppc} = 400$ and $N_\mathrm{D} \approx 8800$) had reasonably low $\nu_e \tau_e  = 0.20$ at $t=0$, by the simulation's end this parameter increased to $3.3$ and $9.5$  in the high- and low-$\beta$ regions, respectively. Consequently, the numerical collisionality led to artificial isotropisation and hence shut down of the SFHI in the middle of the box, as shown in Figure~\ref{fig: marginality}. The isotropisation also caused an increase in both $\alpha_\perp$ and $\alpha_\parallel$ (see Figure~\ref{fig: alpha}). However, via higher-resolution local simulations described in the next section, we shall verify that $\alpha_\perp$ and $\alpha_\parallel$ do tend to constant values.

{
\floatstyle{plain}  
\restylefloat{table}
\begin{table}[t]
\centering
\setlength{\tabcolsep}{12pt}
\begin{tabular}{cccccc}
 \hline\hline
 Run & $L_x, \, L_z $ & $N_\text{ppc}$ & $\tau_0 \Omega_0$ & $N_{\mathrm{D}}$ & $\nu_e \tau_e$\\
 \hline
\texttt{baseline} & $20 \, \rho_{e0}$ & 484 & 2000 & 33816 & 0.0408\\
\texttt{bl\_double\_size} & $40 \, \rho_{e0}$ & 484 & 2000 & 33816 & 0.0408\\
\texttt{bl\_ppc\_256} & $20 \, \rho_{e0}$ & 256 & 2000 & 17886 & 0.0771\\
\texttt{bl\_ppc\_961} & $20 \, \rho_{e0}$ & 961 & 2000 & 67143& 0.0204\\
\texttt{bl\_ppc\_1936} & $20 \, \rho_{e0}$ & 1936 & 2000 & 135264 & 0.0102\\
\texttt{bl\_tau\_omega\_500} & $20 \, \rho_{e0}$ & 484  & 500 & 33816 & 0.0102 \\
\texttt{bl\_tau\_omega\_1000} & $20 \, \rho_{e0}$ & 484  & 1000 & 33816 & 0.0204\\
\texttt{bl\_tau\_omega\_4000} & $20 \, \rho_{e0}$ & 484  & 4000 & 33816& 0.0815 \\
 \hline
\end{tabular}
\caption{Parameters of the simulations reported in Appendix~\ref{sec: Numerical Convergence Studies}, along with their number of computational particles per Debye area \eqref{eq: N_D definition} and value of $\nu_e \tau_e$ given by~\eqref{eq: PIC collisons}, at the start of the simulation. The simulations use a square domain, $L_x = L_z$.}
\label{tab: sim}
\end{table}
}

\subsection{Numerical Convergence Studies}
\label{sec: Numerical Convergence Studies}
The total computational cost of the simulation presented in Section~\ref{sec: Particle-in-Cell Simulations} was 1.6 million CPU hours on the MareNostrum 5 supercomputer. A self-consistent simulation with a value of $N_\mathrm{D}$ large enough to capture the SCI dynamics over the number of cooling times needed to see the true shut-off of the SFHI while avoiding the numerical isotropisation in the cool region, would be at least four times as expensive. As we cannot afford this, we instead verify the key conclusions reached in Section~\ref{sec: Particle-in-Cell Simulations} regarding the values of $\alpha_\perp$ and $\alpha_\parallel$ by carrying out smaller simulations of a homogeneous plasma, without any SCI modes. The values of these $\alpha$ parameters are important because they govern the macroscopic evolution of the SCI (see Section~\ref{sec: Synchrotron Cooling Instability}), but they can be established locally.

\begin{figure}
\centering
\begin{subfigure}{\linewidth}
    \centering
\includegraphics[width=\linewidth]{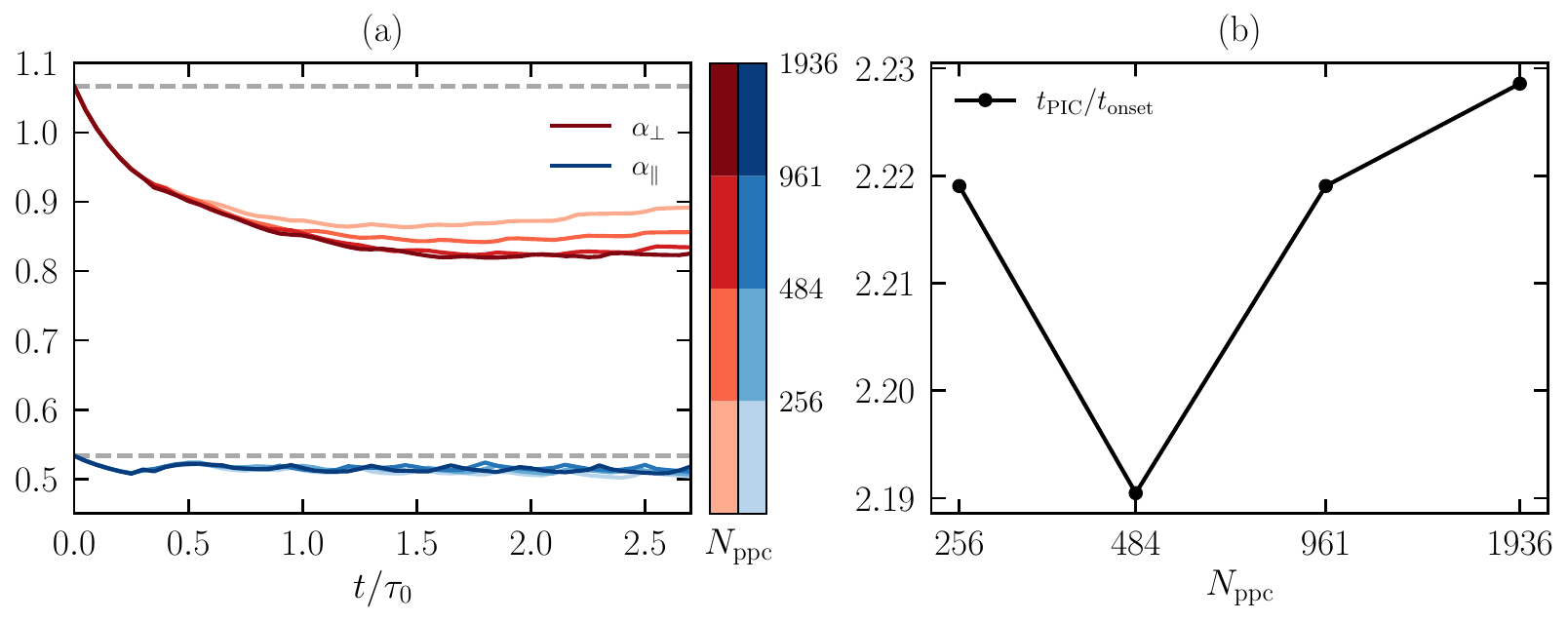}
\end{subfigure}
\\[1ex]
\begin{subfigure}{\linewidth}
    \centering
\includegraphics[width=\linewidth]{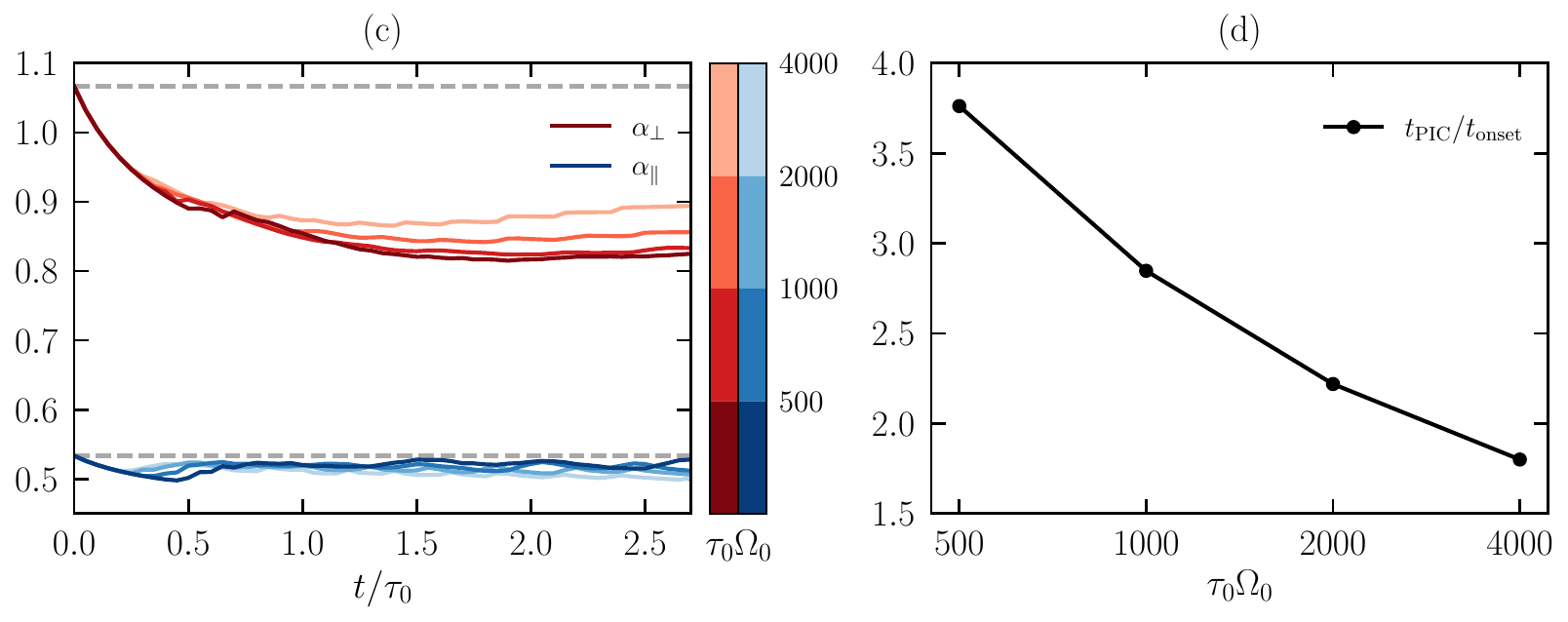}
\end{subfigure}
\caption{(a) The coefficients $\alpha_\perp$ (red lines) and $\alpha_\parallel$ (blue lines), defined in \eqref{eq: alpha def}, computed as functions of time during the PIC simulations of Appendix~\ref{sec: Numerical Convergence Studies}, with $\tau_0 \Omega_0 = 2000$. Shades of colours indicate the values of the $N_\text{ppc}$ used. The grey dashed lines correspond to the Maxwell--Jüttner values, $\alpha_\perp = 16/15$ and $\alpha_\parallel = 8/15$. (b) The ratio of the onset time of the SFHI measured in the PIC simulations and the theoretical prediction~\eqref{eq: onset time} for different values of $N_\text{ppc}$. 
(c) and (d) are the same as (a) and (b), but varying $\tau_0 \Omega_0$ (colour shade) at fixed $N_{\text{ppc}} = 484$.}
\label{fig: convergence scan}
\end{figure}

Our simulation setup is analogous to that of Section~\ref{sec: Numerical Setup} apart from a few key parameters that we list below. To reduce computational costs, we choose a lower value of $\beta_0$ than in the main simulation. The values of $B_0$ and $n_0$ are chosen so that $\beta_{e0} = \beta_{p0} = \beta_0/2 = 12.5$ and $\tau_0 \Omega_0 = 2\times 10^3$. The simulation domain has size $L_x \times L_z = 20 \, \rho_{e0} \times 20 \, \rho_{e0}$, which is sufficient to capture the fastest-growing SFHI mode at $k\rho_{e0} \approx 0.6$ \citep{Zhdankin_etal-2023} (we also ran simulations at $L_x \times L_z = 40 \, \rho_{e0} \times 40 \, \rho_{e0}$ and noticed no drastic change to the measured quantities). We use $N_x \times N_z  = 1276 \times 1232$ cells, which gives us a resolution of $\Delta x \approx \Delta z \approx 0.016 \, \rho_{e0}$, ensuring both the Larmor radius and Debye length \eqref{eq: lambda de} are well resolved throughout the simulation. To satisfy the CFL condition, we select $\Delta t = 0.007 \, \Omega_{0}^{-1}$. The duration of our simulations is $2.7 \, \tau_0$ to allow the SFHI to reach a saturated phase. All our simulations are summarised in Table~\ref{tab: sim}.

As expected, we observe the excitation of the SFHI, demonstrated by an increase in $\delta B_x$ and $\delta B_y$, at time $t \approx 0.23 \, \tau_0$, and the subsequent pinning of pressure anisotropy to the value $\Delta \approx - 1/\beta_\parallel$. Both the onset time and pinned pressure anisotropy slightly deviate from the theoretical expectations in Section~\ref{sec: Synchrotron Firehose Instability}, $t_\text{onset} = 0.11 \, \tau_0$ [see \eqref{eq: onset time}] and $\Delta = -1.4/\beta_\parallel$. This discrepancy is likely caused by the limited scale separation between the fast SFHI growth and slow cooling, quantified by the value of~$\tau_0 \Omega_0$. In order to investigate this, as well as the isotropising effect of finite $N_\mathrm{D}$ discussed in Section~\ref{sec: Fluid Simulations} and Appendix~\ref{Appendix: Finite-Particle-Number Effects in PIC Simulations}, we conducted two parameter-space scans: one varying $N_\text{ppc}$ and the other varying $\tau_0 \Omega_0$, as detailed in Table~\ref{tab: sim}. In all the simulations, we measured three quantities: the SFHI onset time $t_\text{PIC}$, $\alpha_\perp$ and~$\alpha_\parallel$. 

Our scan in $N_\text{ppc}$ is summarised in the top panels of Figure~\ref{fig: convergence scan}, where we plot $\alpha_\perp$ and $\alpha_\parallel$ as functions of time [panel (a)], as well as the measured onset time $t_\text{PIC}$ compared to the theoretical value \eqref{eq: onset time}, here equal to $t_\text{onset}  = 0.11 \, \tau_0$ [panel (b)]. We discover that when $N_\text{ppc}$ is large enough, $\alpha_\perp$ and $\alpha_\parallel$ approach constant values at late times, and these asymptotic, late-time values converge with $N_\text{ppc}$. This is in line with the discussion of Appendix~\ref{Appendix: Finite-Particle-Number Effects in PIC Simulations}, since larger values of $N_\text{ppc}$ reduce the effect of numerical collisions as per~\eqref{eq: PIC collisons}. The saturated values of $\alpha_\perp = 0.82$ and $\alpha_\parallel = 0.51$ are close to those in the macroscale simulations shown in Figure~\ref{fig: alpha} ($\alpha_\perp = 0.9$ and $\alpha_\parallel = 0.48$). We  suspect the small discrepancy is due to a higher value of initial $\nu_e\tau_e$ in the macroscale simulations, and so more substantial influence from numerical effects. We also find that the excitation of the SFHI occurs at almost the same time for all our values of~$N_\text{ppc}$, just a factor of two later than $t_\text{onset}$.

An analogous study of our $\tau_0 \Omega_0$ scan is summarised in the bottom two panels of Figure~\ref{fig: convergence scan}. While we see an eventual growth in $\alpha_\perp$ and $\alpha_\parallel$, Figure~\ref{fig: convergence scan}(c) shows that sufficiently small values of $\tau_0 \Omega_0$ inhibit this behaviour, resulting in approximate saturation of the coefficients at late times. This is not surprising given the way these two parameters enter~\eqref{eq: PIC collisons}; lower values of $\tau_0 \Omega_0$ imply stronger cooling and hence faster system evolution, which leaves numerical scattering, whose rate, in units of $\Omega_0$, is independent of the cooling strength, too little time to accumulate. Additionally, we find that increasing $\tau_0 \Omega_0$ hastens the excitation of the SFHI compared to the expected time \eqref{eq: onset time} [see Figure~\ref{fig: convergence scan}(d)].

To summarise, following a series of small local simulations with an initially homogeneous magnetic field, we found that a higher number of particles per Debye area $N_\mathrm{D}$ and lower values of $\tau_0 \Omega_0$ suppress late-time growth of $\alpha_\perp$ and $\alpha_\parallel$, thus allowing these coefficients to remain at their saturated values for longer. Ensuring that $\alpha_\perp$ and $\alpha_\parallel$ remain constant in the PIC simulations, as well as knowing their exact values is required to capture accurately the evolution of the SCI with the model described in Section~\ref{sec: Synchrotron Cooling Instability}. We also found that high values of $\tau_0\Omega_0$ result in the SFHI being excited at a time closer to the theoretical prediction~\eqref{eq: onset time}. 

\section{Details of Fluid Simulations}
\label{Appendix: Details of Fluid Simulations}

In this Appendix, we outline the method used to simulate the fluid model of two-phase, synchrotron-cooling plasmas discussed in Section \ref{sec: Emergent Two-Phase Structure of Synchrotron Cooling Plasmas}. We restrict ourselves to the setup with $\mathbf{B} = B\hat{\mathbf{z}}$, $\mathbf{u} = u_x\hat{\mathbf{x}}$, and one-dimensional variation of all quantities restricted to the $x$-axis. The radiative anisotropic fluid equations that we solve numerically, which are essentially a modified version of the equations derived in Section~\ref{sec: Relativistic, Cooling Plasmas}, in normalised form,~are:
\begin{gather}
    \label{eq: A fluid 1}
     \frac{\partial n}{\partial t} + \frac{\partial (n u_x)}{\partial x} = 0, \\
    \label{eq: A fluid 2}
    \frac{\partial B}{\partial t} + \frac{\partial (B u_x)}{\partial x} = 0, \\
    \frac{\partial}{\partial t} \left[\left( 3P_\perp + P_\parallel + {B^2}\right)u_x \right ] +  \frac{\partial}{\partial x} \left[\left( 3P_\perp + P_\parallel + {B^2}\right)u_x^2 \right ]= \nonumber
    \\-\frac{\partial}{\partial x} \left( P_\perp + \frac{B^2}{2\gamma_u^2} \right)- \frac{\alpha_m}{\tau_0\Omega_0} \frac{\beta_0}{2 \theta_0^2} \frac{u_xB^2 P_{\perp}^2}{n} 
    \label{eq: A fluid 3}, \\
    \label{eq: A fluid 4}
    P_{\perp } \frac{d}{dt} \ln \left[
    \frac{P_{\perp }}{B^{2}} \left( \frac{B^3}{n^2} \right)^{2 - \Gamma_\perp}
    \right]
    = - \frac{\alpha_\perp}{\tau_0\Omega_0}\frac{\beta_0}{2 \theta_0^2} \frac{B^2 P_{\perp}^2}{n} - \nu\left( P_\perp - P_\parallel\right), 
    \\
    \label{eq: A fluid 5}
    P_{\parallel} \frac{d}{dt} \ln \left[
    \frac{P_{\parallel} B}{n^{2}} \left( \frac{B^3}{n^2} \right)^{1 - \Gamma_\parallel}
    \right]
    = - \frac{\alpha_\parallel}{\tau_0\Omega_0}\frac{\beta_0}{2 \theta_0^2} \frac{B^2 P_{\perp} P_{\parallel}}{n} + 2\nu\left( P_\perp - P_\parallel \right),
\end{gather}
where $\beta_0$ is the plasma beta, $\tau_0$ the radiative cooling time, and $\theta_0$ the dimensionless temperature of the initial state, as defined in Section~\ref{sec: Synchrotron Firehose Instability}. Time and space are normalised, respectively, by the inverse Larmor frequency $\Omega_0^{-1}$ and the Larmor radius $\rho_{e0} = c/\Omega_0$ of the initial state. Velocities are expressed in units of~$c$, density in $n_0$, pressure in $n_0 m_e c^2$, and magnetic field in $\sqrt{4 \pi n_0m_e c^2}$. 

We set $(\Gamma_\perp, \Gamma_\parallel) = (8/5, 4/5)$ and $(\alpha_m, \alpha_\perp, \alpha_\parallel) = (8/3, 16/15, 8/15)$. The values of the latter correspond to the isotropic Maxwell--Jüttner distribution (see Section \ref{sec: Relativistic, Cooling Plasmas}). Note that we use these fixed values of $(\Gamma_\perp, \Gamma_\parallel)$, as the pressure anisotropy throughout the simulations does not reach levels at which the anisotropic corrections for these two parameters derived by \cite{Wierzchucka_etal-2026} start to have a significant effect. 

To capture the pinning of pressure anisotropy to the firehose threshold \eqref{eq: threshold} in regions unstable to the SFHI, we introduce an effective collision rate in~\eqref{eq: A fluid 4} and~\eqref{eq: A fluid 5}, given~by
\begin{equation}
    \label{eq: nu}
    \nu = \bar{\nu} \ H\left( P_\parallel - P_\perp -  C_\text{th}\frac{B^2}{2}\right),
\end{equation}
where $\bar{\nu}$ is the scattering frequency modelling the effect of the SFHI, $C_\text{th} = 1.4$ \citep{Zhdankin_etal-2023}, and $H(x)$ is the Heaviside step function. In our simulations, we set 
\begin{equation}
    \label{eq: nu x}
    \bar{\nu}(x, t) =  \tau_e^{-1/3} \Omega_e^{2/3} = \left(\frac{B}{B_0}\right)^{4/3} \left(\frac{T}{T_0}\right)^{-1/3}\tau_0^{-1/3} \Omega_0^{2/3},
\end{equation}
which is the instantaneous scattering frequency of fully developed, nonlinear SFHI turbulence \citep{Zhdankin_etal-2023}. However, we find that any scattering frequency substantially higher than the cooling rate $\tau_e^{-1}$, but such that $\bar{\nu}^{-1} \gg \Delta t$ (the time step), yields similar results. Prescription \eqref{eq: nu}--\eqref{eq: nu x} ensures that the dominant contribution to~\eqref{eq: A fluid 4} and~\eqref{eq: A fluid 5} on the unstable side of the firehose threshold is the scattering term, and so the $P_\perp$ and $P_\parallel$ evolution equations reduce to the threshold condition~\eqref{eq: threshold}. The particular form of the scattering terms in~\eqref{eq: A fluid 4} and~\eqref{eq: A fluid 5} ensures that the total energy is conserved as per~\eqref{eq: energy}. Thus,~\eqref{eq: A fluid 4} and~\eqref{eq: A fluid 5} are equivalent to applying~\eqref{eq: energy} and~\eqref{eq: threshold} in firehose-unstable regions, while evolving~\eqref{eq: pressure evolv perp} and~\eqref{eq: pressure evolv parallel} in SFHI-free regions. Consequently, the system~\eqref{eq: A fluid 1}-\eqref{eq: A fluid 5} corresponds to the fluid model~\eqref{eq: continuity},~\eqref{eq: induction},~\eqref{eq: KMHD 1},~\eqref{eq: pressure evolv perp}, and~\eqref{eq: pressure evolv parallel}, supplemented by scattering-induced modifications to the pressure-evolution equations. Note that this formalism does not impose the two-phase structure of the SCI by itself, but allows it to emerge naturally once cooling can no longer drive anisotropy rapidly enough to excite the SFHI, as discussed in Section~\ref{sec: Shut-Off of the Synchrotron Firehose Instability}.

To match the delayed onset of the SFHI in our PIC simulation, which arises due to a finite value of $\tau_0 \Omega_0$, we delay the initial excitation of the SFHI until the onset time of the SFHI, as measured in the PIC simulations, i.e., we turn off the scattering term in \eqref{eq: A fluid 4} and \eqref{eq: A fluid 5} until $t_\text{PIC}  = 0.17 \, \tau_0 $. Additionally, after the SFHI onset, we change the values of $\alpha_\perp$ and $\alpha_\parallel$ to $0.9$ and $0.48$, respectively, as found in our PIC simulations (see Figure~\ref{fig: alpha}). 

As discussed in Sections~\ref{sec: Particle-in-Cell Simulations} and \ref{sec: Emergent Two-Phase Structure of Synchrotron Cooling Plasmas}, we initialise a stationary fluid with $\theta_0 = 100$, $\beta_0 = 40$, and $\tau_0 \Omega_{0} = 2 \times 10^3$. These parameters correspond to $P_0 = \theta_0 = 100$, $B_0 = \sqrt{2 \theta_0 /\beta_0} = \sqrt{\sigma_{c0}}= \sqrt{5}$ and, naturally as a result of our normalisation, $n_0 = 1$. Our box has length $L_x = 300 $ and periodic boundaries. We include an initial perturbation of the magnetic field and density $\delta B_z/B_0 = \delta n/n_0 = -0.3\cos(kx)$, where $k = 2\pi/L_x$, and a corresponding pressure perturbation
\begin{equation}
\frac{\delta P}{P_0} = \frac{1}{\beta_0} \left[ 1 - \left(\frac{B_z}{B_0}\right)^2 \right]
\end{equation}
to ensure initial total pressure balance. This set of parameters guarantees that at $t=0$, the state of our fluid matches that of the PIC simulations in Section \ref{sec: Particle-in-Cell Simulations}. We use a grid of $3600$ points, so that $\Delta x = 0.08 $. Our simulations have a CFL number of $C = c\Delta t/\Delta x = 0.065$, but we have also conducted runs with $C = 0.325$, which ended up producing near identical results. 

To solve the fluid equations~\eqref{eq: A fluid 1}–\eqref{eq: A fluid 5} numerically, we use a two-step Lax–Wendroff method, with the cooling and scattering terms in~\eqref{eq: A fluid 4} and~\eqref{eq: A fluid 5} advanced at half time steps. As expected, we find the inertial and the cooling terms in~\eqref{eq: A fluid 3} to be negligible. We run the fluid simulation until $t = 20 \, \tau_0$, much longer than the PIC simulations of Section~\ref{sec: Particle-in-Cell Simulations}, beyond which $T_\perp/m_e c^2 \approx 2$ and the full equations outside the ultra-relativistic limit would be required \citep{Wierzchucka_etal-2026, Ley_etal-2026}. 

\bibliographystyle{jpp}
\bibliography{references}

\end{document}